\documentclass[11pt]{article}
\usepackage{latexsym,amsthm,amssymb,amsfonts,amsbsy,calrsfs,multibox,wasysym,stmaryrd}

\usepackage[dvips]{graphicx,epsfig}

\csname @addtoreset\endcsname{equation}{section}

\def\mso{\mathfrak{so}}
\def\miso{\mathfrak{iso}}

\def\msl{\mathfrak{sl}}
\def\mgl{\mathfrak{gl}}
\def\msp{\mathfrak{sp}}

\def\musp{\mathfrak{usp}}

\def\mg{\mathfrak{g}}\def\ms{\mathfrak{s}}
\def\mm{\mathfrak{m}}
\def\mh{\mathfrak{h}}
\def\mD{\mathfrak{D}}
\def\mR{\mathfrak{R}}

\def\mC{\mathfrak{C}}

\def\mI{\mathfrak{I}}

\def\mS{\mathfrak{S}}
\def\mT{\mathfrak{T}}

\def\mpe{\mathfrak{p}}
\def\mk{\mathfrak{k}}
\def\ml{\mathfrak{l}}

\def\Real{{\mathbb R}}
\def\Comp{{\mathbb C}}
\def\integ{{\mathbb Z}}
\def\1{1\hspace{-4pt}1}
\def\j1{\widetilde{1\hspace{-4pt}1}}
\def\bec{\begin{center}}
\def\ec{\end{center}}
\def\a{\alpha}

\def\b{\beta}  
\def\c{\gamma} 

\def\d{\delta} 
\def\D{\Delta}
\def\e{\epsilon}

\def\k{\kappa}
\def\l{\lambda}
\def\L{\Lambda}

\def\n{\nu}
\def\r{\rho}
\def\s{\sigma}
\def\S{\Sigma}
\def\t{\tau}
\def\th{\theta} 
\def\Th{\Theta}
\def\x{\xi}

\def\O{\Omega}
\def\o{\omega}

\def\nn{\nonumber}
\newcommand{\eq}[1]{(\ref{#1})}

\def\be{\begin{equation}}
\def\ee{\end{equation}}
\def\bea{\begin{eqnarray}}
\def\eea{\end{eqnarray}}
\def\ba{\begin{array}}
\def\ea{\end{array}}

\def\mx#1#2#3#4{\left#1\begin{array}{#2} #3 \end{array}\right#4}

\def\ft#1#2{{\textstyle{{\scriptstyle #1}
\over {\scriptstyle #2}}}}

\def\ket#1{|#1\rangle}
\def\bra#1{\langle#1|}

\usepackage{calc} 

\newsavebox{\carre}
\savebox{\carre}(3,3)[bl]{ \put(0,0){\line(0,1){3}}
\put(0,0){\line(1,0){3}} \put(3,3){\line(0,-1){3}}
\put(3,3){\line(-1,0){3}} }

\newsavebox{\carrebold}
\savebox{\carrebold}(3,3)[bl]{ \put(0,0){\line(0,1){3}}
\put(0,0){\line(1,0){3}} \put(3,3){\line(0,-1){3}}
\put(3,3){\line(-1,0){3}} \put(0.1,0.1){\line(0,1){2.8}}
\put(0.1,0.1){\line(1,0){2.8}} \put(2.9,2.9){\line(0,-1){2.8}}
\put(2.9,2.9){\line(-1,0){2.8}} \put(0.2,0.2){\line(0,1){2.6}}
\put(0.2,0.2){\line(1,0){2.6}} \put(2.8,2.8){\line(0,-1){2.6}}
\put(2.8,2.8){\line(-1,0){2.6} \put(-2.7,-2.7){$\blacksquare$}} }

\newcounter{long}

\newcommand{\YoungU}[1]{
    \setcounter{long}{#1}
    \setcounter{long}{\value{long}*3}
    \begin{picture}(\value{long},3)
        \multiput(0,0)(3,0){#1}{\usebox{\carre}}
    \end{picture}
}

\newcommand{\YoungD}[2]{
    \setcounter{long}{#1}
    \setcounter{long}{\value{long}*3}
    \begin{picture}(\value{long},6)
        \multiput(0,0)(3,0){#1}{\usebox{\carre}}
        \multiput(0,-3)(3,0){#2}{\usebox{\carre}}
    \end{picture}
}

\newcommand{\YoungUdash}[1]{
    \setcounter{long}{#1}
    \setcounter{long}{\value{long}*3}
    \begin{picture}(\value{long},3)
        \multiput(0,0)(3,0){#1}{\multiput(0,0)(.5,0){6}{\line(1,0){.3}}\multiput(0,0)(0,0.5){6}{\line(0,1){.3}}
        \multiput(3,3)(-.5,0){6}{\line(-1,0){.3}}\multiput(3,3)(0,-0.5){6}{\line(0,-1){.3}}}
    \end{picture}
}

\newcommand{\YoungDdash}[2]{
    \setcounter{long}{#1}
    \setcounter{long}{\value{long}*3}
    \begin{picture}(\value{long},6)
        \multiput(0,0)(3,0){#1}{\multiput(0,0)(.5,0){6}{\line(1,0){.3}}\multiput(0,0)(0,0.5){6}{\line(0,1){.3}}
        \multiput(3,3)(-.5,0){6}{\line(-1,0){.3}}\multiput(3,3)(0,-0.5){6}{\line(0,-1){.3}}}
        \multiput(0,-3)(3,0){#2}{\multiput(0,0)(.5,0){6}{\line(1,0){.3}}\multiput(0,0)(0,0.5){6}{\line(0,1){.3}}
        \multiput(3,3)(-.5,0){6}{\line(-1,0){.3}}\multiput(3,3)(0,-0.5){6}{\line(0,-1){.3}}}
    \end{picture}
}

\begin{document}

\begin{center}

\vspace*{30pt}

{\huge\sc Unfolding Mixed-Symmetry Fields in AdS\\and the BMV Conjecture: \\[15pt] II. Oscillator Realization}

\vspace{20pt}

{\Large Nicolas Boulanger\footnote{Work
supported by a ``Progetto Italia'' fellowship. F.R.S.-FNRS
associate researcher on leave from the {\it{Service de M\'ecanique
et Gravitation, Universit\'e de Mons-Hainaut, Belgium.}}}, Carlo
Iazeolla\footnote{Previous address:
{\it{Dipartimento di Fisica, Universit\`{a} di Roma ``Tor
Vergata'' and INFN, Sezione di Roma ``Tor Vergata'', Via della
Ricerca Scientifica 1, 00133 Roma, Italy}}} and Per Sundell}
\\[10pt]
{\it Scuola Normale Superiore\\
Piazza dei Cavalieri 7, 56126 Pisa, Italy} \vspace{15pt}


 {\sc\large Abstract}\end{center}

Following the general formalism presented in \texttt{0812.3615} --- referred to as Paper I --- we derive the unfolded equations of motion for tensor fields of arbitrary shape and mass in constantly curved backgrounds by radial reduction of Skvortsov's equations in one higher dimension. The complete unfolded system is embedded into a single master field, valued in a tensorial Schur module realized equivalently via either bosonic (symmetric basis) or fermionic (anti-symmetric basis) vector oscillators. At critical masses the reduced Weyl zero$\,$-form modules become indecomposable. We explicitly project the latter onto the submodules carrying Metsaev's massless representations. The remainder of the reduced system contains a set of St\"uckelberg fields and dynamical potentials that leads to a smooth flat limit in accordance with the Brink--Metsaev--Vasiliev (BMV) conjecture. In the unitary massless cases in $AdS$, we identify the Alkalaev--Shaynkman--Vasiliev frame-like potentials and explicitly disentangle their unfolded field equations.\\[10pt]
\phantom{aaa}
\setcounter{page}{1}

\pagebreak
\tableofcontents

\newpage

\section{\sc \large Introduction}\label{sec:Into}

In a companion paper \cite{Boulanger:2008up} referred to as Paper I, 
we introduced some
general tools and notation adapted to the unfolded description of
tensor fields propagating in constantly curved backgrounds. Such an
analysis was initiated by Alkalaev, Shaynkman and Vasiliev (ASV), who proposed an action 
in frame-like formalism for mixed-symmetry fields in $(A)dS_D$ 
spacetimes~\cite{Alkalaev:2003qv}, and was recently
performed in the case of
Minkowski spacetime by Skvortsov~\cite{Skvortsov:2008vs}, who provided the corresponding 
unfolded field equations.
In the present Paper II, we use these tools together with an
oscillator formulation of Schur modules in order to effectively
write down unfolded field equations for arbitrary tensor fields
freely propagating in $AdS_D$ spacetime. Metric-like and partially
gauge-fixed equations have previously been given by Metsaev
in~\cite{Metsaev:1995re,Metsaev:1997nj}. For some recent works on
mixed-symmetry fields in $AdS\,$,
see~\cite{Fotopoulos:2008ka,Zinoviev:2008ve,Buchbinder:2008kw,Reshetnyak:2008sf}
and references therein.

As in Paper I, we use the unfolded
formalism~\cite{Vasiliev:1988xc,Vasiliev:1988sa,Vasiliev:1992gr}
whereby the concepts of spacetime, dynamics and observables are
derived from free differential
algebras~\cite{Sullivan,D'Auria:1982nx,D'Auria:1982pm,vanNieuwenhuizen:1982zf}.
The key features are that (i) equations of motion, Bianchi identities as well as definitions of auxiliary fields are encoded into flatness conditions on complete sets of generalized curvatures, including in general an infinite
set of zero$\,$-forms called \emph{Weyl zero$\,$-forms}; (ii) the diffeomorphism invariance is manifest (this symmetry is then broken spontaneously by given solutions); and (iii) the gauge invariance is ensured by consistency conditions on the coupling constants in the curvatures that can be solved using algebraic techniques for deformations of associative algebras (including Lie algebras) and their representations. This powerful framework is instrumental in controlling the field content and symmetries of higher-spin gauge theory and underlies Vasiliev's fully nonlinear field equations for totally symmetric gauge fields~\cite{Vasiliev:1990en,Vasiliev:2003ev}. It is therefore likely to be helpful also in addressing the challenging issue of interacting mixed-symmetry gauge fields.

Tensor fields of mixed symmetry exhibit, already at the free level, peculiarities that are absent in the ``rectangular'' case, including symmetric tensor fields and ordinary $p$-forms.
Such fields must be considered in flat spacetime as soon as
$D\geqslant 6$ and in constantly curved spacetime as soon as
$D\geqslant 4$ (in accordance with the analysis done
in~\cite{Brink:2000ag}, unitary massless mixed-symmetry two-row
tensor fields in $AdS_4$ decompose in the flat limit into
topological dittos plus one massless field in $\mathbb{R}^{1,3}\,$).

As far as free tensor gauge fields in flat spacetime of dimension
$D\geqslant 4\,$ are concerned, a Lagrangian formulation was
proposed some time ago by Labastida \cite{Labastida:1987kw}. That
the corresponding equations of motion indeed propagate the proper
massless degrees of freedom was understood later
\cite{Bekaert:2003zq} --- see~\cite{Bekaert:2006ix} for a review and
references. The proof of the propagation of the proper massless
physical degrees of freedom crucially relies on the properties of
the generalized curvature $K$ and its traceless part, defined
in~\cite{Bekaert:2002dt}. The local wave equation proposed
in~\cite{Bekaert:2002dt,Bekaert:2003az} for an arbitrary tensor
gauge field in flat spacetime can be seen as the generalization to
arbitrary dimensions of the Bargmann--Wigner
equation~\cite{Bargmann:1948ck} proposed in $D=4\,$, and were
therefore called ``generalized Bargmann--Wigner equations''
in~\cite{Bekaert:2003az} and henceforth. 
Let us finally mention that a trace-unconstrained version
of Labastida's formulation has appeared in \cite{Campoleoni:2008jq}, 
though we shall not make direct contact with this off-shell formulation here.

In Paper I, we reviewed and extended the generalized Bargmann--Wigner equations to constantly
curved spacetimes, translating them into the unfolding language
which facilitates their integration whereupon $p\,$-form variables
arise that generalize the vielbein and Lorentz-connection of spin-2
theory. The results, that complete the analysis of the pioneering
work~\cite{Alkalaev:2003qv}, are given here, comprising the complete
infinite-dimensional Weyl zero$\,$-form module as well as the
finite-dimensional $p\,$-form modules. 

As we mentioned previously, the unfolded presentation of
Labastida's formalism was given recently by
Skvortsov~\cite{Skvortsov:2008vs,Skvortsov:2008sh} and results in a
system consisting of $p\,$-forms ($p\geqslant 0$) that are traceless
Lorentz tensors of various symmetry types determined by the Young
diagram of the massless metric-like field. The $p\,$-forms with
fixed $p$ constitute on-shell $\miso(1,D-1)$-modules that are
finite-dimensional for $p>0$ and infinite-dimensional for $p=0\,$
--- the aforementioned Weyl zero$\,$-form module. The first-order
action~\cite{Skvortsov:2008sh} directly generalizes Vasiliev's
first-order action~\cite{Vasiliev:1980as} for Fronsdal fields in
flat space~\cite{Fronsdal:1978rb} to arbitrarily-shaped gauge
fields. In the present paper we review and reformulate Skvortsov's
unfolded equations in terms of master fields taking their values in
generalized Schur modules realized explicitly using oscillators and
Fock spaces. We use this reformulation in order to extend
Skvortsov's formulation to $AdS_D\,$, thereby making contact with
the equations and the $p$-form module proposed by ASV in
\cite{Alkalaev:2003qv,Alkalaev:2005kw,Alkalaev:2006rw}, see also
\cite{Alkalaev:2003hc}.

The present analysis in $AdS_D$ allows us to unfold a conjecture due to 
Brink, Metsaev and Vasiliev. The
BMV conjecture~\cite{Brink:2000ag} anticipates a field-theoretic
realization of an $AdS$ mixed-symmetry gauge field with shape
$\Th\,$, $\varphi(\L;\Th)\,$, in terms of an ``unbroken'' gauge
field plus a set of St\"uckelberg fields
$\left\{\chi(\L;\Th')\right\}$ that break the gauge symmetries
associated with all blocks but one, in such a way that the combined
system has a smooth flat limit --- in the sense that the number of
local degrees of freedom is conserved --- given by the direct sum
$\varphi(\L\!\!=\!0;\Th)\oplus\bigoplus_{\Th'}\chi(\L\!\!=\!0;\Th')$
of irreducible gauge fields in $\Real^{1,D-1}$. More precisely, the
set $\{\Th'\}$ should be given by the reduction of the
$\mso(D-1)$-tensor of shape $\Th$ under $\mso(D-2)$ subject to the
condition that one block, the one associated with the leftover gauge
invariance, must remain untouched. In the unitary case, 
that block must be the uppermost one.

The partially massive nature of mixed-symmetry gauge fields in
$AdS_D$~\cite{Metsaev:1995re,Metsaev:1997nj} and the dimensional
reduction leading to $\{\Th'\}$ suggest that the St\"uckelberg
fields can be incorporated explicitly via a suitable radial
reduction of an unbroken gauge field in $(D+1)$-dimensional flat
ambient space with signature $(2,D-1)\,$. In this paper, we carry
out this procedure using the unfolded language, which is readily adapted to dimensional reductions as ``world'' and ``fiber'' indices are treated separately from the outset.
We stress that our treatment accommodates any combinations of ambient and tangent space signatures, and that the radial reduction allows for arbitrary values of the mass parameter, introduced by constraining the radial derivatives of all $p$-forms in the unfolded system (see eq. \eq{constr1}). In particular, the reduction allows for general ``critical masses'' (see items (i)-(iv) in Section I.4.3.4), though we shall focus mainly on the case of Metsaev's massless fields in $AdS_D$, leaving a number of details in other special cases for future work.

The paper is organized as follows: The general formalism underlying
the analysis in this paper is contained in Paper I. 
(We recall some of our notation in Appendix
\ref{App:0}.) In Section \ref{Sec:Flat} we review Skvortsov's
unfolded equations in $\Real^{1,D-1}$ and then cast them into a
master-field form suitable for radial reduction using oscillator
realizations of Young diagrams. Finally, in Section \ref{Sec:AdS} we
derive the unfolded equations for general tensor fields in $AdS_D$,
analyze critical limits for the mass parameter and show the
resulting smoothness of the flat limit in accordance with the BMV
conjecture~\cite{Brink:2000ag}. In particular, see equations \eq{0-form1} and \eq{0-form2} for the zero-forms. The appropriate projection to Metsaev's critical cases  is given in Eq. \eq{construu}, and the corresponding value of the mass parameter in Eq. \eq{claim}. Finally, the unfolded equations for the unitary  ASV potential are \eq{ASVeq}. Our conclusions and an outlook are
presented in Section \ref{Sec:Conclusions}. Appendix \ref{App:A}
contains a review of Howe duality in the context of classical Lie
algebras. Appendix \ref{App:reduc} details the radial reduction of
the background fields in $\Real^{2,D-1}$. Appendix \ref{App:q+g=3}
lists shapes occurring in the computation of the
$\sigma^-$-cohomology groups for ASV potentials with $h_1=1\,$. (We
note that the general sigma-minus construction was introduced in
\cite{Shaynkman:2000ts}.) Appendix \ref{App:Singleton} shows that
some $AdS_D$-massless lowest-weight unitary representations may
arise in tensor products of $P$ bosonic singletons only if $P=2\,$.
Besides, the Metsaev's mixed-symmetry that may appear have at most
six blocks, the first of height one, and are therefore associated
with a one-form ASV potential.
%
\section{\sc \large Tensor Gauge Fields in Flat Spacetime}
\label{Sec:Flat}
%
In this Section we first review Skvortsov's unfolded formalism for
free tensor gauge fields in flat spacetime
\cite{Skvortsov:2008vs,Skvortsov:2008sh}. We then cast them into a
compact master-field form using an oscillator realization of Young
tableaux.

\subsection{\sc{Skvortsov's unfolded equations}}\label{Sec:Skvortsov}

The unfolding in $D$-dimensional Minkowski spacetime of an on-shell
tensor gauge field $\varphi(\Th)$ sitting in the $\mm$-type
\footnote{In the following, we shall frequently suppress the labels
$s_{_0}$, $s_{_{B+1}}$, $h_{_0}$ and $h_{_{B+1}}$, in the
presentation of Young diagrams associated to dynamical fields.}
\bea
\Th&=&\Big([s_{_0};h_{_0}],[s_{_1};h_{_1}],\dots,[s_{_B};h_{_B}],[s_{_{B+1}};
h_{_{B+1}}]\Big)\ ,\\[5pt]
s_{_0}&:=&\infty\ >\ s_{_1}\ >\ \cdots\ >\ s_{_B}\ >\ s_{_{B+1}}\ :=0\ ,\\[5pt] h_{0}&:=0&\ ,\quad h_{_1}\ \geqslant\ 1\ ,\quad
h_2\ \geqslant\ 1\ ,\ \dots\ ,\ h_{_{B+1}}\ :=\ \infty\eea
results in a triangular $\mg_{_0}$-module
$\mT(\Th)=\bigoplus_{q\in\integ}\mR_q(\Th)$ with indecomposable
structure
\bea \mR_q|_{\mg_{_0}}&= & \mR_q^{{\bf p}_B+{\bf q}}\supsetplus
\mR_q^{{\bf p}_{_{B-1}}+{\bf
q}}\supsetplus\cdots\supsetplus\mR_q^{{\bf h}_1+{\bf q}}
\supsetplus\mR_q^{\bf q}\ , \eea
where $p_{_I}=\sum_{J=1}^{I} h_{_J}\,$, ($I=1,\dots,B\,$),
$p_{_0}:=0\,$. The submodules are given by
\bea \mR_q^{{\bf p}_{I}+{\bf q}}&=&\O^{{\bf p}_{I}+{\bf
q}}(U)\otimes {\cal T}^-_{(p_I+1)} (\Th^-_{[p_I]})\ ,
\\[5pt]
\Th^-_{[p_I]}&=& \Big(\underbrace{[s_{_1}-1;h_{_1}], \dots ,
[s_{_I}-1;h_{_I}]}_{\tiny \mbox{cut one column}},
\underbrace{[s_{_{I+1}};h_{_{I+1}}+1]}_{\tiny\mbox{add one
row}},[s_{_{I+2}};h_{_{I+2}}], \dots,[s_{_B};h_{_B}]\Big)\
,\label{Thetaminus}\eea
which vanishes trivially if $p_{_I}+q<0\,$. For $I>0$ the
submodules are finite-dimensional and one has
\bea I\geqslant 1&:& {\cal T}^-_{(p_I+1)}(\Th^-_{[p_I]})\ \cong \
{\cal T}^+_{(p_I+1)}
(\Th^+_{[p_I]})\ ,\\[5pt]&&
\Th^+_{[p_I]}\ =\ \Big(\underbrace{[s_{_1}-1;h_{_1}],\dots,
[s_{_{I-1}}-1;h_{_{I-1}}],
\underbrace{[s_{_{I}}-1;h_{_I}+1]}_{\tiny\mbox{add one
row}}}_{\tiny\mbox{cut one column}},[s_{_{I+1}};h_{_{I+1}}],\dots,
[s_{_B};h_{_B}]\Big)\ .\label{Thetaplus}\eea
For $I=0$ the submodule is infinite-dimensional and defines the
twisted-adjoint $\mg_{_0}$-module
\bea I=0&:&\mR^{\bf q}_q\ :=\ \O^{\bf q}(U) \otimes {\cal
T}(\L\!\!=\!0;M^2\!\!=\!0;\overline\Th)\ , \quad {\cal
T}(\L\!\!=\!0;M^2\!\!=\!0;\overline\Th)\ :=\ {\cal
T}^{-}_{(1)}(\Th^-_{[p_0]})\ ,
\\[5pt]
&&\overline\Th\ =\ \Th^-_{[p_0]}\ =\
\Big([s_{_1};h_{_1}+1],[s_{_2};h_{_2}],\dots,
[s_{_B};h_{_B}]\Big)\ .\eea
Upon defining
\bea s_{_{B+1}}\ :=0\ \ ,\quad s_{_0}\ :=\ \infty\ ,\quad
s_{_{I,J}}\ :=\ s_{_I}-s_{_J}\ , \quad \a\ := \ k_{_I}+s_{_{I+1,1}}\ , \quad
k_{_I}\ \in \ \{0,\dots,s_{_{I,I+1}}-1\}\quad\eea
one has
\bea \left.\mR_q\right.|_{\mm}&=& \bigoplus_{\a=-s_{_1}}^\infty
\O^{[{\bf p}_\a+{\bf q}]}(U)
\otimes \Th_{[p_\a];\a}\ ,\\[5pt]
\Th_{[p_{_I}];\a}&=&\Big(\underbrace{[s_{_1}-1;h_{_1}],
\dots,[s_{_I}-1;h_{_I}]}_{\tiny\mbox{cut one column}},
\underbrace{[s_{_{I+1}}+k_I;1]}_{\tiny\mbox{insert one
row}},[s_{_{I+1}};h_{_{I+1}}], \dots,[s_{_B};h_{_B}]\Big)\ ,
\label{Y0k}\eea
that is, for fixed $I\in\{0,\dots,B\}$, the set
$\left\{\Th_{[p_{_I}];\a}\right\}$ is obtained from $\Th$ by first
deleting one column from each of the first $I$ blocks of $\Th$ and
then inserting one extra row of variable length between the $I$th
and $(I+1)$th blocks in compliance with row order (with
$s_{_{B+1}}:=0$ and $s_{_0}:=\infty\,$). In particular, the form
of highest degree ${\bf p}_B$ sits in the smallest Lorentz type
$\widetilde \Th:=\Th_{[{\bf p}_B];-s_1}$ given by $\Th$ minus its first
column, and the smallest zero$\,$-form is the primary Weyl tensor
sitting in $\overline \Th=: \Th_{[0];0}$ given by $\Th$ plus one
extra first row of length $s_{_1}\,$. The global
$\mathbb{N}$-grading of $\mR_{_0}$ is given by the one-to-one map
$g:\mR\mapsto \mathbb{N}$ defined by
$g(\Th_{[p_I];\a})=\a+s_{_1}\,$. It has the property that if
$g(\Th_\a)>g(\Th_\b)$ then ${p}_\a\leqslant {p}_\b$ and
$|\Th_\a|>|\Th_\b|\,$.

The representation of $\mg_{_0}$ in $\mR_q$ takes the form
\bea \rho_q&=& \left[\ba{ccccccc} (\r_q)^{{\bf p}_B+{\bf q}}_{{\bf p}_B+{\bf q}} &(\rho_q)^{{\bf p}_B+{\bf q}}_{{\bf p}_{_{B-1}}+{\bf
q}}(e)&0&\cdots\cdots\cdots&\!\!\!\!\!\!\!\!\!\!\!\!
\cdots\cdots\cdots&&\\[5pt]
0 &(\r_q)^{{\bf p}_{_{B-1}}+{\bf q}}_{{\bf p}_{_{B-1}}+{\bf
q}}&(\rho_q)^{{\bf p}_{_{B-1}}+{\bf q}}_{{\bf p}_{_{B-2}}+{\bf q}}
(e)&0&\cdots\cdots&&
\\[0pt] \ba{c}\vdots\\[-10pt]\vdots\\[-10pt]\vdots\ea&0&
(\r_q)^{{\bf p}_{_{B-2}}+{\bf q}}_{{\bf p}_{_{B-2}}+{\bf
q}}&(\rho_q)^{{\bf p}_{_{B-2}}+{\bf q}}_{{\bf p}_{_{B-3}}+{\bf q}}
(e)&0&\cdots\cdots&\\[5pt]
&\ba{c}\vdots\\[-10pt]\vdots\ea&\ddots\qquad\quad\ddots&\qquad\qquad
\ddots&&\qquad\ddots&\ea\right]\ , \hspace{1cm}\eea
where the diagonal blocks are $e^a$-independent representations on
submodules and the off-diagonal blocks are $e^a$-dependent
Chevalley-Eilenberg cocycles that are activated by the
translations $P_a\in \miso(1,D-1)$ and depend on $q$ via phase
factors. The representations of $P_a$ within the submodules read
($\a\equiv k_{_I}+s_{_{I+1,1}}$,
$k_{_I}=0,\dots,s_{_{I,I+1}}-1\,$)
\bea \xi^a \left[(\r_q)^{{\bf p}_I+{\bf q}}_{{\bf p}_I+{\bf
q}}(P_a)\right]^{\a}_{\b} X_{q}^{{\bf p}_I+{\bf
q}}(\Th^{\ast\b}_{[p_I]})&=&\left\{\ba{ll} \xi_{(p_I+1)}
X_{q}^{{\bf p}_I+{\bf q}}
(\Th^{\ast\,\a+1}_{[p_I]})&\mbox{if $k_{_I}<s_{_{I,I+1}}-1$}\\[5pt]
0&\mbox{if $k_{_I}=s_{_{I,I+1}}-1$}\ea\right.\ ,\eea
where $\xi_{(i)}$ denotes the operation of contracting $\xi^a$ into
the $i$th row of a tensor followed by Young projection onto the
shape with one less cell in that row. In terms of this operation,
the Chevalley-Eilenberg cocycles have representation matrices
($\a\equiv k_{_I}+s_{_{I+1,1}}\,$,
$k_{_I}=0,\dots,s_{_{I,I+1}}-1\,$)
\bea &&\x^a \left[(\r_q)^{{\bf p}_I+{\bf q}}_{{\bf p}_{I-1}+{\bf
q}}(P_a|e)\right]^{\a}_{\b} X_{q}^{{\bf p}_I+{\bf
q}}(\Th^{\ast\b}_{[p_{I-1}]})
\nn \\
[5pt]&&=\left\{\ba{ll}0 &\mbox{if $k_{_I}<s_{_{I,I+1}}-1$}\\
[5pt](-1)^{q(h_I+1)} e_{(p_{I-1}+1)} \cdots e_{(p_{I})} \x_{(p_I
+1)} X_{q}^{{\bf p}_{I-1}+{\bf
q}}(\Th^{\ast\,s_{I,1}}_{[p_{I-1}]})& \mbox{if
$k_{_I}=s_{_{I,I+1}}-1$}\ea\right.\ . \eea
Integrating the above representation matrix and setting the
integration constant to zero, yields the operator
$\s^-_q:\mR_q\rightarrow\mR_{q+1}$ given by
\bea \s^-_q&=& -i \int_0^1 dt ~e^a \rho_q(P_a| te)\
,\label{intcocycle}\eea
with the following key property ($\nabla:=d-\ft{i}2 \o^{ab}
\rho(M_{ab})$ and $\nabla e^a=0$)
\bea
(\nabla+\s^-_{q+1})(\nabla+\s^-_q)&\equiv&0\quad\Leftrightarrow\quad\nabla^2\
\frac{}{}\equiv\ \nabla\sigma^-_{q}+\s^-_{q+1}\nabla\ \equiv\
\sigma^-_{q+1}\s^-_q\ \equiv\ 0\ ,\eea
which is equivalent to the closure of the
$\mg_{_0}$-transformations
\bea \d_{\x,\L} X^\a&=& \ft{i}2 \L^{ab}\rho_q(M_{ab})X^\a+i \x^a
\left[\rho_q(P_a|e) \right]^\a_{\a+1}X^{\a+1}\ ,\quad
\d_{\x,\L}(e+\o)\ =\ 0\ .\eea

The Skvortsov equations are now the generalized curvature
constraints
\bea R^\a&:=& [(\nabla+\s^-_{_0})X]^\a\ =\ \nabla X^\a +
\left[\sigma^-_{_0}\right]^\a_{\a+1} X^{\a+1}\ \approx\ 0\ .\eea
The first levels of Bianchi identities and gauge transformations
take the form
\bea Z^\a&:=&[(\nabla+\s^-_{_1})R]^\a\ =\ \nabla R^\a+
\left[\sigma^-_{_1}\right]^\a_{\a+1}R^{\a+1}\ \equiv\ 0\ ,
\\[5pt] \delta_\e X^\a&:=&[(\nabla+\s^-_{_{-1}})\e]^\a\
=\ \nabla\e^\a+\left[\sigma^-_{_{-1}}\right]^\a_{\a+1} \e^{\a+1}\
.\eea
The cohomology of $\s^{-}$ in the triangular module $\mT$
determines the on-shell content of the Skvortsov equations. In
particular, the non-trivial content of $H^\ast(\s^-|\mT)\cap
\mR_{_0}$ is the (trace-constrained) Labastida gauge field
\bea \varphi(\Th)&=& {\mathbb P}^{\msl(D)}_{\Th}\;
i_{\th^{a_1}}\cdots i_{\th^{a_H}} X^{{\bf p}_B} (\widetilde \Th^\ast)\
.\eea
The Labastida field equation is the non-trivial content of
$H^\ast(\s^-|\mT)\cap \mR_{_1}\,$. The restriction of the
triangular module $\mT$ to its submodule $\mT_{\rm Weyl}$
consisting of states with $p_\a=0$ yields the primary Weyl tensor
$C(\overline\Th)$ as the non-trivial content of
$H^\ast(\s^-|\mT_{\rm Weyl})\cap \mR^{\bf 0}_{_0}\,$.

As realized early in \cite{Vasiliev:1990bu} (see
\cite{Vasiliev:1999ba,Bekaert:2005vh} for reviews) and also pointed
out later in \cite{Bekaert:2003zq,Bekaert:2006ix}, the local degrees
of freedom are encoded in the Weyl zero-form module and may be put
in correspondence with the massless $\mg_{_0}$-irrep
$\mD(M^2\!\!=\!0;\Th)$ through harmonic expansion. Thus, for the
purpose of counting the local on-shell degrees of freedom carried by
$\varphi(\Th)$ it suffices to analyze $C(\overline\Th)$ and it is
not necessary to actually extract the precise form of the Labastida
operator.

\subsection{\sc{Interlude: Oscillator realization of the Young tableaux}}
\label{Sec:Cellop}

In order to study the integrability of Skvortsov's equations and
more generally to describe tensor fields of arbitrary shapes, one
may adopt the notion of a generalized Schur module and related
hyperform
complex~\cite{Olver:1983,Olver:1987,DuboisViolette:1999rd,DuboisViolette:2001jk,Bekaert:2002dt}
and to give these an explicit oscillator
realization~\cite{Metsaev:1995re,Metsaev:1997nj}

The general properties of the cell operators presented in Section
\ref{Sec:GenProp} suffice for handling the unfolded master-field
equations in flat spacetime as well as the generic massive
master-field equations in $AdS_D$. However, in order to examine
the critically massless cases in $AdS_D\,$ (namely in analyzing
the projection \eq{construu} of the reducible Weyl zero$\,$-form)
it appears that a more explicit expression for the cell operators
is needed as was realized by
Metsaev~\cite{Metsaev:1995re,Metsaev:1997nj}. Such an expression
is rederived here and will be crucial to our analysis in Section
\ref{Sec:AdS} --- more precisely, for our derivation of
\eq{claim}.

\subsubsection{\sc Howe duality and Schur states}

The decomposition of tensor products of finite-dimensional
representations of the classical matrix algebras, $\mm$ say, using
manifestly symmetric $(+)$ and anti-symmetric $(-)$ bases leads to
the notion of Howe dual algebras $\widetilde \mm^\pm$ and
associated generalized Schur modules ${\cal S}^\pm$ as described
in Appendix \ref{App:A}. Using bosonic $(+)$ and fermionic $(-)$
oscillator realizations, the Lie algebra $\widetilde
\mm^\pm$ arises as a subalgebra of the infinite-dimensional Lie
algebra of canonical transformations of the oscillator algebra and
is identified with the maximal finite-dimensional subalgebra that commutes
with $\mm\,$. The corresponding ${\cal S}^\pm$ are by definition
the subspaces of the Fock modules ${\cal F}^\pm$ consisting of
states $\ket{\D}^\pm$ that are annihilated by a Borel subalgebra
of $\widetilde\mm^\pm\,$. Using $\nu_\pm$ oscillator flavors, say
$\{\a_i^a,\bar\a^i_a\}_{i=1}^{\nu_\pm}$, leads to
finite-dimensional Howe dual algebras, namely $\msl(\n_\pm)$ for
$\msl(D)$ tensors, and $\msp(2\n_+)$ and $\mso(2\n_-)$ for
$\mso(D)\,$.

If $\mm=\msl(D)$ then the Schur states can be chosen to obey
\bea (N^i_j-\d^i_j \widetilde \l^\pm_i)\ket{\D}^\pm&=& 0\ ,\quad
1\leqslant i\leqslant j\leqslant\nu_\pm\ ,
\qquad\label{Schurstates}\eea
where $N^i_j\in \msl(\n_\pm)$. If $\mm=\mso(D)$ then the Schur
states also obey the tracelessness condition
\bea \mm=\mso(D)&:& \left\{\ba{ll}T_{(11)}\ket{\D}^+\ =\
0&\mbox{in ${\cal F}^+$}\\T_{[12]}\ket{\D}^-\ =\ 0&\mbox{in ${\cal
F}^-$}\ea\right.\label{leadingtraces}\eea
where in a three-graded splitting (see \eq{A1grading})
$T_{(11)}\in \left[\msp(2\n_+)\right]^{(-1)}$ and
$T_{[12]}\left[\mso(2\n_-)\right]^{(-1)}$, taking the leading
traces of Schur states such that \eq{Schurstates} and
\eq{leadingtraces} imply $T_{(ij)}\ket{\D}^+=0$ and
$T_{[ij]}\ket{\D}^-=0$. In both cases one can show that $
\widetilde w^\pm_{_1}\geqslant \cdots\geqslant\widetilde
w^\pm_{\nu^\pm}\geqslant 0$ where $\widetilde
w^\pm_i=\widetilde\l^\pm_i\mp\ft{D}2$, and that $\ket{\D}^\pm$
contains exactly one copy of the $\mm$-irrep with highest weight
given by $\{\widetilde w^\pm_i\}_{i=1}^{\n_\pm}\,$. Moreover, in
the limit $\n_\pm\rightarrow \infty$ arise the universal Howe-dual
algebras $\widetilde\mm^\pm\cong \msl(\infty)$ for $\msl(D)$
tensors, and $\widetilde\mm^+\cong\msp(2\infty)$ and
$\widetilde\mm^-\cong\mso(2\infty)$ for $\mso(D)$ tensors, such
that
\bea \n^\pm\ \rightarrow\ \infty&:& {\cal S}^\pm\ \cong\ {\cal
S}^\mp\ .\eea
%


\subsubsection{\sc Cell operators: General definitions and properties}
\label{Sec:GenProp}

From now on we consider the general classical matrix algebras denoted here by
$\mm=(\msl(D),\mso(D),\msp(D))$ and parameterized by
$\e(\mm)=(0,+1,-1)$, and use the notation of Appendix \ref{App:A} otherwise.

The oscillator formalism can be used to define the cell
operators~\cite{Olver:1983,Olver:1987,Metsaev:1995re,Metsaev:1997nj}
$\left\{\beta_{\pm(i),a},\bar\beta^{\pm(i),a}\right\}_{i=1}^{\n_\pm}$
as a set of operators on the oscillator module ${\cal M}^\pm$ that
induces a non-trivial and regular action on the corresponding
Schur modules ${\cal S}^\pm$ obeying: (i) the amputation and
generation properties
\bea (N^i_j-\d^i_j (\widetilde\l^\pm_i-1))\b_{\pm(i),a}\ket{\D}=
0\ ,\quad (N^i_j-\d^i_j (\widetilde \l^\pm_i+1))\bar
\b^{\pm(i),a}\ket{\D} = 0\ ,\quad 1\leqslant i\leqslant
j\leqslant\nu_\pm \qquad \label{A1hom}\eea
for $\ket{\D}\in{\cal S}^\pm$; and (ii) the conjugation rule
\bea
\bar\beta^{\pm(i),a}&=&\pi\left(\beta_{\pm(\nu_\pm-i+1),a}\right)\
, \label{A1conj}\eea
where $\pi:=\pi_{(\nu_\pm,\dots,1)}$ with $(\nu_\pm,\dots,1)$
denoting the reverse permutation in $S_{\nu_{\pm}}$ and $\pi_\s$
($\s \in S_{\nu_\pm}$) being the linear automorphisms of the
oscillator algebra defined for arbitrary composite operators $f$
and $g$ by
\bea \pi_\s(fg)&=&\pi_\s(f)\pi_\s(g)\ ,\qquad
\pi_\s\left(f(\a_{i,a}, \bar\a^{j,b})\right)\ =\
f(\bar\a^{\s(i),a},\mp \a_{\s(j),b})\ .\eea
The amputation property amounts to that
\bea N^i_j\, \beta_{\pm(k),a}&=&\sum_{m<n}\c^{i,n}_{j,k,m}N^m_n\quad
\mbox{if $i<j$}\ ,\label{A1nij}\eea
for some operators $\c^{i,n}_{j,k,m}\,$. Thus, the conjugation
rule \eq{A1conj} is well-defined since
\bea \pi_\s(N^i_j)&=&\mp N^{\s(j)}_{\s(i)}\ ,\eea
together with \eq{A1nij} imply that if $i<j$ then
\bea N^i_j\, \pi(\beta_{\pm(k),a})\ket{\D}&=& \mp\,\pi\left(
N^{\nu_\pm-j+1}_{\nu_\pm-i+1} \beta_{\pm(k),a}\right)\ket{\D}\nn\\[5pt]
&=&\sum_{m<n}\pi\left(
\c^{\nu_\pm-j+1,n}_{\nu_\pm-i+1,k,m}\right)
N^{\nu_\pm-n+1}_{\nu_\pm-m+1}\ket{\D}\ =\ 0\ .\eea
The amputation and generation properties imply that
\bea \beta_{\pm(i),a}\ket{\D}=0=\bar\beta^{\pm(i+1),a}\ket{\D}\
\qquad \mbox{if $\widetilde w^\pm_{i}=\widetilde w^\pm_{i+1}$}\
.\eea
In the Fock space realization, where the Schur modules decompose
into Young tableaux, this means that $\beta_{\pm(i),a}$ and
$\bar\beta^{\pm(i),a}$, respectively, add and remove cells from
the $i$th row $(+)$ or column ($-)$ of $\D$ in accordance with row
and column order, \emph{viz.}
\bea \beta_{\pm(i),a}\ket{\D}=0=\bar\beta^{\pm(i+1),a}\ket{\D}\
\qquad \mbox{if}
\quad \mx{\{}{ll}{w_i=w_{i+1}&+\ ,\\[5pt]h_i=h_{i+1}&-\ .}{.}
\label{compatibi} \eea
[See under \eq{A1li}--\eq{A1li-} for the definitions of $w_i$ and $h_i\,$.]
Thus, the Schur modules ${\cal S}^\pm_{D;\nu_\pm} \subset {\cal
F}^\pm_{D;\nu_\pm}$ are generated by acting on $\ket{0}$ with
row-ordered $(+)$ or column-ordered $(-)$ strings of
$\bar\beta^{\pm(i),a}$ operators that in addition need to be taken
to be $J$-traceless (see Appendix \ref{App:A}) if $\e(\mm)=\pm 1$.

Conversely, if
\bea{\cal O}^\pm_\xi&=&\xi^{a^1_1,a^1_2,\ldots
,a^1_{m_1}\!,a^2_{_1},\dots,a^r_{m_r}} \prod_{\ell=1}^r
\prod_{j=1}^{m_{\ell}} \beta_{\pm(i_{\ell}),a^{\ell}_j}\ , \eea
where $\xi$ is a reducible tensor of rank $R$ and the product is
left-ordered, then
\bea {\cal O}^\pm_\xi \ket{\D} &=& \ket{\D'}\ ,\qquad \widetilde
w^{\pm '}_i \ =\ \widetilde w^\pm_i - \sum_{\ell =1}^r
m_{\ell}\,\delta_{i_{\ell},i}\ .\eea
The amputation property implies that ${\cal O}^\pm_\xi $ preserves
$J$-tracelessness in case $\e(\mm)=\pm 1$,
\emph{i.e.}\footnote{Eq. \eq{faithful} can also be checked
directly using the explicit expressions \eq{A1beta} and
\eq{A1barbeta} for $\beta_{\pm(i),a}$. The latter actually imply
the stronger property $[T_{11},\beta_{\pm(i),a}]=0$ in the case
$\e=+1$.}
\bea T_{ij}\ket{\Delta}&=&0\quad \Rightarrow\quad
T_{ij}\beta_{\pm(k),a}\ket{\Delta}=0\ .\label{faithful}\eea
One can decompose the tensor $\xi$ into irreducible
representations
\bea \xi^{a^1_1,a^1_2,\ldots ,a^1_{m_1}\!,a^2_1,\dots,a^r_{m_r}}
&=& \sum_{\widetilde\Delta} \sum_{\t_{\widetilde\Delta}} ({\mathbb
P}_{\t_{\widetilde\Delta}}\xi) ^{a^1_1,a^1_2,\ldots
,a^1_{m_1}\!,a^2_1,\dots,a^r_{m_r}} \ ,\eea
where we sum over the different inequivalent Young tableaux
$\t_{\widetilde\Delta}$ with rank-$R$ shape $\widetilde\Delta\,$.
Thus
\bea \ket{\D'}&=& \sum_{\D_\x\in \D/\D'} {\cal O}^\pm_{\D_\xi}
\ket{\D}\ ,\eea
where ${\cal O}^\pm_{\D_\xi}$ gathers together the contribution to
${\cal O}^\pm_\xi$ from all inequivalent tableaux $\t_{\D_\xi}$
corresponding to the diagram $\D_\xi$, and $\D/\D'$ is the set of
Young diagrams $\D_\xi$ of rank $R$ such that the outer product
$\D_\xi \otimes \D'$ contains $\D$ with multiplicity ${\rm
mult}(\D|(\D_\xi \otimes \D'))\geqslant 1\,$. More precisely, ${\cal
O}^\pm_{\D_\xi}|\Delta\rangle$ is the sum
\begin{eqnarray}
{\cal O}^\pm_{\D_\xi}|\Delta\rangle = \sum_{\t_{\Delta_{\xi}}}
({\mathbb P}_{\t_{\Delta_{\xi}}}\xi) ^{a^1_1,a^1_2,\ldots
,a^1_{m_1}\!,a^2_1,\dots,a^r_{m_r}} \prod_{\ell=1}^r
\prod_{j=1}^{m_{\ell}} \beta_{\pm(i_{\ell}),a^{\ell}_j}
\,|\Delta\rangle \quad
\end{eqnarray}
so that one has \bea \ket{\D'}&=& \left( \sum_{\D_\x\in \D/\D'}
\sum_{~~\t_{\Delta_{\xi}}} ({\mathbb P}_{\t_{\Delta_{\xi}}}\xi)
^{a^1_1,a^1_2,\ldots ,a^1_{m_1}\!,a^2_1,\dots,a^r_{m_r}} \right)
\left[ \prod_{\ell=1}^r \prod_{j=1}^{m_{\ell}}
\beta_{\pm(i_{\ell}),a^{\ell}_j} \,|\Delta\rangle \right]\ . \quad
\eea
Depending on the symmetries of $|\Delta\rangle\,$ not all the
\emph{diagrams} $\Delta_{\xi}$ need to contribute to the above
expression. However, if one diagram $\Delta_{\xi}$ contributes
nontrivially, then all the tableaux $\t_{\Delta_{\xi}}$ will
contribute if $\xi$ has no definite symmetry property. If $\xi$
already possesses some symmetry properties in some of its indices,
then several tableaux with the same shape will give the same
contributions (up to an overall coefficient).

A special case, which ensures the integrability of the various
master-field equations, is when $R=m+n$ with $\beta_{\pm (i)}^{a}$
appearing twice, so that the sequence of cell operators is as
follows
\bea \beta_{\pm(i-n+1)}^{b_1}\,\beta_{\pm(i-n+2)}^{b_2}\,\dots
\,\beta_{\pm(i-1)}^{b_{n-1}}
\,\beta_{\pm(i)}^{b_n}\,\beta_{\pm(i)}^{a_1}\,\beta_{\pm(i+1)}^{a_2}\,\dots
\,\beta_{\pm(i+m-2)}^{a_{m-1}}\,\beta_{\pm(i+m-1)}^{a_{m}}
\nonumber\;, \eea
and when $\D$ contains a block of height $h = n$ between the
$(i-n+1)$th and the $i$th rows on top of a block of height $h' =
m-1$ between the $(i+1)$th and the $(i+m-1)$th rows. Then
\bea \{\D_\xi\}&=&\{[m+n-p,p]\}_{p=1}^{\min(m,n)}\
,\label{A1deltaxi}\eea
where we note that ${\rm mult}(\D|(\D'\otimes [m+n-1,1]))=1$ (higher
multiplicities arise for $p\geqslant 2$). For $m=n=1$, the above
reduces to
\bea [\beta^{\pm(i),a},\beta^{\pm(i),b}]&=&0\ ,\qquad
[\bar\beta^{\pm(i),a}, \bar\beta^{\pm(i),b}]\ =\ 0\ ,
\label{commutbeta} \eea
and for $m=1$ and $h=n>1$, $\ket{\D}$ containing a block of height
$h$ between the $(i-h+1)$th and $i$th rows, then
\bea \prod_{\ell=1}^h \beta_{(i-h+l)}^{a_\ell}\ket{\D} &=&
\beta_{(i-h+1)}^{[a_1}\cdots \beta_{(i)}^{a_h]}\ket{\D}\ \eea
and only the two-column diagram $\Delta_{\xi}=[h,1]$ will
contribute to $|\Delta'\rangle\,$. \vspace*{.5cm}

We note that \eq{A1hom} together with with the Casimir formula
\eq{A1cas} yield
\bea [C_{_2}[\mm],\beta_{\pm(k),a}]&=&
(-D+\e\mp(2N^k_k+2-2k))\beta_{\pm(k),a}\ ,
\\[5pt]
[C_{_2}[\mm],\bar\beta^{\pm(k),a}]&=&
(D-\e\pm(2N^k_k-2k))\bar\beta^{\pm(k),a}\ .\eea
In the case of $\e=\pm1$, these commutators imply the following
anti-commutators:
\bea
\{M_{ac},\beta^c_{\pm(k)}\}&=&i[C_{_2}[\mm],\beta_{\pm(k),a}]\ =\
i(-D+\e\mp(2N^k_k+2-2k))\beta_{\pm(k),a}\ , \label{anti1}
\\[5pt]\{M_{ac},
\bar\beta^{\pm(k),c}\}&=&i[C_{_2}[\mm],\bar\beta^{\pm(k)}_a]\ =\
i(D-\e\pm(2N^k_k-2k))\bar\beta^{\pm(k)}_a\ . \label{anti2} \eea
They also imply that
\bea
\bar\beta^{\pm(i),a}\beta_{\pm(j),a}&=&\beta_{\pm(j),a}\bar\beta^{\pm(i),a}\
=\ 0\quad\mbox{if $i\neq j$}\ .\eea
The solution space to \eq{A1hom} and \eq{A1conj} is invariant
under rescalings of the form
\bea \beta_{\pm(i),a}&\rightarrow& \beta_{\pm(i),a}~f_{\pm(i)}\ =\
t_{_1}(i)f_{\pm(i)} ~\beta_{\pm(i),a}\ ,\label{A1rescale}\eea
where $f_{\pm(i)}=f_{\pm(i)}(N^1_1,\dots,N^{\nu_\pm}_{\nu_{\pm}})$
are functions that are regular and non-vanishing on ${\cal
S}^\pm$, and we use the notation
\bea t_x(i) f& :=&f(\dots,N^i_i+x,\dots)\quad\mbox{for
$f=f(N^1_1,\dots,N^{\nu_\pm}_{\nu_{\pm}})$}\ .\eea
This ambiguity can be removed partially by considering normalized
cell operators $(\gamma_{\pm(i),a},\bar\gamma^{\pm(i),a})$ obeying
$\sum_{i=1}^{\nu_\pm}[\gamma_{\pm(i),a},\bar\gamma^{\pm(i),a}]=M^a_b$
which fixes the scale factors up to constant rescalings at least
for $\nu=2\,$. However, at the level of the free master-field
equations, the normalization is immaterial since the rescalings
\eq{A1rescale} amount to non-singular redefinitions of auxiliary
fields.

\subsubsection{\sc Cell operators: Explicit oscillator realization}
\label{Sec:Expli}

The explicit form of the cell operators can be found by an
iterative procedure based on the assumption that
$\beta_{\pm(i),a}$ only depends on $\a_{j,a}$ and $N^j_k$ with
$j\geqslant k\geqslant i\,$. Then
$N^j_k\beta_{\pm(i),a}\ket{\D}=0$ for $j<k<i$ and it remains to
solve $N^j_k\beta_{\pm(i),a}\ket{\D}=0$ for $i\leqslant
j<k\leqslant \nu_\pm\,$. From $N^j_k\beta_{\pm(i+1),a}\ket{\D}=0$
for $i+1\leqslant j\leqslant \nu_\pm$ it follows that
$N^j_k\check{\beta}_{\pm(i),a}\ket{\D}=0$ for $i\leqslant
j\leqslant \nu_\pm-1$ where
$\check{\beta}_{\pm(i),a}=\beta_{\pm(i),a}|_{(\a_{i',a},
\bar\a^{i',a})\rightarrow (\a_{i'-1,a},\bar\a^{i'-1,a})}\,$. Thus,
by the assumption,
\bea \beta_{\pm(i),a}&=& \check{\beta}_{\pm(i),a}
g_{(i)}+\a_{\nu_\pm,a}N^{\nu_\pm}_i f_{(i,\nu_\pm)}
+\sum_{p=1}^{\nu_\pm-i-1}\sum_{i<j_1\cdots<j_p<\nu_\pm}
\a_{\nu_\pm,a}N^{\nu_\pm}_{j_p} \cdots N^{j_1}_i
f_{(i,j_1,\dots,j_p,\nu_\pm)}+\cdots\ , \nonumber \eea
where $g_{(i)}$, $f_{(i,\nu_\pm)}$ and
$f_{(i,j_1,\dots,j_p,\nu_\pm)}$ are functions of
$(N^i_i,\dots,N^{\nu_\pm}_{\nu_{\pm}})$ to be determined from
\bea N^{j}_{j+1}\beta_{\pm(i),a}&=&0\qquad \mbox{for
$j=i,\dots,\nu_\pm-1$}\ ,\eea
and the initial condition
\bea \beta_{\pm(\nu_{\pm}),a}&=&\a_{\nu_\pm,a}\ .\eea
One solution, which is actually regular on ${\cal M}^\pm$, is
\bea g_{(i)}&\simeq &P(i,\nu_\pm)\ ,\quad f_{(i,\nu_\pm)}\ \simeq\
{\prod_{j=i+1}^{\nu_\pm} P(i,j)\over P(i,\nu_\pm)}\ ,\quad
f_{(i,j_1,\dots,j_p,\nu_\pm)}\ \simeq\ {\prod_{j=i+1}^{\nu_\pm}
P(i,j) \over \prod_{q=1}^p P(i,j_q)}\ ,\eea
that is
\bea \beta_{\pm(i),a}&\simeq & \left[ \alpha_{i,a} +
\sum_{i<j_1<\cdots<j_p\leqslant\nu_\pm}
\a_{j_p,a}N^{j_p}_{j_{p-1}}\cdots N^{j_1}_i~\right]
{\prod_{j=i+1}^{\nu_\pm}P(i,j) \over \prod_{q=1}^p P(i,j_q)}
\nn\\[5pt]
&=& \left[ \alpha_{i,a} + \sum_{i<j_1<\cdots<j_p\leqslant\nu_\pm}
N^{j_1}_i{1\over P(i,j_1)}\cdots N^{j_p}_{j_{p-1}}{1\over
P(i,j_p)}\a_{j_p,a}~ \right] \prod_{j=i+1}^{\nu_\pm}(P(i,j)+1)\
,\qquad \label{betaused}\eea
where $\simeq$ refers to the ambiguity residing in rescalings of
the form \eq{A1rescale}, and
\bea P(i,j)&=&N(i,j)+j-i-1\ ,\qquad N(i,j)\ =\ N^i_i-N^j_j\ .\eea
Correspondingly,
\bea \bar\beta^{\pm(i),a}&\simeq &\pi(\beta_{\pm(\nu_\pm-i+1),a})
\\[5pt]&=& \left[
\bar\alpha^{i,a}+\sum_{1\leqslant j_p<\cdots<j_1<i} (-1)^p
N_{j_1}^i{1\over P(j_1,i)}\cdots N_{j_p}^{j_{p-1}}{1\over
P(j_p,i)}\bar\a^{j_p,a}\right] ~\prod_{j=1}^{i-1} (P(j,i)+1) \eea
The overall factors $\prod_{j=i+1}^{\nu_\pm}(P(i,j)+1)$ and
$\prod_{j=1}^{i-1} (P(j,i)+1)$ as well as the inverses of $P(i,j)$
and $P(j,i)$ are regular and non-vanishing in ${\cal S}^\pm\,$.
Thus, as long as regularity in ${\cal M}$ is not of any concern,
one may rescale the cell operators, and work with
\bea \beta_{\pm(i),a}&= & \a_{i,a} +
\sum_{i<j_1\cdots<j_p\leqslant\nu_\pm} N^{j_1}_i{1\over
P(i,j_1)}\cdots N^{j_p}_{j_{p-1}}{1\over P(i,j_p)}\a_{j_p,a} \ ,
\label{A1beta}\\
[5pt]\bar\beta^{\pm(i),a}&= & \bar\a^{i,a} + \sum_{1\leqslant
j_p<\cdots<j_1<i} (-1)^p N_{j_1}^i{1\over P(j_1,i)}\cdots
N_{j_p}^{j_{p-1}}{1\over P(j_p,i)} \bar\a^{j_p,a} \ .
\label{A1barbeta} \eea
%

\subsubsection{\sc{Equivalent bosonic and fermionic universal Schur modules}}
\label{Sec:Equiv}
%
By definition, the Fock spaces ${\cal F}^\pm_{D;\nu_\pm}$ and the
corresponding generalized Schur modules ${\cal
S}^\pm_{D;\nu_\pm}\subset {\cal F}^\pm_{D;\nu_\pm}$ consist of
states $\ket{\Psi}=\Psi(\bar\a^{i,a})\ket{0}$ and
$\ket{\D}=\D(\bar\a^{i,a})\ket{0}$, respectively, generated by
$\Psi(\bar\a^{i,a})$ and $\D(\bar\a^{i,a})$ that are arbitrary
polynomials. Acting on $\ket{\D}$ with the cell operators
$\b_{\pm(i),a}\ket{\D}$ and $\bar\b^{\pm(i),a}\ket{\D}$, given in
\eq{A1beta} and \eq{A1barbeta}, yields states
$\b_{\pm(i),a}\ket{\D}$ and $\bar\b^{\pm(i),a}\ket{\D}$ that
remain arbitrary polynomials (finite sums) for arbitrary
$\nu_\pm\,$. Thus, the cell operators have a well-defined action
in ${\cal S}^\pm_{D;\nu_\pm}$ in the limit
$\nu_\pm\rightarrow\infty\,$. From the expressions \eq{A1mult+},
\eq{A1mult-} and \eq{A1mult2} for the multiplicities, it follows
that the bosonic and fermionic oscillator realizations are on
equal footing in the sense that
\bea {\cal S}^+_{D;\nu_+}&\simeq&{\cal S}^-_{D;\infty}
\quad\mbox{for $\nu_+\geqslant D$}\ ,\label{A1iso}\eea
and, taking into account the fact that $h_{w_i}\geqslant i$ and
$h_{w_{i+1}}\leqslant i-1$, one finds

\bea \beta_{(i),a}\ket{\D}&\simeq& \sum_{j=1}^\infty
\beta_{[j],a}\d_{h_{j},i}\ket{\D}\ ,\qquad
\bar\beta^{(i),a}\ket{\D}\ \simeq\ \sum_{j=1}^\infty
\bar\beta^{[j+1],a}\d_{h_{j},i-1}\ket{\D}\ ,\label{Aliso3}\eea
where we use the notation
\bea \beta_{(i),a}&:=&\beta_{+(i),a}\ ,\qquad \beta_{[i],a}\ :=\
\beta_{-(i),a}\ , \label{A1iso2}\eea
\emph{idem} $\bar\beta\,$.

For example, for $D=1$ the ground states are
$\ket{\D}=\ket{(n)}=\ket{[\underbrace{1,\dots,1}_{\footnotesize
\mbox{$n$ columns}}]}$, and
\bea \beta_{[i]}\ket{(n)}&=&\delta_{in}\ket{(n-1)}\ ,\qquad
\bar\beta^{[i]}\ket{(n)}\ =\ \d_{i,n+1}\ket{(n+1)}\ ,\eea
and the above map takes the form
\bea \b&:=&\b_{(1)}\ =\ \sum_{i=1}^\infty \b_{[i]}\ =\
\a{1\over\sqrt{\bar \a \a}}\ ,\qquad \bar\b\ :=\ \bar\b^{(1)}\ =\
\sum_{i=1}^\infty \bar\b^{[i]}\ =\ {1\over\sqrt{\bar \a \a}}\bar
\a\ ,\eea
where $\a:=\a_{1}$ and $\bar\a:=\bar\a^{1}$ obey $[\a,\bar \a]=1$
and we note that $\{\b,\bar\b\}=1\,$.

Roughly speaking, the correspondence between the bosonic and
fermionic oscillators is the result of ``gauging'' on the one hand
$\widetilde \mm^-=\mgl(\infty)$ in ${\cal F}^-_{D;\infty}$, and on
the other hand $\widetilde \mm^+=\mgl(\nu_+)$ in ${\cal
F}^+_{D;\nu_+}$ for $\nu_+\geqslant D\,$. Thus, in the limit
$\nu_+\rightarrow \infty$,
\bea {\cal S}^+_{D;\infty}&\cong&{\cal S}^-_{D;\infty}\
,\label{A1iso3}\eea
where both sides are $\mgl(\infty)$-gauged oscillator spaces.

%
%
\subsection{\sc{Master-field reformulation of Skvortsov's equations}}
\label{Sec:Master}
%
The master field
\bea \mathbf{X}&:=&\sum_{p=0}^{\infty}\mathbf{X}^{\bf p}\ \in\
\mR\ =\ \bigoplus_{p\geqslant 0} \mR^{\bf p}\ ,\quad \mR^{\bf p}\
:=\ \O^{\bf p}(U)\otimes {\cal S}\ ,\eea
where ${\cal S}$ is the Schur module described in the previous
Section. The Skvortsov equations amount to subjecting $\mathbf{X}$
to: i) curvature constraints; and ii) mass-shell and irreducibility
conditions. The curvatures and irreducibility conditions can be
examined at the level of the $\msl(D)$ Schur module, while the
mass-shell condition breaks $\msl(D)$ down to $\mso(1,D-1)$.

\subsubsection{\sc Bosonic oscillators (symmetric basis)}\label{Sec:Bos}

\begin{center}{\it Curvature constraints}\end{center}

The generalized curvature constraints can be written using
symmetric conventions as
\bea \mathbf R&:=& \Big(\nabla+\s^-_0\Big)\mathbf{X}\ \approx\ 0\ ,\qquad \s^-_{_0}\ :=\ \sum_{p\geqslant p'}(\s_{_0})^{\bf{p+1}}_{\bf p'}\ ,\\[5pt]
(\s_{_0})^{\bf{p+1}}_{\bf p'}&:=&-ie_{(p'+1)}\cdots e_{(p+1)}{\mathbb
P}(p+1,p'+1)\ ,\eea
where $\nabla:=d-\frac{i}2 \o^{ab}M_{ab}$, $e_{(i)}:=e^a
\beta_{(i),a}$ and ${\mathbb P}(p+1,p'+1):\mR\rightarrow \mR^{\bf
p}$ is a projector defined by
\bea {\mathbb P}(p+1,p'+1)\mathbf{X}&:=& \left\{\ba{ll}\d\left\{N(p'+1,p'+2),N(p'+2,p'+3),\dots,N(p,p+1)\right\} \mathbf{X}^{\bf p'}&\mbox{for $p>p'$}\\[5pt]\mathbf{X}^{\bf p}&\mbox{for $p=p'$}\ea\right.\ ,\label{bosproj}\hspace{1cm}\eea
where $\d\{\l_1,\dots,\l_k\}:=\d_{\l_1,0}\cdots \d_{\l_k,0}$ for
$\l_i\in\integ$, $i=1,\dots,k$. The corresponding triangular
module has the generalized curvatures ($q\in\integ$)
\bea \mathbf Z_{q+1}&:=& (\nabla+\s^-_q)\mathbf Z_q\ ,\\[5pt]
\s^-_q&=&(-1)^{q(1+\s^-_\circ)}\s^-_{_0}\ =\ \sum_{p\geqslant
p'}(-1)^{q(p-p')}(\s_{_0})^{\bf{p+1}}_{\bf p'}\ .\eea
The Cartan integrability amounts to the identity
\bea 0&\equiv&-\mathbf Z_{_2}\ :=\ -\left[(-1)^{1+\s^-_\circ}\s^-_{_0}\right]\s^-_{_0} \mathbf X\\[5pt]
&=& \sum_{p\geqslant p'\geqslant r'} (-1)^{p-p'}e_{(p'+1)}\cdots e_{(p+1)}{\mathbb P}(p+1,p'+1)e_{(r'+1)}\cdots e_{(p'+1)}{\mathbb P}(p'+1,r'+1)\mathbf X\\[5pt]
&=& e_{(1)}e_{(1)}\mathbf X^{\bf 0}+ \left(e_{(2)} e_{(2)} \mathbf X^{\bf 1}+e_{(2)}e_{(1)}e_{(2)}{\mathbb P}(2,1)\mathbf X^{\bf 0}-e_{(1)}e_{(2)}{\mathbb P}(2,1)e_{(1)}\mathbf X^{\bf 0}\right)\nn\\[5pt]
&&+e_{(3)}e_{(3)}\mathbf X^{\bf 2} + e_{(3)}e_{(2)}e_{(3)}{\mathbb
P}(3,2)\mathbf X^{\bf 1}
+ e_{(3)}e_{(1)}e_{(2)}e_{(3)}{\mathbb P}(3,1)\mathbf X^{\bf 0}- e_{(2)}e_{(3)}{\mathbb P}(3,2)e_{(2)}\mathbf X^{\bf 1}\nn\\[5pt]
&&- e_{(2)}e_{(3)}{\mathbb P}(3,2)e_{(1)}e_{(2)}{\mathbb
P}(2,1)\mathbf X^{\bf 0} + e_{(1)}e_{(2)}e_{(3)}{\mathbb
P}(3,1)e_{(1)}\mathbf X^{\bf 0}+\cdots \eea
where the first term is the Bianchi identity for the 0-form
constraint, the second group of terms is the Bianchi identity for
the 1-form constraint and last two lines is the Bianchi identity for
the 2-form constraint. The terms of the form
$e_{(p+1)}e_{(p+1)}\mathbf X^{\bf p}$ vanish by virtue of
$[\b_{(p+1),a},\b_{(p+1),b}]=0\,$ where the commutator is induced by
the anti-commutativity of $e^a\,$. The terms cubic in $e^a$ vanish
because of (\ref{A1deltaxi}). For example,
$\beta_{(p+2),a}\beta_{(p+1),b}\beta_{(p+2),c}{\mathbb
P}(p+2,p+1)\mathbf X^{\bf p}$ contains a hooked Young tableau in the
indices $(a,b,c)\,$ on which the totally anti-symmetric projection
enforced by $e^a\,e^b\,e^c$ vanishes. Similarly,
$e_{(p+1)}e_{(p+2)}{\mathbb P}(p+2,p+1)e_{(p+1)} \mathbf X^{\bf p}$
projects on types having $w_{_{p+1}}=w_{_{p+2}}+1$ (using the
notation in \eq{A1li} and \eq{A1li+}) but then, by
(\ref{A1deltaxi}), this results in hooked shape that gives zero. The
first term in the last line, which is quartic in $e^a$, can be
non-zero only if the types in $\mathbf X^{\bf 0}$ have the symmetry
property $w_{_1}=w_{_2}=w_{_3}+1\,$, because of the presence of the
two projectors ${\mathbb P}(2,1)$ and ${\mathbb P}(3,2)\,$.
Likewise, the projectors in the last term enforces
$w_{_1}=w_{_2}+1=w_{_3}+1\,$. However, by (\ref{A1deltaxi}) again,
the resulting hooked shapes are incompatible with the total
antisymmetry enforced by the four vielbeins.

In the general case, ${\mathbb P}(p+1,p'+1)$ and ${\mathbb
P}(p'+1,r'+1)$ force the flat indices of the cell operators to be
projected on different two-column Young tableaux associated with
the shapes given in (\ref{A1deltaxi}), where $m=p-p'+1$ and
$n=p'-r'+1$, with maximal height $m+n-1=p-r'+1\,$. However, there
are $m+n=p-r'+2$ vielbeins whose flat indices are to be contracted
with the ones of the cell operators, which yields zero.

\begin{center}{\it Mass-shell and irreducibility conditions}\end{center}

The mass-shell and irreducibility conditions are not unique at the
free-field level. Two natural models are: (1) the minimal
trace-constrained Skvortsov system defined by
\begin{eqnarray}
N^i_j \mathbf X^{\bf p} &=& T_{_{11}}\mathbf X^{\bf p} =
0\qquad\mbox{for $i<j$ and $\forall p$}\ ; \label{TrK}
\end{eqnarray}
and (2) the non-minimal trace-unconstrained system, \emph{viz.}
\bea N_i^j \mathbf X^{\bf p}&=&0\qquad\mbox{for $i<j$ and $\forall p$}\ ,\\
T_{_{11}}\mathbf X^{\bf 0}&=&0\ .\eea
Both systems carry the same physical degrees of freedom, namely one
massless particle for each $\overline\Th$ in $({\rm Ker} N(1,2))\cap
{\cal S}_D\,$. The minimal system suffices for constructing
first-order Skvortsov--Vasiliev--Weyl-type actions. The non-minimal
system contains additional St\"uckelberg potentials that could turn
out to be useful in constructing first-order actions that are
equivalent to the unconstrained metric-like formulation of
mixed-symmetry fields~\cite{Campoleoni:2008jq}.

At the non-linear level, the spectrum is to be determined by some
nonabelian extension of $\miso(1,D-1)\,$. Non-linearities are also
sensitive to whether the constraints are imposed strongly, as
above, or weakly by means of multiplication by a projector, or
more generally, by means of a suitable BRST operator.

\subsubsection{\sc Fermionic oscillators (anti-symmetric basis)}\label{Sec:Fer}

The equivalence between the bosonic and fermionic oscillator
realizations of the universal Schur module discussed in Section
\ref{Sec:Equiv} can be used to cast the manifestly symmetric
master-field formulation into a manifestly anti-symmetric ditto
obtained by substituting
\bea \beta_{a,(i)}&\rightarrow&\sum_{j=1}^\infty
\beta_{a,[j]}\,\d\{N^j_j+\ft D2-i\}\ ,\eea
where $\d\{\l\}=\d_{\l,0}$ for $\l\in\integ$ and the eigenvalues
of $N^j_j$ are given by $n_j-\ft D2$ where $n_j$ is the height of
the $j$th column. The $\s^-$-operator now takes the form
\bea \s^-_0&=&-i\sum_{p\geqslant p'}
\sum_{j_{p'},\dots,j_p=1}^\infty \d_{n_{j_{p'}},p'+1}\cdots
\d_{n_{j_{p}},p+1}e_{[j_{p'}]}\cdots e_{[j_{p}]}{\mathbb
P}(p+1,p'+1)\ ,\eea
with ${\mathbb P}(p+1,p'+1)$ defined by \eq{bosproj}. This
expression can be rearranged into the manifestly anti-symmetric
form
\bea \s^-_0 \mathbf X&=&-i\sum_{p\geqslant p'}\sum_{i=1}^\infty
\left(e_{[i]}\right)^{p-p'+1}\d\{N^i_i+\ft {D-2}2-p\}\mathbf
X^{\bf p'}\ .\eea
%
\section{\sc Tensor Fields in $AdS_D$}
\label{Sec:AdS}

This Section contains the derivation of the unfolded equations of
motion for arbitrary tensor gauge fields in $AdS_D\,$ by radial
reduction of Skvortsov's equations in $\Real^{2,D-1}$. We use the
master-field formulation given in Section~\ref{Sec:Master} and the
foliation lemmas of Section I.3.7, and follow the step-by-step
procedure outlined in Section~I.4.5 whereby one\begin{itemize}
\item[1)] decomposes the variables and generalized curvatures into
components parallel and transverse to the radial vector field;
\item[2)] constrains the radial derivatives in terms of a massive
parameter $f$ (\emph{cf.} item (i) of Section I.3.7); \item[3)]
shows that a generic value for $C_{_2}[\mg_\l]$ corresponds to two
``dual'' values $f^\pm$ of $f$ obeying $f^+\geqslant f^-$ and $f^+
+ f^-=D-1\,$, and that in our parametrization turn out to be
$f^+=e_0\,$, the lowest energy of the physical lowest-weight
space, and $f^-=\widetilde e_0\,$, the lowest energy of its
shadow; \item[4)] examines the critical limit where $f=f^+_I$ approaches
Metsaev's massless values $e_{_0}^I\,$, for which we claim (and
prove in a subset of all cases) that $f_{_I}^-$ (given by
\eq{claim}) is consistent with a \emph{projection} of the radially reduced
Weyl zero$\,$-form onto its massless sector (\emph{cf.} item (ii)
of Section I.3.7), whose complement thus constitutes an ideal; \item[5)] shows
that the potential module, as defined in Section I.4.4.4, is
trivial except in the unitary massless case $I=1$ where it
consists of the ASV potential; \item[6)] shows the smoothness of
the flat limit of the projected massless system, and how the BMV
conjecture is realized in an enlarged setting with extra
topological fields arising in the flat limit. The latter represent
the unfolded ``frozen'' St\"uckelberg fields of the $I$th block
whose Weyl$\,$zero form is set to zero in the aforementioned
projection of the zero$\,$-form.\end{itemize}

\subsection{\sc Transverse and parallel components in
$\Real^{2,D-1}$}\label{Sec:D+1}

Skvortsov's equations in a flat $(D+1)$-dimensional spacetime
$\widehat{\cal M}_{D+1}$ with signature $(2,D-1)$ read
\bea \widehat{\mathbf{T}}&:=&
\Big(\widehat{\nabla}+\widehat\s^-_{_0}\Big) \widehat{\mathbf{W}}
\approx 0 \ ,\quad \widehat\s^-_{_0}\ :=\ -i\sum_{p\geqslant p'}
\,\widehat{E}_{(p'+1)} \cdots \widehat{E}_{(p+1)} \widehat{\mathbb
P}(p+1,p'+1)\ ,\eea
with $\widehat\nabla:=d-\frac i2 \widehat \O^{AB} \widehat
M_{AB}\,$, $\,\widehat E_{(i)}:=\widehat E^A \widehat
\b_{A,(i)}\,$ and
$\widehat{\mathbf{W}}\in\widehat\mR=\bigoplus_{p\geqslant
0}\O^{\bf p}$ $(\widehat U)\otimes \widehat{\cal S}_{D+1}\,$,
where $\widehat U$ is a region of $\widehat{{\cal M}}_{D+1}$ that admits a
foliation with $AdS_D$ leaves and $\widehat{\cal S}_{D+1}$ is the
generalized $\widehat\mm\cong \mso(2,D-1)$ Schur module consisting
of all possible tensorial $\widehat\mm$-types $\widehat\Th_\a$,
each occurring with multiplicity one. In the module $\widehat{\cal
S}_{D+1}\,$, the following relations hold true:
\begin{eqnarray}
\widehat\b^A_{(1)}\widehat\b_{A,(1)} &=& 0 \ , \qquad
\widehat\x^B\left\{\widehat M_{B}{}^A,\widehat\b_{A,(1)}\right\} \
= \ -i\,(2\widehat N^1_1+D)\widehat\x_{(1)} \ \label{Schur1}
\end{eqnarray}
where eq. \eq{anti1} is used for the second equality.

If $\x=\x^M\partial_M$ denotes the radial vector field in $\widehat
U$ obeying $\x^2=-1\,$, where $\widehat \x^A:=\widehat
E^A_M\x^M\,$ and $N:=dL$ denotes the corresponding normal one-form,
then $\widehat E^A=\widehat e^A+N\widehat\x^A$ and $\widehat
\O^{AB}=\widehat\o^{AB}+N\widehat \L^{AB}$ where $i_\xi\widehat
e^A=0=i_\xi\widehat \o^{AB}\,$. The local
$\widehat\mm$-symmetry can be used to set $d\widehat\x^A=0$ and
$\widehat\L^{AB}=0$, that are preserved under residual local
$\mm$-transformations on the $AdS_D$ leaves. As described in
Appendix \ref{App:reduc}, the transverse components $\widehat e^A$
and $\o^{AB}:= \widehat\o^{AB}+\l(\widehat
e^A\widehat\x^B-\widehat\x^A \widehat e^B)$ then obey
$\widehat\x_A\widehat e^A=0$ and $\widehat \x_A\o^{AB}=0\,$. Thus,
if $i_{L}:AdS_D(L)\rightarrow \widehat U$ denotes the embedding of
the $AdS_D$ leaf of radius $L=1/\lambda$ into $\widehat U\,$, then
the vielbein and $\mso(1,D-1)$-valued connection on $AdS_D(L)\,$ are
given by $e^a:=i^\ast_{L}{\mathbb{P}}{}^a{}_{\!\! A} \widehat e^A$
and $\o^{ab}:=i^\ast_L{\mathbb{P}}{}^a{}_{\!\! A}$
${\mathbb{P}}{}^b{}_{\!\! B}\omega^{AB}\,$, where
${\mathbb{P}}{}^a{}_{\!\! A}\widehat\x^A \equiv 0\,$. As a result,
the canonical $AdS_D(L)$ connection
$i^*_{L}\widehat{\omega}^{AB}=:\O^{AB}=(\o^{ab},\l e^a)$ obeys
$d\O^{AB}+\O^{AC}\O_{C}{}^B=0\,$, that is, $\nabla e^a=0$ and
$d\o^{ab}+\o^{ac}\o_{c}{}^b+\l^2 e^a e^b=0\,$. The radial reduction
can also be analyzed directly on $\widehat U$, where one has
\bea \widehat\nabla&=&d-\frac i2 \widehat\o^{AB}\widehat M_{AB}\
,\quad \widehat E_{(i)}\ =\ \widehat e_{(i)}+N\widehat\xi_{(i)}\
,\quad \widehat\nabla\widehat E_{(i)}\ =\ 0\ ,\quad
\widehat\nabla^2\ =\ 0\ , \label{useful1}
\\[5pt]
\widehat\nabla \widehat e_{(i)}&=&\l N\widehat e_{(i)}\ ,\quad
\widehat\nabla \widehat\x_{(i)}\ =\ \l\widehat e_{(i)}\ ,\quad
\widehat\nabla \l\ =\ -\l^2 N\ .\label{useful2} \eea
The foliation also induces a splitting of $\widehat{\mathbf{W}}$
into transverse and parallel components, say
\bea \widehat{\mathbf{W}}^{\bf p} &:=& \widehat{\mathbf{X}}^{\bf
p}+N\,\widehat{\mathbf{Y}}^{\bf{p-1}}\ \in\ \mR_\perp\oplus
\mR_\parallel\ ,\label{decW}
\\[5pt]
i_\x \widehat{\mathbf{X}}^{\bf p}&:=&0\ ,\quad i_\x
\widehat{\mathbf{Y}}^{\bf{p-1}}\ :=\ 0\ ,\eea
and a corresponding decomposition
$\widehat{\mathbf{T}}=\widehat{\mathbf{R}}+N\,\widehat{\mathbf{S}}$
where $i_\x \widehat{\mathbf{R}}:=0$ and
$i_\x\widehat{\mathbf{S}}:=0\,$.

\noindent It follows that
\bea \widehat{\mathbf{R}}^{\bf{p+1}}&=& \left(\widehat\nabla-
N{\cal L}_\xi - i\,\widehat
e_{(p+1)}\right)\widehat{\mathbf{X}}^{\bf p}+\sum_{p\geqslant
p'+1} ({\widehat{\s}}^{-}_{_0})^{\bf{p+1}}_{\bf p'}
\widehat{\mathbf{X}}^{\bf{p'}} \ ,
\label{Ttrans1}\\
[5pt] \widehat{\mathbf{S}}^{\bf p}&=&\left(\widehat\nabla-N{\cal
L}_\x - i\,\widehat
e_{(p+1)}\right)\widehat{\mathbf{Y}}^{\bf{p-1}}+
\widehat{\mathbf{Z}}^{\bf p}+ \sum_{p\geqslant p'+1} {\scriptstyle
(-1)}^{p-p'} ({\widehat{\s}}^{-}_{_0})^{\bf{p+1}}_{\bf p'}
\widehat{\mathbf{Y}}^{\bf{p'-1}}\ ,\hspace{1cm}\label{Tpara1}\eea
where $({\widehat{\s}}^{-}_{_0})^{\bf{p+1}}_{\bf p'}$
$:=-i\,\widehat{e}_{(p'+1)} \cdots \widehat{e}_{(p+1)}$
$\widehat{\mathbb P}(p+1,p'+1)$ and ($p\geqslant1$)
\bea \widehat{\mathbf{Z}}^{\bf p}&:=& (-{\cal
L}_\x+i\widehat\xi_{(p+1)})\widehat{\mathbf{X}}^{\bf p}
+i\sum_{p\geqslant p'+1} (p-p'+1) \widehat\xi_{(p'+1)} \widehat
e_{(p'+2)}\cdots \widehat e_{(p+1)} \widehat {\mathbb P}(p+1,p'+1)
\widehat{\mathbf{X}}^{\bf{p'}}\ . \label{Zeta}\hspace{1cm}\eea
%
\subsection{\sc{Radial reduction}}\label{Sec:Lieder}
%
\subsubsection{\sc{Radial Lie derivatives and unfolded mass terms}}

Upon constraining the radial derivatives to be scaling dimensions,
\emph{i.e.}
\bea ({\cal L}_\xi + \l\D_{[p]} )\widehat{\mathbf{X}}^{\bf p}
&\approx& 0\ ,\qquad ({\cal L}_\xi + \l\Upsilon_{[p]})
\widehat{\mathbf{Y}}^{\bf{p-1}} \ \approx\ 0 \ ,\label{constr1}
\eea
where $\D_{[p]}=\D_{[p]}(\{\widehat N^i_i\}_{i=1}^\nu)$
\emph{idem} $\Upsilon_{[p]}\,$, the reduced curvatures
$\widehat{\mathbf{R}}$ and $\widehat{\mathbf{S}}$ form a closed
subsystem with variables $\widehat{\mathbf{X}}$ and
$\widehat{\mathbf{Y}}\,$. Its Cartan integrability (on
$\widehat{\cal M}_{D+1}$) fixes the scaling dimensions. {}From
\bea \widehat\nabla\widehat{\mathbf{R}}^{\bf{p+1}}&\approx&
\lambda\, N\Big(i\left([\widehat e_{(p+1)},\D_{[p]}]-\widehat
e_{(p+1)}\right) \widehat{\mathbf{X}}^{\bf p}\phantom{\sum_p}
\nn\\[5pt]
&+&\sum_{p\geqslant p'+1}\left((\D_{[p]}+1+p)
({\widehat{\s}}^{-}_{_0})^{\bf{p+1}}_{\bf p'}
-({\widehat{\s}}^{-}_{_0})^{\bf{p+1}}_{\bf
p'}(\D_{[p']}+p')\right)
\widehat{\mathbf{X}}^{\bf{p}'}\Big)\qquad\eea
it follows that $\widehat{\mathbf{R}}\approx 0$ is integrable iff
\bea &&\D_{[p]}\ =\ \D^f_{[p]}\ := \ \widehat N^{p+1}_{p+1}
+f_{[p]}(\{\widehat N^i_i\}_{i=1,i\neq p+1}^\nu)\ ,\label{Deltap}\\[5pt]&&
\widehat{\mathbb P}(p+1,p'+1)\Big(t_{-1}(p'+1)\cdots t_{-1}(p)
f_{[p]}+p-f_{[p']}-p'\Big)\ =\ 0\ . \eea
The last relation determines $f_{[p]}$ recursively in terms of a
single function $f\,$,
\bea f_{[p]}&=&-p+f\left(\widehat N^1_1+1,\dots,\widehat
N^p_p+1,\widehat N^{p+2}_{p+2},\dots,\widehat N^\nu_\nu\right) \
\Rightarrow \quad f_{[0]}\ =\
f(\widehat{N}^2_2,\widehat{N}^3_3,\ldots)\;, \label{f[p]}\eea
where the eigenvalues of $f_{[0]}$ are directly related to the
lowest energy $e_{_0}$ of the $\mso(2,D-1)$ lowest-weight space
carried by the constrained system (see \eq{energies} below). The
above form of $\D^f_{[p]}$ also implies that ($p\geqslant 1$)
\bea \Big(\widehat \nabla+\l N(\D^f_{[p]}+1)-i\widehat
e_{(p+1)}\Big) \widehat{\mathbf{Z}}^{\bf p}+\sum_{p\geqslant p'+1}
({\widehat{\s}}^{-}_{_0})^{\bf{p+1}}_{\bf p'}
\widehat{\mathbf{Z}}^{\bf{p'}}& \approx 0 \ .
\label{zetaconstr}\eea
Finally, one has ($p\geqslant 1$)
\bea\nabla \widehat{\mathbf{S}}^{\bf p}&\approx& \l N \Big[
i\,\left([\widehat e_{(p+1)},\Upsilon_{[p]}]-\widehat
e_{(p+1)}\right) \widehat{\mathbf{Y}}^{\bf{p-1}}
+(\Upsilon_{[p]}-\D^f_{[p]}-1)\widehat{\mathbf{Z}}^{\bf p}
\phantom{\sum_p}
\nn\\[5pt]
&+&
\sum_{p\geqslant p'+1}(-1)^{p-p'}\left( (\Upsilon_{[p]}+1+p)
({\widehat{\s}}^{-}_{_0})^{\bf{p+1}}_{\bf p'} -
({\widehat{\s}}^{-}_{_0})^{\bf{p+1}}_{\bf p'}
(\Upsilon_{[p']}+p')\right)\widehat{\mathbf{Y}}^{\bf{p'-1}}
\Big] \ ,\hspace{1.5cm}\eea
and hence $\widehat{\mathbf{S}}\approx0$ is integrable iff
\bea \Upsilon_{[p]}&=& \D^f_{[p]}+1\ ,\label{Upsilonp}\eea
as one may also deduce from dimensional analysis based on
\eq{decW} and $N=dL\,$.

In summary, after radial reduction and constraining the radial
derivatives we have
\bea &\widehat{\mathbf{R}}^{\bf p+1}:= \left(\widehat\nabla+
\lambda\,N\Delta^f_{[p]}-i\,\widehat e_{(p+1)}\right)
\widehat{\mathbf{X}}^{\bf p}+\sum_{p\geqslant p'+1}
({\widehat{\s}}^{-}_{_0})^{\bf{p+1}}_{\bf p'}
\widehat{\mathbf{X}}^{\bf{p}'} \approx 0 \quad ,&
\label{Ttrans}\\
[5pt] &\widehat{\mathbf{S}}^{\bf p}:= \left(\widehat\nabla +
\lambda\,N\,(\D^f_{[p]}+1) -i\,\widehat
e_{(p+1)}\right)\widehat{\mathbf{Y}}^{\bf{p-1}} +
\widehat{\mathbf{Z}}^{\bf p}+ \sum_{p\geqslant p'+1} {\scriptstyle
(-1)}^{p-p'} ({\widehat{\s}}^{-}_{_0})^{\bf{p+1}}_{\bf p'}
\widehat{\mathbf{Y}}^{\bf{p'-1}}\approx 0&
\hspace{1cm}\label{Tpara}\eea
where
\begin{eqnarray}
\widehat{\mathbf{Z}}^{\bf p}:=
(\lambda\,\D^f_{[p]}+i\,\widehat\xi_{(p+1)})
\widehat{\mathbf{X}}^{\bf p}+i\sum_{p\geqslant p'+1} (p-p'+1)
\widehat\xi_{(p'+1)} \widehat e_{(p'+2)}\cdots \widehat e_{(p+1)}
\widehat {\mathbb P}(p+1,p'+1)
\widehat{\mathbf{X}}^{\bf{p}'}\,,\label{Zeta2}
\end{eqnarray}
which we note obeys \eq{zetaconstr}. We denote the resulting
module
\bea \mR_f&:=&\mR_{f,\perp}\oplus \mR_{f,\parallel}\
,\label{mRf}\eea
where $\mR_{f,\perp}\ni \widehat{\mathbf{X}}$ and $
\mR_{f,\parallel}\ni \widehat{\mathbf{Y}}\,$. The variables
$(\widehat{\mathbf{Z}}(\widehat{\mathbf{X}}),\widehat{\mathbf{Y}})$
coordinatize a massively contractible cycle $\mS_f\subset\mR_f$
for all values of $f\,$.
%
\subsubsection{\sc Initial comments on criticality/reducibility}
%
We recall from Section I.4.4.4 that the potential submodule
$\widetilde\mR$ of an unfolded module $\mR$ with Weyl
zero$\,$-form module $\mC^{\bf 0}$ is the maximal chain
$\widetilde\mR:=\widetilde \mR^{\bf p}\supsetplus
\cdots\supsetplus \widetilde \mR^{\bf p'}\subset \mR$ with $p>0$
whose elements cannot be set to zero for non-trivial Weyl
zero$\,$-forms. Thus $\mR=\mR'\oplus \mS$ where
$\mR'=\widetilde\mR\supsetplus\mC^{\bf 0}$ and $\mS$ is massively
contractible (\emph{cf.} the example of massive spin-$1$ in flat
spacetime discussed in Section I.4.4.3).

For generic values of $f$, the map
$(\widehat{\mathbf{X}},\widehat{\mathbf{Y}})\rightarrow
(\widehat{\mathbf{X}}^{\bf 0},\widehat{\mathbf{Z}}
(\widehat{\mathbf{X}}),\widehat{\mathbf{Y}})$ is an invertible
(triangular) change of coordinates, \emph{i.e.}
\bea \mbox{generic $f$}&:& \left.\mR_f\right|_{\mg_\l}\ =\
\mS_{f}\oplus \mR^{\bf 0}_{f,\perp}\ ,\eea
where $\mC^{\bf 0}_f:=\mR^{\bf 0}_{f,\perp}$ is a massive Weyl
zero$\,$-form module coordinatized by $\widehat{\mathbf{X}}^{\bf
0}\,$, and $\mS_{f}$ is a massively contractible cycle
coordinatized by $\{\widehat{\mathbf{Z}}^{\bf
p},\widehat{\mathbf{Y}}^{\bf{p-1}}\}_{p>0}$. From \eq{Zeta2} it
follows that non-trivial potential modules arise iff $f$ assumes
critical values $\widetilde f$ such that
\bea \mbox{non-trivial $\widetilde\mR_{\widetilde
f}$}&\Leftrightarrow &{\rm
Ker}(\l\D^f_{[h_1]}+i\x_{(h_1+1)})~\cap~\mR^{\bf h_1}_{\widetilde
f,\perp}\ \neq\ \emptyset\label{statement}\eea
\emph{and} the elements of ${\rm Ker}(\l\D^f_{[h_1]}+i\x_{(h_1+1)})$
are directly sourced by Weyl zero-forms (\emph{i.e.}, if they have
maximal grade $\a=-1$).

On the other hand, as discussed in Section I.4.3.4, $\mC^{\bf 0}_f$
becomes reducible for the critical values $f^\pm_{I,N}$ of $f$
corresponding to the critical masses $\overline M{}^2_{I,N}$ where
primary Bianchi identities arise, and where $f^\pm$ refers to the
two solutions of the characteristic equation (see \eq{casimircalc}
below). A subset of these, that we denote by $f^\pm_I\,$, correspond
to critically massless fields with critical masses $\overline
M{}^2_I$ defined in item (iii) of I.4.3.4 for which the system
carries massless representations with a singular vector at the first
level reached from the $I^{\rm th}$ block of the spin.

As we shall see, interestingly enough, there is only one
critically massless  $\widetilde f\,$, and it is given by
\bea \widetilde f&=& f^-_{_1}\ ,\eea
and $\widetilde \mR_{f^-_{_1}}$ consists of the unitary ASV
potential\footnote{At non-unitary critical values, namely
$f^-_{p_I}$ for $p_{_I}=\sum_{J=1}^I h_{_J}$ with $I>1$, the
potential module of $\mR_{f^-_{p_I}}$ (as defined in Section
I.4.4.4) vanishes. It is still possible, however, to define a
non-unitary ASV-like potential by partially gauge-fixing the
massively contractible cycle. We thank E.~Skvortsov for
illuminating discussions on this point.}. 

The fact that in $AdS$, differently from the flat-space case, the only two modules that can be glued together are a $h_1$-form module and the infinite-dimensional Weyl zero-form module is a direct consequence of Weyl's complete reducibility theorem, which forbids indecomposable finite-dimensional modules for a semi-simple Lie algebra (see also the comments in Section I.3.4).

\subsection{\sc Radially reduced Weyl zero$\,$-form}\label{Sec:Mass}

\subsubsection{\sc Twisted-adjoint module and mass formula}

The radially reduced Weyl zero$\,$-form obeys
\begin{eqnarray}
\widehat{\mathbf{R}}^{\bf 1}:= \left(\widehat\nabla+\l N\D^f_{[0]}-i\,\widehat
e_{(1)}\right) \widehat{\mathbf{X}}^{\bf 0} & \approx & 0 \ ,\qquad
 \left( {\cal{L}}_\xi + \l\Delta^f_{[0]} \right)
\widehat{\mathbf{X}}^{\bf 0} \ \approx \ 0 \
\label{radialconstraint}
\end{eqnarray}
in $D+1$ dimensions. The pull-back of the latter to $AdS_D$ leaves with radius
$L=\lambda^{-1}$ can be obtained using
\bea {\cal L}_\xi \widehat{\mathbf{X}}^{\bf 0}&=&i_\xi d \widehat{\mathbf{X}}^{\bf 0}\
=\ i_\xi \widehat\nabla \widehat{\mathbf{X}}^{\bf 0}\ \approx \ i\,\widehat
\x_{(1)}\widehat{\mathbf{X}}^{\bf 0}\ , \eea
and that of $\widehat{\mathbf{R}}^{\bf 1}\approx 0$ can be
computed using $i^\ast_{L} \widehat e_{(i)}=e^a\widehat\b_{a,(i)}$
and $i^\ast_{L}\widehat\nabla=\nabla-i\,\lambda
\,e^a\widehat\x^B\widehat M_{Ba}$ with $\nabla:=d-\ft i 2\o^{ab}
\widehat M_{ab}\,$, where $\widehat M_{AB}$ and $\widehat M_{ab}$
act canonically on $\widehat\mm$-types and their $\mm$-subtypes.
Thus, at fixed $\lambda$ one has
\bea &\mathbf{R}^{\bf 1}\ :=\ [\nabla-i\,e^a \rho(P_a)]
\mathbf{X}^{\bf 0}\ \approx\ 0\ ,\quad \rho(P_a)\ =\
\lambda\widehat \x^B\widehat M_{Ba}+\widehat \b_{a,(1)}
\ ,&\label{0-form1}\\[5pt]
&(\lambda\Delta^f_{[0]}+i\,\widehat\x_{(1)})\widehat{\mathbf{X}}^{\bf
0}\ \approx\ 0\ ,\quad \Delta^f_{[0]}\ =\ \widehat N^1_1
+f_{[0]}(\widehat N^2_2,\dots,\widehat N^\nu_\nu)\ .&\label{0-form2}\eea

Let us restrict $\widehat{\mathbf{X}}^{\bf 0}$ to an irreducible
twisted-adjoint $\miso(2,D-1)$-module
\bea \widehat{\cal T}(\L\!\!=\!\!0;\overline M{}^2\!\!=\!\!0;
\widehat{\overline{\Th}})|_{\widehat\mm}&=&\bigoplus_{\a=0}^\infty\widehat\Th_{[0];\a}\
,\qquad \widehat\Th_{[0];\a}\ =\
\Big([s_{_1}+\a;1],[s_{_1};h_{_1}],\dots,[s_{_B};h_{_B}]\Big)
\label{widehatmmtype}\ ,\eea
where the $\widehat\mm$-types are realized in ${\cal S}_{D+1}$ and
descend from the smallest $\widehat\mm$-type
\bea \widehat{\overline{\Th}}&:=& \widehat\Th_{[0];0}\ =\
\Big([s_{_1};h_{_1}+1],[s_{_2};h_{_2}],\dots,[s_{_B};h_{_B}]\Big)\
, \eea
corresponding to the primary Weyl tensor of a tensor gauge field
$\widehat\varphi(\L\!\!=\!0;M^2\!\!=\!0;\widehat\Th)$ in
$\Real^{2,D-1}$ sitting in the $\widehat\mm$-type with shape
$\widehat\Th=\Big([s_{_1};h_{_1}],\dots,[s_{_B};h_{_B}]\Big)\,$.

The constraint \eq{radialconstraint} yields a $\mg_\l$-module
\bea \widehat {\cal S}(\L;f;\widehat{\overline\Th})&:=
&\left\{\widehat {\mathbf C}\in\widehat{\cal T}(\L\!=\!0;\overline
M{}^2\!=\!0; \widehat{\overline{\Th}})\ :\
\left(\l\D^f_{[0]}+i\widehat\x_{(1)}\right)\widehat{\mathbf C}\
\approx\ 0\right\}\ ,\label{calT}\eea
that is irreducible for generic values of $f$ and reducible with
an indecomposable structure for critical values of $f\,$,
determined by the value of
\bea C_{_2}\left[\mg_\l|\widehat {\cal
S}(\L;f;\widehat{\overline\Th})\right]&\equiv&\left.
\left({C}_{_2}[\mm]-L^2 \rho(P^2)\right)\right|_{\widehat {\cal
S}(\L;f;\widehat{\overline\Th})}\ ,\label{C2S}\eea
as discussed in Section I.4.3.4 and below. The operator
\bea \hspace{-1.5cm}
-\rho(P^2)&=&-\left(\l^2\widehat\x^B\widehat\x^C \widehat
M_{B}{}^a\widehat M_{Ca}+\l\widehat\x^B \left\{\widehat M_{B}{}^a,
\widehat\b_{a,(1)}\right\}+\widehat\b^a_{(1)}\widehat\b_{a,(1)}\right)\\[5pt]
&=& \l^2 \left(\frac12 \widehat M^{AB}\widehat M_{AB}-\frac12\,
\widehat M^{ab}\widehat M_{ab}\right)-\lambda\,\widehat\x^B
\left\{\widehat M_{B}{}^A,\widehat\b_{A,(1)}\right\}
-\widehat\b^A_{(1)}\widehat\b_{A,(1)}-(\widehat\x_{(1)})^2\ .\eea
Using the relations (\ref{Schur1}), it follows that
$-\rho(P^2)=\l^2(C_{_2}[\widehat\mm]- C_{_2}[\mm])+
i\,\l(2\widehat N^1_1+D)\widehat\x_{(1)}-(\widehat\x_{(1)})^2$
where $C_{_2}[\widehat\mm]:=\ft12 \,\widehat M^{AB}\widehat
M_{AB}$ and $C_{_2}[\mm]:=\ft12 \,\widehat M^{ab}\widehat M_{ab}$
are invariants for the action of $\widehat\mm$ and $\mm$ on
$\widehat\mm$-types and their $\mm$-subtypes in $\widehat{\cal
S}_{D+1}\,$. Further simplifications follow from
\bea i\l(2\widehat N^1_1+D)\widehat\x_{(1)}&\approx&
-\l^2(2\widehat N^1_1+D)\D^f_{[0]}\ ,\quad -(\widehat\x_{(1)})^2 \
\approx \ \l^2(\D^f_{[0]}+1)\D^f_{[0]}\ ,\eea
that hold in $\widehat {\cal S}(\L;f;\widehat{\overline\Th})$.
Hence $-\left.L^2\rho(P^2)\right|_{\widehat {\cal
S}(\L;f;\widehat{\overline\Th})}=C_{_2}[\widehat\mm]- C_{_2}[\mm]-
(\widehat N^1_1+f_{[0]})(\widehat N^1_1+D-1-f_{[0]})$, and
\eq{C2S} takes the simplified form
\bea C_{_2}\left[\mg_\l|\widehat {\cal
S}(\L;f;\widehat{\overline\Th})\right]&=&\left.\left(
C_{_2}[\widehat\mm]- (\widehat N^1_1+f_{[0]})(\widehat
N^1_1+D-1-f_{[0]}) \right)\right|_{\widehat {\cal
S}(\L;f;\widehat{\overline\Th})}\ .\label{casimircalc}\eea

It follows that any given value $\mu$ of
$C_{_2}\left[\mg_\l|\widehat {\cal
S}(\L;f;\widehat{\overline\Th})\right]\,$  corresponds to two mass
operators $f^{\pm}_{[0],\mu}$ given by
\begin{eqnarray}
f^\pm_{[0],\mu}\ :=\ \e_0+1\pm \sqrt{(\widehat
N^1_1+\e_0+1)^2+\mu- C_{_2}[\widehat\mm]}\ \equiv \
f_{\mu}^\pm(\widehat{N}^2_2,\widehat{N}^3_3,\ldots) \,,
\end{eqnarray}
where the last identity can be seen by expanding the Casimir
$C_{_2}[\widehat\mm]\,$ --- which is a nontrivial and non-constant operator in the
module ${\widehat {\cal S}(\L;f;\widehat{\overline\Th})}\,$.
{}From \eq{widehatmmtype} it can be seen that
\begin{eqnarray}
f^\pm_{[0],\mu}\ket{\widehat\Th_{[0];\a}}_{D+1}\equiv
f^\pm_{\mu}(\Th)\ket{\widehat\Th_{[0];\a}}_{D+1}\qquad
\forall\quad \alpha
\end{eqnarray}
where the $\a$-independent massive parameter $f^\pm_{\mu}(\Th)$ is
given by
\bea
f^\pm_{\mu}(\Th)&:=&f^\pm_{\mu}(\underbrace{s_1,\dots,s_1}_{\mbox{$h_{_1}$
entries}},\dots,\underbrace{s_{_B},\dots,s_{_B}}_{\mbox{$h_{_B}$
entries}},0,\dots)\ .\label{massf}\eea

Decomposing $\widehat {\cal S}(\L;f;\widehat{\overline\Th})$ under
$\mm$ the resulting smallest $\mm$-type is given by
\bea &\Th~=~ \Big([s_{_1};h_{_1}],\dots,[s_{_B};h_{_B}]\Big)\
,&\label{Theta}\eea
This shape is represented in the Schur module $\widehat {\cal
S}_{D+1}$ by the state
\bea \ket{\Th}_{D+1}&=&
\prod_{J=1}^B(\widehat\x_{(p_J+1)})^{s_{J,J+1}}
\ket{\widehat{\overline\Th}}_{D+1} \label{5.40} \ \eea
belonging to the subspace ${\cal S}_D\subset \widehat{\cal
S}_{D+1}\,$. The action of $\mg_\l$ on this state generates a
$\mg_\l$-module. Removing the ideals (as we shall see, at most one
non-trivial ideal arises) leaves an irreducible $\mm$-covariant
$\mg_\l$-module with smallest type $\Th$, \emph{viz.}
\bea {\cal T}(\L;\overline M{}^2;\Th)&:=&\bigoplus_{\a_r}
\Th_{\a_r}\ ,\qquad |\Th_{\a_r}|\ =\ |\Th|+\a\ ,\quad \Th_0\ =\
\Th\ .\eea
%

\subsubsection{\sc Proposition for indecomposability in the critical cases}
\label{Sec:Prop}

We claim that, if $f_{\mu}^\pm$ denotes the two roots of the
characteristic equation (\ref{casimircalc}) for a fixed value
$\mu=C_{_2}[\mg_\l]\,$, then
\bea \mbox{non-critical $f=f_{\mu}^\pm$}&:& \widehat {\cal S}
(\L;f;\widehat{\overline\Th})\
=\ {\cal T}(\L;\overline M{}^2_{f};\Th)\ ,\label{noncrit}\\[10pt]
\mbox{critical $f=f^\pm_{\mu}$}&:& \widehat {\cal S}
(\L;f;\widehat{\overline\Th})\ =\ \left\{\ba{ll}{\cal
T}(\L;\overline M{}^2_{f};\Th)\subsetplus {\cal T}(\L;\overline
M{}^{\prime 2}_{f};\overline\Th_{f})& \mbox{for $f=f^-_{\mu}$}
\\[5pt]
{\cal T}(\L;\overline M{}^2_{f};\Th) \supsetplus {\cal
T}(\L;\overline M{}^{\prime 2}_{f}; \overline\Th_{f})&\mbox{for
$f=f_\mu^+$}\ea\right. \label{indecomposable}\eea
where
\begin{itemize}
\item[(i)] for non-critical $f_{\mu}^\pm\,$, ${\cal
T}(\L;\overline M{}^2_{f_{\mu}^+};\Th)\cong$ ${\cal
T}(\L;\overline M{}^2_{f_{\mu}^-};\Th)\,$ (that is, $\overline
M{}^2_{f_{\mu}^+}\equiv\overline M{}^2_{f_{\mu}^-}\,$) is a
generically massive twisted-adjoint (irreducible) $\mg_\l$-module
(see Section I.4.3.1); and \item[(ii)] at critical $f_{\mu}^\pm\,$,
two dual indecomposable structures arise: The representation
matrices are transposed upon exchanging $f_{\mu}^-$ with
$f_{\mu}^+\,$.
\end{itemize}

In Section \ref{Sec:GenMass} we prove \eq{noncrit} in general.

In Section \ref{Sec:Irr} we then prove a part of the claim
\eq{indecomposable}, namely that in the critically massless cases
(see item (iii) I.4.3.4) it follows that
\bea \mbox{critically massless $f=f_I^-$}&:&
 {\cal T}(\L;\overline M{}^{\prime 2}_{f_I^-};
\overline\Th_{f_I^-})\ =\ {\cal T}(\L;\overline
M{}^{2}_{I};\overline\Th_I)\ ,
\label{masslessf-}\\[5pt]
&&{\cal T}(\L;\overline M{}^2_{f_I^-};\Th)\ =\ {\cal T}
(\L;\overline M{}^2_{I,S_{I,I+1}};\Th)\ ,\eea
where
\bea \overline\Th_I&:=&
\Big([s_{_1};h_{_1}],\dots,[s_{_{I-1}};h_{_{I-1}}],[s_{_I};h_{_I}+1],
[s_{_{I+1}};h_{_{I+1}}-1],[s_{_{I+2}};h_{_{I+2}}],\dots,[s_{_B};h_{_B}]\Big)\
. \label{overlineThI}\eea
We identify the above two modules, respectively, as the
twisted-adjoint representations of the primary Weyl tensors
$C_{\varphi_{_I}}$ and $C_{\chi_{_I}}$ of Metsaev's critically
massless gauge fields $\varphi_{_I}$, and of the corresponding
St\"uckelberg fields $\chi_{_I}$ associated with massive gauge
symmetries in the $I$th block \emph{i.e.}
\bea &C_{\varphi_{_I}}(\L;\overline M{}^{2}_{I};\overline\Th_I)~
\stackrel{\tiny{\rm integrate}}{\rightsquigarrow}~
\varphi_{{_I}}(\L;M^2_{I};\Th)\ , \qquad
C_{\chi_{_I}}(\L;\overline M{}^2_{I,S_{I,I+1}};\Th)~
\stackrel{\tiny{\mbox{integrate}}}{\rightsquigarrow}~
\chi_{{_I}}(\L;M^{\prime 2}_{I};\Th'_I)\ ,&\eea
where $\Th'_I$ is obtained by deleting one cell from the $I$th
block of $\Th$, \emph{viz.}
\bea & \Th'_I\ =\
\Big([s_{_1};h_{_1}],\dots,[s_{_{I-1}};h_{_{I-1}}],[s_{_{I}};h_{_{I}}-1],[s_{_I}-1;1],
[s_{_{I+1}};h_{_{I+1}}],\dots,\dots,[s_{_B};h_{_B}]\Big)\ .&\eea
These St\"uckelberg fields $\chi_{{_I}}(\L;M^{\prime
2}_{I};\Th'_I)$ are partially massless\footnote{Actually, setting
$I=2=B\,$, one obtains shapes $\widehat{\overline{\Th}}=(s,s,t)$
corresponding to a partially massless St\"uckelberg field
$\chi(s,t-1)$ having gauge invariance
$\delta\chi(s,t-1)=(\overline\nabla^{(1)})^{s-t+1}\epsilon(t-1,t-1)\,$.
This field reduces to a non-generic partially massless symmetric
tensor of \cite{Deser:2001us} iff $t=1\,$.}, in accordance with
our general definition in item (iv) of Section I.4.3.4, whenever
any block $I=2,\dots,B$ of $\Th$ is of height one
--- while the case $h_{_1}=1=I$ instead gives cut twisted-adjoint
modules, as defined in item (ii) of Section I.4.3.4. In the latter
case, these cut modules actually arise from factoring out a
tensorial $\mg_{\l}$-module (see item (i) of Section I.4.3.4) from
the Weyl zero-form module generated from a primary Weyl tensor of
the same shape as $\chi_{_I}(\Th^\prime_I)\,$.

The fact that $\chi_{{_I}}(\L;M^{\prime 2}_{I};\Th'_I)$ can be
factored out is a manifestation of the fact that there is
enhancement of gauge symmetry in the $I$th block: The radially
reduced $(D+1)$-dimensional gauge field
$\widehat\varphi(\widehat\Th)$ (with constrained radial
derivatives) decomposes into
$$\widehat\varphi(\widehat\Th)~\rightarrow~\varphi(\Th)\cup
\left\{\chi(\Th'_I)\right\}_{I=1}^B\cup\left\{\chi(\Th^{\prime\prime}_{I,J})
\right\}_{I,J=1}^B\cup\cdots\ ,$$ where $\Th'_I$ is obtained by
deleting one cell from the $I$th block of $\Th\,$,
$\Th^{\prime\prime}_{I,J}\equiv \Th^{\prime\prime}_{J,I}$ is
obtained by deleting one cell from the last row of the $J$th block
of $\Th'_I\,$, and so on. For generic mass all St\"uckelberg
fields are ``eaten'' by the massive field $\varphi(\L;M^2;\Th)\,$.
To examine the critical limit $\zeta^2_I:=(M^2-
M^2_I)\rightarrow0$ (fixed $I$) one may arrange the reduced field
content as follows:
\begin{eqnarray}
\widehat\varphi(\widehat\Th) &~\rightarrow~ &
\underbrace{\left\{\varphi(\Th) \cup
\left\{\chi(\Th'_J)\right\}_{J=1;J\neq I}^B\cup
\left\{\chi(\Th^{\prime\prime}_{J,K})\right\}_{J,K=1; J,K\neq I}^B
\cup\cdots\ ,\right\}}_{=: \widehat\varphi_{_I}}
\\
&&\cup  \quad \underbrace{\left\{\chi(\Th'_I)\cup
\left\{\chi(\Th^{\prime\prime}_{I,J}) \right\}_{J=1}^B
\cup\cdots\right\}}_{=:\widehat\chi_{_I}} \ .
\end{eqnarray}
In the limit $\zeta^2_I\rightarrow0$ there is enhancement of gauge
symmetry in the $I$th block which means that the
$\widehat\varphi_{_I}$ system decouples from $\widehat\chi_{_I}$
that becomes an independent --- generically partially massless ---
field system. One may remove $\widehat\chi_{_I}$ from the
equations of motion/action by fixing the gauge
$\widehat\chi_{_I}\stackrel{!}{=}0$ for $\zeta^2_I\neq 0$ (which
involves division by $\zeta^2_I$) and then send $\zeta^2_I$ to
zero. The equations of motion/action remain smooth in this limit
though the number of degrees of freedom change.
%
\subsection{\sc The generically massive case}\label{Sec:GenMass}

Let us show that the generic Weyl zero$\,$-form module $\mC^{\bf
0}(\L;\overline M{}^2;\Th)$ carries the massive representations
$\mD(e_{_0};\Th)$ with
\bea e_{_0}&=& \left\{\ba{l} f(\Th)\\[5pt] D-1-f(\Th)\ea\right.\ .\label{energies}
\eea\\[-15pt]
%
\subsubsection{\sc{Harmonic expansion}}
%
To this end we first construct the harmonic map
\cite{Iazeolla:2008ix}\footnote{The map extends to real Weyl tensors
in $({\cal S}^-_{\cal T}\mD^+_{D-1})\oplus ({\cal S}^-_{\cal
T}\mD^-_{D-1})$ where $\mD^\pm (\pm e_{_0};\Th)$ are lowest-weight
(+) and highest-weight ($-$) spaces.}
\bea {\cal S}_{\cal T}&:&\ \mD^+_{D-1}(e_{_0};\Th)\rightarrow\
\mC^{\bf 0}(\L;\overline M^2;\Th)\ ,\label{harmmap1}\eea
where $\mD^+_{D-1}(e_{_0};\Th):= \left[\mD^+(e_{_0};\Th)\otimes
{\cal S}_{D-1}\right]_{\rm diag}$ is the subspace of
$\mD^+(e_{_0};\Th)\otimes {\cal S}_{D-1}$ consisting of states
that are invariant under $\ms_{\rm
diag}=(\ms^+\oplus\widehat\ms)_{\rm diag}$ generated by
\bea M^{\rm diag}_{rs}&=&M_{rs}+\widehat M_{rs}\ ,\eea
where $M_{rs}$ act in $\mD^+(e_{_0};\Th)$ and $\widehat M_{rs}$
act in ${\cal S}_{D-1}\,$. The diagonal states are
\bea \ket{e_{_0}+m+n;\th}^+_{D-1}&:=& L^{+,(j_1)}\cdots
L^{+,(j_m)} L^+_{(k_1)}\cdots L^+_{(k_n)}\ket{e_{_0};\Th}^+_{D-1}\
,\label{A4exc} \eea
where $L^{+,(j)}:=\bar\beta^{r,(j)}L^+_r$ modulo traces~terms and
$L^+_{(j)}:=\beta^r_{(j)}L^+_r$ with
$\overrightarrow{w}(\th)=\overrightarrow{w}(\Th)+\overrightarrow{w}\,$
such that $w_j=\sum_{l=1}^m\d_{j,j_l}-\sum_{l=1}^n\d_{j,k_l}\,$,
$j_l\geqslant j_{l+1}\,$, $k_l\leqslant k_{l+1}\,$. The diagonal
ground state obeys
\bea L^-_r \ket{e_{_0};\Th}^+_{D-1}&=&0\ ,\quad
(E-e_{_0})\ket{e_{_0};\Th}^+_{D-1}\ =\ 0\ .\eea
For generic $e_{_0}$ there are no singular vectors.

Decomposing under $\ms^+$ yields
\bea \left.\mD(e_{_0};\Th)^+_{D-1}\right|_{\ms^+}&=&
\bigoplus_{\th\in{\cal S}(e_{_0};\Th)}\bigoplus_{n=0}^\infty
\Comp\otimes \left(x^n \ket{e(\th);\th}^+_{D-1}\right)\ ,\quad
x:=\delta^{rs}\,L^+_r L^+_s \ ,\eea
where ${\cal S}(e_{_0};\Th)$ is the set of $\ms^+$-types arising
in $\mD^+(e_{_0};\Th)\,$. This set contains a unique minimal
$\ms^+$-type $\th'_{_0}\,$. The lowest-spin state
$\ket{e^{\prime}_{_0};\th'_{_0}}^+_{D-1}$ is defined to be the
state of minimal $\ms^+$-type that minimizes the energy (see Fig.
\ref{fig1}). By its definition this state obeys
\bea L^+_{(j)}\ket{e'_{_0};\th'_{_0}}^+_{D-1}&=&0\qquad \forall j\
.\eea
Under the assumption that there are no singular vectors, it
follows that ($p_{_J}=\sum_{K=1}^J h_{_K}$)
\bea \ket{e^{\prime}_{_0};\th'_{_0}}^+_{D-1}&=& \prod_{J=1}^B
(L^{+}_{(p_J)})^{s_{J,J+1}}\ket{e_{_0};\Th}^+_{D-1}\ ,\quad
e^{\prime}_{_0}=e_{_0}+s_{_1}\ .\eea

To show \eq{harmmap1} it suffices to map $\mD^+_{D-1}(e_{_0};\Th)$
to the primary (massive) Weyl tensor $C(\Th)\in\mC^{\bf
0}(\L;\overline M^2;\Th)\,$. This tensor belongs to ${\cal S}_{D}$
due to \eq{5.40}
and decomposes under $\widehat\ms$ as follows:
\bea C(\Th)|_{\widehat\ms}&=&\sum_{\th\in \Th|_{\widehat \ms}}
\prod_{J=1}^B(\bar\beta_{_0}^{(p_J+1)})^{n_J(\th|\Th)} C(\th|\Th)\
, \quad C(\th|\Th)\ \in\ \O^{\bf 0}(U)\otimes \th\ ,\eea
where $n_{_J}(\th|\Th)$ is the number of boxes which are removed
from the $J^{\rm{\footnotesize{th}}}$ block of $\Th$ in order to
obtain $\th\,$. It follows that the smallest $\widehat\ms$-type of
$C(\Th)$, \emph{i.e.} its most electric component, is given by
$C(\th'_{_0}|\Th)\,$. Let us seek a harmonic expansion given by
the Ansatz (\emph{cf.} totally symmetric massless tensors
\cite{Iazeolla:2008ix})
\bea C(\th'_{_0}|\Th)&=&
\sum_{(e,\th)} {}^+\bra{C^{\ast}_{(e,\th)}} L^+\ket{\th'_{_0}| \Th}\ ,\\[5pt]
\ket{\th'_{_0}|\Th}&:=& \psi_{\th'_{_0}|\Th}(x)
\ket{e^{\prime}_{_0};\th'_{_0}}\ ,\quad \psi_{\th'_{_0}|\Th}(x)\
:=\ \sum_{n=0}^\infty x^n\psi_{n;\th'_{_0}|\Th}\ ,\eea
where:
\begin{itemize}
\item[i)]
${}^+\bra{C^{\ast}_{(e,\th)}}:=C^+_{(e,\th)}{}^+\bra{e;\th}\in
\left[\mD^+(e_{_0};\Th)\right]^\ast$ are states with fixed energy
and spin; \item[ii)] $L^+$ is a coset representative of $AdS_D$
acting in $\mD^+(e_{_0};\Th)$; and \item[iii)] the embedding
function $\psi_{\th'_{_0}|\Th}(x)$ is determined by demanding
$\ket{\th'_{_0}|\Th}$ to be an $\ms$-type in $\Th$, \emph{i.e.}
\bea \bar
\beta^{r,(1)}M_{0r}\ket{\th'_{_0}|\Th}-\mbox{traces}&=&0\quad
\mbox{where $M_{0r}=\ft12(L^+_r+L^-_r)\,$.}\eea
\end{itemize}
The latter condition amounts to that
\bea &M_{0\{r_1}\psi_{\th'_{_0}|\Th}(x)\ket{e^{\prime}_{_0};
\th'_{_0}}_{r_1(s_1)\}, \dots,r_{h_1}(s_1);\ldots;t_1(s_B),\dots,
t_{h_B}(s_B)} \ = \ 0& .\eea
Using the commutation relations \eq{algd-1} which yield the useful
relation
$$[L^-_r, x^n]=4nx^{n-1}(iL^+_s M_{rs}+L^+_r(E+n-\e_{_0}-1))\ ,$$
the embedding condition can be rewritten as
\bea
&L^+_{\{r_1}D_{_2}\psi_{\th'_{_0}|\Th}(x)\ket{e^{\prime}_{_0};
\th_{_0}}_{r_1(s_1)\}, \dots,r_{h_1}(s_1);\ldots;t_1(s_B),\dots,
t_{h_B}(s_B)} \ = \ 0& ,\eea
where $\{\cdots\}$ denotes symmetric and traceless projection, and
\bea D_{_2}\ :=\
4x\frac{d^2}{dx^2}+4(e_{_0}-\e_{_0})\frac{d}{dx}+1\ .\eea
It follows that there exists a regular embedding function given by
the rescaled Bessel function
\bea \psi_{\th'_{_0}|\Th}(x) \ = \
(\sqrt{x})^{-\nu}\,J_{\nu}(\sqrt{x}) \ , \qquad \nu \ = \
e_{_0}-\e_{_0}-1 \ . \eea\\[-25pt]
%
\subsubsection{\sc Characteristic equation}

Finally, the values \eq{energies} of the lowest energy $e_{_0}$
are determined by the characteristic equation
\bea
C_{_2}\left[\mg_\l|\mD(e_{_0};\Th)\right]&=&C_{_2}[\mg_\l|{\cal T}
(\L;\overline M{}^2;\Th)]\ ,\label{char}\eea
where
$C_{_2}[\mg_\l|\mD(e_{_0};\Th)]=e_{_0}(e_{_0}-2\e_{_0}-2)+C_{_2}[\ms|\Th]$
with $\ms=\mso(D-1)\,$, and the right-hand side is given by
\eq{casimircalc}. Using the parametrization of $\widehat\mm$-types
given in \eq{widehatmmtype}, one finds
\bea C_{_2}[\widehat\mm|\widehat\Th_{\a_i}]&=&
C_{_2}[\ms|\Th]+(s_{_1}+\a)(s_{_1}+\a+2\e_{_0}+2)\ ,\eea
leading to the following form of the characteristic equation
\eq{char}:
\bea (e_{_0}-\e_{_0}-1)^2&=&(\e_{_0}+1)^2+(s_{_1}+\a)
\left(s_{_1}+\a+2\e_{_0}+2\right)-\left(s_{_1}+\a+f(\Th)\right)
\left(s_{_1}+\a+2\e_{_0}+2-f(\Th)\right)\nn\\[5pt]
&=& (f(\Th)-\e_{_0}-1)^2\ ,\eea
with the roots \eq{energies}.
%
\subsection{\sc The critically massless case}\label{Sec:Irr}

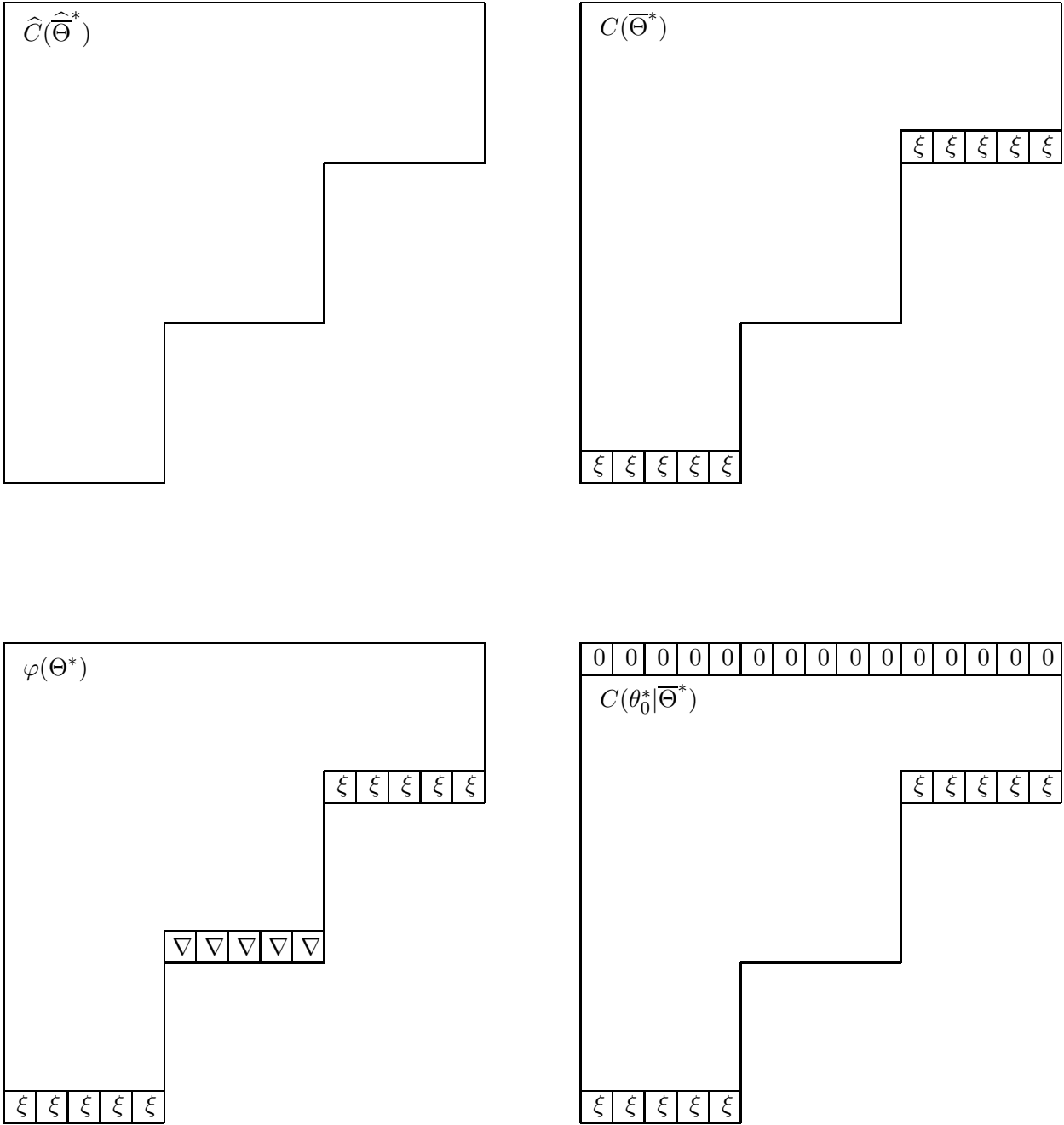
\begin{figure}
\begin{picture}(200,200)(0,0)
\multiput(0,200)(90,0){2}{\multiput(0,0)(0,-100){2}{
\put(0,0){\line(1,0){75}} \put(0,0){\line(0,-1){75}}
\put(50,-25){\line(1,0){25}} \put(50,-25){\line(0,-1){25}}
\put(25,-50){\line(1,0){25}} \put(25,-50){\line(0,-1){25}}
\put(75,0){\line(0,-1){25}} \put(0,-75){\line(1,0){25}} }}
\put(3,194){$\widehat C(\widehat{\overline \Th}^\ast)$}
\multiput(140,175)(5,0){5}{ \put(0,0){\line(0,1){5}}
\put(0,0){\line(1,0){5}} \put(5,5){\line(0,-1){5}}
\put(5,5){\line(-1,0){5}}\put(2,1.7){$\xi$}}
\multiput(90,125)(5,0){5}{ \put(0,0){\line(0,1){5}}
\put(0,0){\line(1,0){5}} \put(5,5){\line(0,-1){5}}
\put(5,5){\line(-1,0){5}}\put(2,1.7){$\xi$}}
\put(93,195){$C(\overline\Th^\ast)$}
\multiput(0,100)(90,0){2}{ \multiput(50,-25)(5,0){5}{
\put(0,0){\line(0,1){5}} \put(0,0){\line(1,0){5}}
\put(5,5){\line(0,-1){5}}
\put(5,5){\line(-1,0){5}}\put(2,1.7){$\xi$}}
\multiput(0,-75)(5,0){5}{ \put(0,0){\line(0,1){5}}
\put(0,0){\line(1,0){5}} \put(5,5){\line(0,-1){5}}
\put(5,5){\line(-1,0){5}}\put(2,1.7){$\xi$}}}
\multiput(25,50)(5,0){5}{ \put(0,0){\line(0,1){5}}
\put(0,0){\line(1,0){5}} \put(5,5){\line(0,-1){5}}
\put(5,5){\line(-1,0){5}}\put(1.3,1.3){$\nabla$}}
\put(3,95){$\varphi(\Th^\ast)$}
\multiput(90,95)(5,0){15}{\put(0,0){\line(0,1){5}}
\put(0,0){\line(1,0){5}} \put(5,5){\line(0,-1){5}}
\put(5,5){\line(-1,0){5}}\put(2,1.5){$0$}}
\put(93,90){$C(\th_0^\ast|\overline\Th^\ast)$}
\end{picture}
\caption{The four shapes associated with (1) the original strictly
massless primary Weyl tensor in $\mathbb{R}^{2,D-1}$; (2) the
reduced, critically massless primary Weyl tensor in $AdS_D\,$; (3)
the corresponding critically massless gauge potential in
$AdS_D\,$; and (4) the most electric component of (2).
}\label{Table:Thetas}
\end{figure}

\subsubsection{\sc Proof of indecomposability for massless cases}

Let us first show \eq{masslessf-}. To this end let us seek the
critical values $f_{_I}$ of $f$ for which the representation
$\rho(P_a)$ in the $\mg_\l$-module $\widehat{\cal
S}(\L;f_{_I};\widehat{\overline\Th})$ defined in \eq{calT} becomes
indecomposable with ideal
\begin{eqnarray}
{\cal T}_{\chi}(\Lambda;\overline{M}^2_{I,S_{I,I+1}};\Th) &:=&
{\rm Im}(\widehat{\xi}_{p_{_I}+1})\cap \widehat{\cal{S}}(\Lambda;f;
\widehat{\overline{\Th}})\ = \ \{ \widehat{\xi}_{(p_{_I}+1)}
\widehat{\mathbf{C}} \ {\rm for}\ \widehat{\mathbf{C}}\in
\widehat{\cal{S}}(\Lambda;f; \widehat{\overline{\Th}}) \}\ .
\end{eqnarray}
where
$$ p_{_I}\ := \ \sum_{J=1}^I h_{_J}\,.$$
Setting this ideal to zero amounts to constraining the Weyl
zero$\,$-form module as follows:
\bea \ba{l}\mbox{$(p_{_I}+1)$-row projection}\\ \mbox{(freezing $(p_{_I}+1)$st row}\\
\mbox{in the primary Weyl tensor)}\ea&:& \widehat\x_{(p_{_I}+1)}
\widehat{X}^{\bf 0}\ \approx\ 0\ . \label{construu} \eea Cartan
integrability of the above constraint, which is equivalent to the
ideal property of
\\ ${\cal T}_{\chi}(\Lambda;\overline{M}^2_{I,S_{I,I+1}};\Th)\,$,
amounts to
\bea \left(\lambda\,\widehat e_{(p_{_I}+1)}+i[\widehat\x_{(p_{_I}+1)},
\widehat e_{(1)}]\right)\widehat X^{\bf 0}\ \equiv
0\qquad\mbox{modulo
$\left(\l\D^f_{[0]}+i\widehat\x_{(1)}\right)\widehat X^{\bf
0}~\approx~0$}\ ,\eea
where $\D^f_{[0]}:=\widehat N^1_1+f_{[0]}$ are the scaling
dimensions appearing in the radial velocity constraints
\eq{constr1}. We claim that this equation has the unique
solution\footnote{One consequence of \eq{claim} is that $\widehat
X^{{\bf 1} a}$ and $\widehat e^a$ have the same scaling
dimensions, \emph{viz.} $({\cal L}_\x -\l)\widehat X^{{\bf 1}
a}=({\cal L}_\x-\l)\widehat e^a=0\,$, so that the ``graviton
field'' $\widehat X^{{\bf 1 }a}$ can consistently deform the
background vielbein $\widehat e^a$ upon switching on
interactions.}
\bea f\ =\ f_{p_{_I}}&:=& p_{_I}+1-\widehat N^{p_{_I}+1}_{p_{_I}+1}\ \Rightarrow\
\D^{f_{p_{_I}}}_{[p]}\ =\ \widehat N^{p+1}_{p+1}+\left\{\ba{ll}p_{_I}+1-p
-\widehat N^{{p_{_I}}+1}_{{p_{_I}}+1}&\mbox{for $p\leqslant p_{_I}-1$}\ ,
\\[5pt] p_{_I}-p-\widehat N^{p_{_I}}_{p_{_I}}& \mbox{for $p\geqslant p_{_I}$}
\ea\right.\ . \label{claim} \eea
We have shown this for $\nu=p_{_I}+1\geqslant 2$ (in which case
$\widehat\x_{(p_{_I}+1)}=\widehat\x_A\a^A_\n$ that simplifies the
calculations somewhat) and $\nu=p_{_I}+2=3$ using the explicit
expression \eq{betaused} for the cell operators. For fixed
$\widehat{\overline\Th}$ it follows that
$\widehat{\cal S}(\L;f_{p_{_I}};\widehat{\overline\Th})$ contains the
proper submodule ${\cal T}(\L;\overline
M{}^2_{I};{\overline\Th}_I)$ with primary type of shape
$\overline\Th_I$ given by \eq{overlineThI} represented in
$\widehat {\cal S}_{D+1}$ by
\bea \ket{\overline\Th_I}_{D+1}&=&
\prod_{\ba{c}{}\\[-30pt]{\scriptstyle J=1}\\[-10pt]
{\scriptstyle J\neq I}\ea}^B(\widehat\x_{(p_J+1)})^{s_{J,J+1}}
\ket{\widehat{\overline\Th}}_{D+1} \ ,
\nn \\
\widehat\x_{(p_J)}\ket{\overline\Th_I}_{D+1} &= 0& \, (J\neq I)\;,
\qquad \widehat\x_{(p_I+1)}\ket{\overline\Th_I}_{D+1}\ =\ 0\ ,
\label{primW}\eea
which means that $\ket{\overline\Th_I}_{D+1}\in{\cal S}_D\subset
\widehat{\cal S}_{D+1}\,$. This embedding implies that
$\rho(P_a)\ket{\overline\Th_I}_{D+1}$ cannot be anti-symmetrized
into the $I$th block. It follows that the generalized Verma module
${\cal V}^\ast(\L;\overline M{}^2_{I};{\overline\Th}_I)$ contains
a singular vector corresponding to the primary Bianchi identity
\bea \overline\nabla^{[s_{I+1}+1]} C(\L;\overline
M{}^2_I;\overline\Th_I)\ =\ 0\ .\eea
Integration yields the gauge field $\varphi(\L;M^2_{I};\Th)$ in
$AdS_D$ sitting in the same $\mm$-type $\Th$ as the generically
massive gauge field $\varphi(\L;M^2;\Th)$, given by \eq{Theta}.
According to the nomenclature of Section I.4.3.4 the field
$\varphi(\L;M^2_{I};\Th)$ is massless except if $I=B$ and $h_B=1$
in which case it is partially massless\footnote{If $B=1=h_{_B}\,$,
namely, only one block of height one (totally symmetric case),
this reduces to the case first investigated in
\cite{Deser:1983mm}, later revisited in~\cite{Deser:2001us}. See
also \cite{Zinoviev:2001dt,Skvortsov:2006at}.}.

Identifying $f_{p_{_I}}\equiv f^-_{p_{_I}}$ it follows that $C_{_2}[\mg_\l]$
assumes the same value in $\widehat{\cal
S}(\L;f_{p_{_I}}^+;\widehat{\overline\Th})$ where
\bea f_{p_{_I}}^+&:=& D-1-f^{p_{_I}}_-\ =\ \widehat N^{1+p_{_I}}_{1+p_{_I}}+D-p_{_I}-2 \ \equiv \ e^I_0\ .\eea
Hence $\widehat{\cal S}(\L;f_{p_{_I}}^+;\widehat{\overline\Th})$ must
consist of the same twisted-adjoint representations as
$\widehat{\cal S}(\L;f_{p_{_I}}^-;\widehat{\overline\Th})\,$. But
$\widehat{\cal S}(\L;f_{p_{_I}}^+;\widehat{\overline\Th})$ does not
contain ${\cal T}(\L;\overline M{}^2_I;\overline\Th_I)$ as an
ideal. We claim that the indecomposable structure of
$\widehat{\cal S}(\L;f_{p_{_I}}^+;\widehat{\overline\Th})$ takes the form
given in eq. \eq{indecomposable} for $f=f^+,$ in other words, the
decomposition order is reversed with respect to that of
$\widehat{\cal S}(\L;f_{p_{_I}}^-;\widehat{\overline\Th})\,$. It appears
to us that the reversed indecomposable structure cannot be
characterized by means of any algebraic subsidiary condition
involving $\widehat\xi^A$ contractions.

Referring to item (iv) in Section I.4.3.4 it is plausible that the following generalization of \eq{construu}:
\bea \ba{l}\mbox{$(p_{_I}+1)$-row projection ($k\geqslant 1$)}\\ \mbox{(reducing $k-1$ cells in $(p_{_I}+1)$st row}\\
\mbox{in the primary Weyl tensor)}\ea :\qquad
(\widehat\xi_{(p_{_I}+1)})^k\widehat{X}^{\bf 0}\ \approx\ 0\ ,
\label{construu2} \eea
which -- as we already have shown -- leads to mixed-symmetry massless fields if $k=1$, will give rise to mixed-symmetry partially massless fields if $k\geqslant 2$, since the projection then creates a block of height one in the primary Weyl tensor. We leave this for future work.

\subsubsection{\sc Harmonic expansion via most electric primary Weyl tensor}

In the critical limit the larger of the two characteristic
energies in \eq{energies} becomes $e_{_0}^I:=s_{_I}+D-2-p_{_I}$
corresponding to the massless lowest-weight irrep
$\mD^+(e^I_{_0};\Th)$ with singular vector
\bea L^{+}_{(p_I)}\ket{e^I_{_0};\Th}^+_{D-1}&\approx& 0\
,\label{singvect}\eea
presented here as a state in the doubled space
$[\mD^+(e^I_{_0};\Th)\otimes {\cal S}_{D-1}]_{\rm diag}\,$, using
the notation of Section \ref{Sec:GenMass}. Let us show that this
irrep is carried by $C(\L;\overline M{}^2_{I};\overline\Th_I)\,$,
\emph{i.e.} that there exists a harmonic map
\bea {\cal S}^{C_{\rm electric}}&:&\left[\mD^+(e^I_{_0};\Th)
\otimes {\cal S}_{D-1}\right]_{\rm diag}\ \rightarrow\ \mC^{\bf
0}(\L;\overline M{}^2_{I};\overline\Th_I)\ ,\label{hmapC}\eea
with reference state given by the most electric component of the
primary Weyl tensor.

To this end we note that the existence of the singular vector
\eq{singvect} implies that
\bea \left.\mD(e^I_{_0};\Th)^+_{D-1}\right|_{\ms^+}&=&
\bigoplus_{\th\in{\cal S}}\bigoplus_{n=0}^\infty \Comp\otimes
\left(x^n \ket{e(\th);\th}^+_{D-1}\right)\ ,\quad x:=\d^{rs}L^+_r
L^+_s\ ,\eea
where ${\cal S}(e^I_{_0};\Th)\,$, the set of $\ms^+$-types arising
in $\mD^+(e^I_{_0};\Th)\,$, is smaller than in the massive case
presented above since the operator $L^+_{(p_I)}$ annihilates
$\ket{e^I_{_0};\Th}^+_{D-1}\,$. In other words, the lowest-spin
state $\ket{e^{I\prime}_{_0};\th^I_{_0}}^+_{D-1}$ is now given by
\bea \ket{e^{I\prime}_{_0};\th^I_{_0}}^+_{D-1}&=& \prod_{J=1,J\neq
I}^B (L^{+}_{(p_J)})^{s_{J,J+1}}\ket{e^I_{_0};\Th}^+_{D-1}\ ,\quad
e^{I\prime}_{_0}=e^I_{_0}+s_{_1}-s_{_I}+s_{_{I+1}}\ .\eea
On the other hand, the primary Weyl tensor
$C(\overline\Th_I)\in\mC^{\bf 0}(\L;\overline
M{}^2_I;\overline\Th_I)\,$ belongs to ${\cal S}_{D}$ due to
\eq{primW}. Therefore, it decomposes under ${\ms}$ as follows:
\bea \left.C(\overline\Th_I)\right|_{\ms}&=&
\sum_{\overline\th\in\overline\Th_I|_{\ms}}
(\bar\beta_{_0}^{(p_I+1)})^{n_I} \prod_{J=1,J\neq
I}^B(\bar\beta_{_0}^{(p_J)})^{n_J} C(\overline\th|\overline\Th_I)\
,\quad C(\overline\th|\overline\Th_I)\ \in\ \O^{\bf 0}(U)\otimes
\overline\th\ .\eea
We then seek a harmonic expansion for the most electric component,
\emph{viz.}
\bea C(\th^I_{_0}|\overline\Th_I)&=& \sum_{(e,\th)}
{}^+\bra{C^{\ast}_{(e,\th)}} L^+\ket{\th^I_{_0}|\overline\Th_I}\ ,
\qquad \ket{\th^I_{_0}|\overline\Th_I}\ :=\
\psi_{\th^I_{_0}|\overline\Th_I}(x)
\ket{e^{I\prime}_{_0};\th^I_{_0}}\ ,\eea
where the embedding function obeys
\bea &M_{0\{r_1}\psi_{\th^I_{_0}|\overline\Th_I}(x)
\ket{e^{I\prime}_{_0};\th^I_{_0}}_{r_1(s_1)\},
\dots,r_{h_1}(s_1);\ldots;t_1(s_B),\dots, t_{h_B}(s_B)} \ = \ 0&
.\eea
This implies the second-order differential equation $D_{_2}
\psi_{\th^I_{_0}|\overline\Th_I}=0$ with
\bea D_{_2}\ =\
4x\frac{d^2}{dx^2}+4(\e_{_0}+1+s_{_{I+1}}-p_{_I})\frac{d}{dx}+1\
,\eea
leading to the regular embedding function
\bea \psi_{\th^I_{_0}|\overline\Th_I}(x) \ = \
(\sqrt{x})^{-\nu_I}\,J_{\nu_I}(\sqrt{x}) \ , \qquad \nu_{_I} \ = \
\e_{_0}+s_{_{I+1}}-p_{_I} \ . \eea
Thus $\mD(e^I_{_0};\Th)$ is carried by $C(\overline\Th_I)$, and
hence by all elements of $\mC^{\bf
0}(\L;\overline{M}^2_I;\overline\Th_I)$. $\Square$

\subsubsection{\sc Harmonic expansion via most magnetic Weyl tensor and shadow}

One can also show that there exists a harmonic map with reference
state given by the most magnetic component of the primary Weyl
tensor as follows:
\bea {\cal S}^{C_{\rm
magn}}&:&\left[\mD^+(\check{e}^I_{_0};\overline\Th_I) \otimes
{\cal S}_{D-1}\right]_{\rm diag}\ \rightarrow\ \mC^{\bf
0}(\L;\overline M{}^2_{I};\overline\Th_I)\ ,\label{hmapC2}\eea
where the lowest-energy is given by
\bea \check{e}^I_{_0}&=&1+p_{_I}-s_{_{I+1}}\ .\eea
This is a direct generalization of the special case $B=1$, $h_1=1$
spelled out for composite massless fields in
\cite{Iazeolla:2008ix}.

The critical limit of the smaller energy eigenvalue in
\eq{energies} is given by $\widetilde e_{_0}^I:=D-1-e^I_{_0}$.
This energy corresponds to the shadow $\mso(2,D-1)$- module
$\mD(\widetilde e^I_{_0};\Th)\,$. This module has a different
pattern of singular vectors. It has no singular vector with rank
smaller than $|\Th|\,$. Hence its lowest-spin state
$\ket{\widetilde e'_0;\th'_0}$ has the same $\ms^+$-spin $\th'_0$
as in the generically massive case analyzed in Section
\ref{Sec:GenMass}. It follows that there exists a harmonic map
\bea {\cal S}^{\varphi}&:&\left[\mD^+(\widetilde
e^I_{_0};\Th)\otimes {\cal S}_{D-1}\right]_{\rm diag}\
\rightarrow\ \varphi(\L;\overline{M}^2_{I};\Th)\ , \eea
so that $\varphi(\L;\overline{M}^2_I;\Th)$ carries
$\mD^+(\widetilde e^I_{_0};\Th)\,$.

\begin{figure}[!h]
\begin{center}
\unitlength=.6mm
\begin{picture}(150,180)(0,-10)
\put(0,0){\vector(1,0){150}} \put(0,0){\vector(0,1){150}}
\put(150,-10){Rank } \put(-25,150){Energy}
\put(60,30){$\bullet$}\put(27,60){$\bigstar$}
\put(60,32){\line(-1,1){30}}\put(60,32){\line(1,1){100}}
\put(30,62){\line(0,1){80}} \multiput(0,60)(5,0){6}{-}
\multiput(0,30)(5,0){12}{-} \multiput(30,0)(0,5){12}{$\bf
\shortmid$} \multiput(60,0)(0,5){6}{$\bf \shortmid$}
\put(-10,60){$e'_{_0}$} \put(-10,30){$e_{_0}$}
\put(30,-10){$|\th'_{_0}|$} \put(60,-10){$|\Th|$}
\end{picture}
\end{center}
\caption{{\small{A lowest-weight module $\mD(e_{_0};\Th)$ with its
lowest-energy state $\ket{e_{_0};\Th}$ and the lowest-spin state
$\ket{e'_{_0};\th'_{_0}}$ indicated by the $\bullet$ and
$\bigstar$, respectively.}}} \label{fig1}
\end{figure}
%
%

\subsection{\sc{Unitarizable ASV potential}}\label{Sec:ASV}

\subsubsection{\sc Occurrence of non-trivial potential module}
%
Let us consider the subsector $\mR_f(\L;\Th)\subset \mR_f$ obtained
by constrained radial reduction of the gauge field
$\widehat\varphi(\L\!\!=\!0;\widehat \Th)$ with
$\widehat\Th=([s_{_1};h_{_1}],\dots,[s_{_B};h_{_B}])$ so that
$f_{[0]}$ can be replaced by its eigenvalue $f(\Th)\,$ and let us
denote its potential module by $\widetilde \mR_f(\L;\Th)\,$. For
generic $f\,$, $\mR^{\bf h_1}_{f,\perp}$ belongs to $\mS_f(\Th)\,$,
that in its turn implies that all $p$-forms with $p>0$ belongs to
$\mS_f(\Th)\,$. This is so even in case ${\rm
Ker}(\l\D^f_{[p_I]}+i\x_{(p_I+1)})\cap\mR^{\bf p_I}_{f,\perp}$ is
non-empty for some $I>1\,$, because higher-degree potentials are not
sourced directly by the Weyl zero$\,$-form. Thus, $\mR_f(\L;\Th)$
contains a non-trivial potential (in the sense explained in Section
I.4.4.4) iff ${\rm
Ker}(\l\D^f_{[h_1]}+i\x_{(h_1+1)})~\cap~\mR^{\mathbf{h_1}}_{f,\perp}\neq
\emptyset\,$, as already stated in \eq{statement}.

Since $\x_{(h_1+1)}$ is nilpotent, the kernel is spanned by the
$\widehat\Th_{[h_{_1}];\a(k_1)}$, $k_{_1}=0,\dots,s_{_{1,2}}-1$,
that obey
\bea \D^f_{[h_1];\a(k_{_1})}&=&0\ ,\qquad
(\D^f_{[h_1]}-\Delta^f_{[h_1];\a(k_{_1})})\widehat\Th_{[h_1];\a(k_{_1})}\
:=\ 0\ .\eea
Restricting our analysis to the critically massless values,
\emph{i.e.}
\bea I=1&:& \D^{f^-_1}_{[h_1]}\ =\ \widehat N^{h_1+1}_{h_1+1}-\widehat N^{h_1}_{h_1}\ ,\quad \\[5pt]
I>1&:& \D^{f^-_I}_{[h_1]}\ =\ \widehat
N^{h_1+1}_{h_1+1}+p_{_I}+1-h_{_1}-\widehat N^{p_I+1}_{p_I+1}\
,\eea
it follows that
\bea I=1&:& \D^{f^-_1}_{[h_1];\a(k_{_1})}\ =\ s_{_2}+k_{_1}-s_{_1}+1\ =\ 0\quad \mbox{iff $k_{_1}=s_{_{1,2}}-1$}\ ,\\[5pt]
I>1&:& \D^{f^-_I}_{[h_1];\a(k_{_1})}\ =\
s_{_2}+k_{_1}+p_{_I}+1-h_{_1}-s_{_I}\ >\ 0\quad\mbox{for
$k_{_1}=0,\dots,s_{_{1,2}}-1$}\ .\eea
Thus it is only the unitarizable critical value $f^-_{_1}$ that
yields a potential module, \emph{i.e.}
\bea \left.\mR_{f^-_{I}}\right|_{\mg_\l}&=&\widehat
\mS_{f^-_{_I}}\oplus \mR'_{f^-_{I}}\ ,\qquad
\left.\mR'_{f^-_{I}}\right|_{\mg_\l} \ =\
\left\{\ba{ll}\widetilde\mR^{\mathbf{h_1}}_{f^-_{_1}}\supsetplus
\mC^{\bf 0}_{f^-_{_1}}&I=1\\[5pt]\mC^{\bf 0}_{f^-_{I}}&
I>1\ea\right.\ .\label{mRfi}\eea
The unitarizable $h_{_1}$-form potential
$\widehat{U}^{\mathbf{h_1}}(\widehat\Th_{[h_1]})$ sits in the
$\mg$-type
\bea \widehat\Th_{[h_1]}&:=&
\left([s_{_1}-1;h_{_1}+1],[s_{_2};h_{_2}],\dots,[s_{_B};h_{_B}]\right)\
,\eea
which we identify as the ASV gauge potential \cite{Alkalaev:2003qv}.

The embedding of $\widehat{U}^{\mathbf{h_1}}(\widehat\Th_{[h_1]})$
into $\widehat{\mathbf{X}}^{{\mathbf{h_1}}}$ is given by
\bea \widetilde\mR^{\mathbf{h_1}}_{f^-_{_1}}\ := \ {\rm
Ker}(\l\D^{f^-_{_1}}_{[h_1]}
+i\x_{(h_1+1)})\cap\mR^{\mathbf{h_1}}_{f^-_{_1},\perp}& \ni&
\widehat{\mathbf{X}}^{{\mathbf{h_1}}}_{\rm ASV}\ =\
e^{i\widehat\x_{(h_1+1)}\over\l}
\widehat{U}^{\mathbf{h_1}}(\widehat\Th_{[h_1]})\
,\label{kernel}\eea
and the resulting generalized curvature constraint takes the
form\footnote{The exponential in \eq{kernel} ``untwists'' the
``twisted'' covariant derivative in the constraint on
$\widehat{\mathbf{X}}^{\mathbf{h_1}}\,$. The zero$\,$-form
constraint cannot be untwisted, however, since
$(\l\D^{f^-_{_1}}_{[0]}+i\,\x_{(1)}) \widehat{\mathbf{X}}^{\bf
0}\approx 0$ implies that $\exp(-{i\x_{(1)}\over\l})
\widehat{\mathbf{X}}^{\bf 0}$ is logarithmically divergent.}
\bea
 \widehat R^{\mathbf{h_1+1}}_{\rm ASV}&:=&
 (\widehat \nabla-iN\widehat\x_{(h_1+1)})\widehat U^{\mathbf{h_1}}
 -i\,\widehat e_{(1)}\cdots \widehat e_{(h_1+1)}
 \widehat {\mathbb P}(h_{_1}+1,1) \widehat{\mathbf{X}}^{\bf 0}\ \approx\
 0\ .\eea
Its pullback to a fixed $AdS_D$ leaf with radius $L$ reads
\bea
 R^{\mathbf{h_1+1}}_{\rm ASV}&:=& i^\ast_L\widehat\nabla U^{\mathbf{h_1}}
 -i \,~i^\ast_L(\widehat e_{(1)}\cdots \widehat e_{(h_1+1)}) \widehat {\mathbb P}
 (h_{_1}+1,1) \mathbf{X}^{\bf 0}\ \approx\
 0\ ,\label{ASVeq}\eea
where $i^\ast_L \widehat e_{(i)}=e^a\widehat\b_{a,(i)}$ and
$i^\ast_L\widehat \nabla=d-\ft i2\O^{AB}\widehat M_{AB}$ with
$\widehat M_{AB}$ acting canonically on $\mg$-types and
$\O^{AB}=(\o^{ab},\l e^a)$ being the flat $\mg$-connection. 

We stress again that, although Weyl's complete reducibility theorem only allows gluing  the infinite-dimensional zero-form module to one module in higher form-degree, the latter need not necessarily be the unitary ASV potential. More precisely, taking different combinations of the fields occurring on the right-hand side of Eq. \eq{Zeta2}\,, possibly together with some zero-forms, it should be possible to find non-unitary ASV potential in form-degree higher that $h_1$ that will appear directly glued to the corresponding Weyl zero-form module in the reduced equation.

\label{Sec:UnitarASV}
\subsubsection{\sc On $\s^-$-$\,$cohomology for unitarizable ASV gauge potential}

The constraints $R^{\bf 1}\approx0$ and $R^{\mathbf{h_1+1}}_{\rm
ASV}\approx 0$ have the form $(\nabla+\s_{_0}) X\approx0$ where
$X\in\mR'_{ASV}:=\mR'_{f^-_{_1}}=
\widetilde\mR_{f^-_{_1}}^{\mathbf{h_1}}\supsetplus \mC^{\bf 0}$,
$\nabla=d-\frac{i}{2}\,\omega^{ab}\widehat M_{ab}$ and
$\s_{_0}=\s^{\bf 1}_{0,\bf 0}+\s^{[h_1+1]}_{0,\bf
0}+\s^{[h_1+1]}_{0,[h_1]}$ with
\bea \s^{\bf 1}_{0,\bf 0}&=&-ie^a\rho^{\bf 0}_{0,\bf 0}(P_a)\ =\ -
ie^a(\l\widehat\x^B\widehat M_{Ba}+\widehat\beta_{a,(1)})\ ,\\[5pt]
\s^{[h_1+1]}_{0,[h_1]}&=&-ie^a\rho^{[h_1]}_{0,[h_1]}(P_a)\ =\ -
ie^a\widehat\x^B\widehat M_{Ba}\ ,\\[5pt]
\s^{[h_1+1]}_{0,\bf 0}&=&-\ft{i}{h_1+1}e^{a}\rho^{[h_1]}_{0,\bf
0}(P_a|e)\ =\ -\ft{i}{h_1+1}e^{a_1}\cdots
e^{a_{h_1+1}}\beta_{a_1,(1)}\cdots \beta_{a_{h_1+1},(h_1+1)}{\mathbb
P}(h_1+1,1)\ .\hspace{1cm}\eea
The corresponding triangular module
$\mT'_{ASV}=\bigoplus_{q\in\integ}\mR'_q$ where
$\mR'_{_0}:=\mR'_{ASV}\,$. If $e^a$ is non-degenerate, then the
maps $\s_q=(-1)^{q(1+\s_{_0})}\s_{_0}$ decompose into
$\s_q=\s^-_q+\s^+_q$ with respect to the ordering
$g:\mR'_q\rightarrow \mathbb N$ defined by
\bea
g\left({\mR^{\prime}}^{\mathbf{p}_{\a_i}+\mathbf{q}}_q(\Th_{[p_{\a_i}];\a_i})\right)\
:=\ g(\a)\ :=\ \a+s_{_{1,2}}\ ,\eea
where the primary type-setting index $\a\in s_{_{2,1}}+\mathbb N$
is defined by
\bea s_{_{2,1}}\leqslant \a\leqslant s_{_2}&:& \Th_{[h_1];\a_i}\in\left.\widehat\Th_{[h_1]}\right|_{\mm}\ ,\quad |\Th_{[h_1];\a_i}|\ :=\ |\widetilde\Th|+\a+s_{_{1,2}}\ ,\\[5pt]
0\leqslant \a&:& \Th_{[0];\a_i}\in\left.{\cal
T}(\overline\Th)\right|_{\mm}\ ,\quad |\Th_{[0];\a_i}|\ :=\
|\overline\Th|+\a\ ,\eea
where $\widetilde\Th$ is the smallest $\mm$-type in
$\widehat\Th_{[h_1]}$, \emph{viz.}
\bea\widetilde\Th&=&([s_{_1}-1;h_{_1}],\Xi)\ ,\quad \Xi\ :=\
([s_{_2}-1;h_{_2}],\dots,[s_{_B}-1;h_{_B}])\ ,\eea
and the secondary type-setting index $i=1,\dots,n_\a$ takes into
the account degeneracies (due to that there are many internal
$\mm$-types of fixed rank). The resulting
$\mR'_q=\bigoplus_{k\in\mathbb N} T'_{k,q}$ where
 \bea T'_{k,q}&:=&g^{-1}(k)\cap \mR'_q\ =\ \bigoplus_{\a,g(\a)=k} \mR^{\prime\mathbf{p_\a+q}}_q(\Th_\a)\ =\ T^{\prime \bf 0}_{k,q}\oplus
 T^{\prime \bf h_1}_{k,q}\ ,\\[5pt]
 T^{\prime \bf p}_{k,q}&=&\left.\left(i_{\th^{a_1}}\cdots i_{\th^{a_{p+q}}}
 \O^{\bf p+\bf q}(U)\otimes R'_{[p];\a(g)}\right)\right|_{\mm}\ ,
 \qquad p\ =\ 0\,,\;h_1\ ,\\[5pt]
 R'_{[p];\a}&=&\bigoplus_{i=1}^{n_\a}\Th_{[p];\a_i}\ .\eea
The space $R'_{[h_1];\a(g)}$ ($g\in\{0,\dots,s_1\})$ is obtained
from $R'_{[h_1];\a(0)}$ by inserting $g$ cells below the first
block while adhering to the rules of Young diagrams. This amounts
to
\bea R'_{[h_1];\a(g)}&=&([s_1-1;h_1],\Xi\diamond (g))\ ,\\[5pt]
\Xi\diamond(g)&:=& \left\{\Xi'\in \Xi\otimes(g)\,:\ |\Xi'|=|\Xi|+g\,,\ {\rm width}(\Xi')\leqslant s_1-1\right\}\ ,\\[5pt]&=&
\left\{\Xi'\in \Xi ~\widetilde{\otimes}(g)\,:\ {\rm
width}(\Xi')\leqslant s_1-1\right\}\ ,\eea
where $\otimes$ is the direct product of $\mm$-tensors and
$\widetilde\otimes$ is the direct product of $\msl(D)$-tensors. It
follows that
\bea g\leqslant s_{_{1,2}}&\Rightarrow & \diamond\ =\
\widetilde\otimes\ .\eea
such that
\bea T'_{[h_1];q,g}&\stackrel{s_{1,2}\geqslant
g}{\cong}&\bigoplus_{k=0}^p\bigoplus_{\tiny\ba{c} k_1\leqslant
h_1\,,\ k_3=0,1\\k_2=k-k_1-k_3\geqslant 0\\p_1\leqslant
h_1\\p_2=p-k-p_1\geqslant 0\ea}
\!\!\!\left[\ba{c}\left[\ba{c}[s_1-1;h_1-k_1]\\ {[s_1-2;k_1]}\ea\right]\widetilde\otimes[p_1]\\
i_{[k_2]}\Xi~\widetilde
\otimes(g-k_3)~\widetilde\otimes[p_2]\ea\right]\ ,\qquad\eea
where $i_{[k_2]}\Xi$ denotes the direct sum of shapes given by the
contraction of $k_{_2}$ anti-symmetric cells from the shape
$\Xi\,$.

In what follows we examine the $\s^-$-$\,$cohomology in more
detail in the cases $h_1=1,2\,$.

\textbf{The example $\Th=(2,1)$}:
\noindent The irreducible module carrying the unitary
representation $\mD(D-1;(2,1))$ is given by
\bea \mR'_{ASV}&=&\left\{\underbrace{U^{\bf
1}\widehat{[3]}}_{g=0,1};\underbrace{C^{\bf 0}[2,2]}_{g=1}
;\underbrace{X^{\bf 0}[3,2],X^{\bf 0}(3,2)}_{g=2};\cdots\right\}\
.\eea
The corresponding triangular module
$\mT'=\mR'_{-1}\oplus\mR'_{_0}\oplus
\mR'_1\oplus\mR'_{_2}\oplus\cdots$, with variables
$\mR'_{_0}=\mR'$, parameters in $\mR'_{-1}=\left\{\e^{\bf
0}\widehat{[3]}\right\}$ with $g=0,1$, constraints in
$\mR'_1=\left\{R^{\mathbf{2}}\widehat{[3]};R^{\bf 1}[2,2];R^{\bf
1}[3,2],R^{\bf 1}(3,2);\cdots\right\}$ with $g\geqslant 0$, and
first level of Bianchi identities in
$\mR'_{_2}=\left\{Z^{[3]}\widehat{[3]};\cdots\right\}$ with
$g\geqslant 0\,$.

The non-trivial $\s^-$-$\,$cohomology for $q\leqslant 1$ is a
parameter $\e[2]$ at $g=0$, two fields $\varphi(2,1)$ and $S(1)$
at $g=0$, two Proca-like field equations at $g=0$ and one
Labastida-like field equation at $g=1\,$. The degree $1$ module is
``glued'' to the degree $0$ module via the Weyl tensor $C(2,1)$ in
$T^{\prime \bf 0}_{q=0,g=1}$ via a constraint in $T^{\prime \bf
1}_{q=1,g=0}\,$.

\textbf{The case $h_{_1}=1$, $B\geqslant 2$ and
$s_{_1}-s_{_2}\geqslant 4$}: In this generic case the triangular
module (see Fig. \ref{fig2})
\bea
\mT'_{ASV}&=&\mR_{-1}\oplus\mR'_{_0}\oplus\mR'_{_1}\oplus\mR'_{_2}
\oplus\mR'_{_3}\oplus\cdots\ \ni\ (\e,X,R,Z,Z_3,\dots)\ ,\eea
where $\e\in \O^{\bf 0}(U)$ for $\a<0$ and $\e\equiv 0$ for
$\a\geqslant 0\,$. For $s_1-s_2\geqslant4$ the lowest
$\s^-$-chains are
\bea g+q=-1&:& 0\hookrightarrow \e^{\bf 0}(R'_{\a(0)})\rightarrow 0\ ,\\[5pt]
g+q=0&:& 0\hookrightarrow \e^{\bf 0}(R'_{\a(1)})\rightarrow X^{\bf 1}(R'_{\a(0)})\rightarrow 0\ ,\\[5pt]
g+q=1&:& 0\hookrightarrow \e^{\mathbf{0}}(R'_{\a(2)})\rightarrow X^{\bf 1}(R'_{\a(1)})\rightarrow R^{\mathbf{2}}(R'_{\a(0)})\rightarrow 0\ ,\\[5pt]
g+q=2&:& 0\hookrightarrow \e^{\mathbf{0}}(R'_{\a(3)})\rightarrow X^{\bf 1}(R'_{\a(2)})\rightarrow R^{\mathbf{2}}(R'_{\a(1)})\rightarrow Z^{\mathbf{3}}(R'_{\a(0)})\rightarrow 0\ ,\\[5pt]
g+q=3&:& 0\hookrightarrow \e^{\mathbf{0}}(R'_{\a(4)})\rightarrow
X^{\bf 1}(R'_{\a(3)})\rightarrow
R^{\mathbf{2}}(R'_{\a(2)})\rightarrow
Z^{\mathbf{3}}(R'_{\a(1)})\rightarrow
Z^{\mathbf{4}}_3(R'_{\a(0)})\rightarrow 0\ ,\hspace{1.3cm}\eea
where the $\mm$-content of the parameters is given by
$\e^{\mathbf{0}}(R'_{\a(g)})\in\mx{[}{c}{s-1\\\Xi~\widetilde\otimes
(g)}{]}\,$. The chain with $g+q=-1$ contains the differential
gauge parameter given by
\bea H_{-1,0}(\sigma^-)&\ni&\e[(s_1-1);\Xi]\ .\eea
The chain with $g+q=0$, where \bea X^{\bf
1}(R'_{\a(0)})&\in&\mx{[}{c}{s_1\\\Xi}{]}\oplus\mx{[}{c}{s_1-1\\\Xi~\widetilde\otimes(1)}{]}\oplus
\mx{[}{c}{s_1-2\\\Xi}{]}\oplus\mx{[}{c}{s_1-1\\i_{[1]}\Xi}{]}\
,\eea
leaves dynamical tensor gauge fields in
\bea H_{0,0}(\sigma^-)&\ni & \varphi[(s_1);\Xi]\oplus
A[(s_1-2);\Xi]\oplus S[(s_1-1);i_{[1]}\Xi]\ \eea where $\varphi$,
$A$ and $S$ denote the three Lorentz-irreps that occur in the
dynamical metric-like field. We shall use similar notation below.
The chain with $g+q=1$, where
\bea X^{\bf
1}(R'_{\a(1)})&\in&\mx{[}{c}{s_1\\\Xi~\widetilde\otimes(1)}{]}\oplus
 \mx{[}{c}{s_1-1\\\Xi~\widetilde\otimes(2)}{]}\oplus\mx{[}{c}{s_1-1\\\Xi~\widetilde\otimes[2]}{]}\oplus
 \mx{[}{c}{s_1-2\\\Xi~\widetilde\otimes(1)}{]}\nn\\[5pt]&&\oplus\mx{[}{c}{s_1-1\\i_{[1]}
 \Xi~\widetilde\otimes(1)}{]}\oplus\mx{[}{c}{s_1-1\\\Xi}{]}\ ,\\[5pt]
R^{\mathbf{2}}(R'_{\a(0)})&\in&
\mx{[}{c}{s_1\\\Xi~\widetilde\otimes(1)}{]}\oplus
\mx{[}{c}{s_1-1\\\Xi~\widetilde\otimes[2]}{]}\oplus
\mx{[}{c}{s_1-2\\\Xi~\widetilde\otimes(1)}{]}\oplus
\mx{[}{c}{s_1-1\\\Xi}{]}\nn\\[5pt]&&\oplus
\mx{[}{c}{s_1-1\\i_{[1]}\Xi~\widetilde\otimes(1)}{]}\oplus
\mx{[}{c}{s_1-2\\i_{[1]}\Xi}{]}\oplus\mx{[}{c}{s_1-1\\i_{[2]}\Xi}{]}\oplus
\mx{[}{c}{s_1\\i_{[1]}\Xi}{]}\ ,\eea
leaves Proca-like field equations in the Lorentz-irreps
\bea H_{1,0}(\sigma^-)&\ni& P_A[(s_1-2);i_{[1]}\Xi]\oplus
P_S[(s_1-1);i_{[2]}\Xi]\oplus P_\varphi[(s_1);i_{[1]}\Xi]\ .\eea
The chain with $g+q=2$, whose content is listed in Appendix
\ref{App:q+g=3}, leaves: i) Labastida-like field equations in
\bea H_{1,1}(\sigma^-)&\ni& F_\varphi[(s_1);\Xi]\oplus
F_A[(s_1-2);\Xi]\ ;\eea
and ii) a Bianchi identity for the Proca-like equations, in
\bea H_{2,0}(\sigma^-)&\ni& B[(s_1);i_{[2]}\Xi]\ .\eea
The chain with $q+g=3$, whose content is listed in Appendix
\ref{App:q+g=3}, leaves Noether\footnote{Strictly speaking, one
should use the terminology \emph{Noether identity} only in case
one has an action principle.}/Bianchi identities in
\bea H_{2,2}(\s^-)&\ni& N^{1}_\varphi[(s_1-1);\Xi]\oplus
N^2_{\varphi}[(s_1);i_{[1]}\Xi]\oplus N_{A}[s_1-2;i_{[1]}\Xi]\
.\eea
Thus the dynamical system consists of a parameter $\e\,$; fields
$\varphi\,$, $A$ and $S\,$; Proca-like equations of motion of the
schematic form
$P_\varphi:=\nabla_{\Xi}\phi+\overline\nabla^{(1)}S\approx 0$,
$P_A:=\nabla_{\Xi}A+\nabla_{(1)}S\approx0$,
$P_S:=\nabla_{\Xi}S\approx0\,$; and Labastida-like field equations
 $F_\varphi\approx0$ and $F_A\approx 0$
containing the d'Alembertians of $\phi$ and $A\,$, respectively.
In the above $\nabla_{\Xi}$ denotes all possible divergencies in
$\Xi\,$. The parameter can be used to gauge away
$\nabla_{(1)}\varphi\,$, so that $\nabla_{(1)}P_\varphi\approx 0$
implies a mass-shell condition for $S\,$. Since all field are now
on-shell, the divergencies $\nabla_{\Xi}\e$ and
$\nabla_{(1)}\e\,$, respectively, of the residual parameter $\e$
can be used to remove $S$ and $A\,$, leaving a transverse on-shell
Lorentz tensor $\phi\,$.

\vspace{0.5cm}

\textbf{The case $h_1=2$, $B\geqslant 2$, $s_1-s_4\geqslant 4$}:
Here the triangular module (see Fig. \ref{fig2})
\bea
\mT'_{ASV}&=&\mR'_{-2}\oplus\mR'_{-1}\oplus\mR'_{_0}\oplus\mR'_{1}\oplus\mR'_{_2}
\oplus\mR'_3\oplus\cdots\ \ni\ (\eta,\e,X,R,Z,Z_3,\dots)\ ,\eea
and one can show that if $s_1-s_{_2}\geqslant 4$ then the
dynamical system contains ($s:=s_1$) parameters $\e(s,s-1;\Xi)$,
$\e_S(s-1,s-1;i_{[1]}\Xi)$ and $\e_A(s-1,s-2;\Xi)$; fields
$\varphi(s,s;\Xi)$, $\varphi_A(s,s-2;\Xi)$,
$\varphi_S(s,s-1;i_{[1]}\Xi)$, $S_A(s-1,s-2;i_{[1]}\Xi)$,
$S(s-1,s-1;i_{[2]}\Xi)$ and $A(s-2,s-2;\Xi)$; Proca-like equations
$P_\varphi(s,s;i_{[1]}\Xi):=\nabla_{\Xi}\varphi+\overline\nabla^{(2)}\varphi_S\approx0$,
$P_{\varphi_A}(s,s-2;i_{[1]}\Xi):=\nabla_\Xi\varphi_A+\nabla_{(2)}\varphi_S+\overline\nabla^{(1)}S_A\approx0$,
$P_{\varphi_S}(s,s-1;i_{[2]}\Xi):=\nabla_\Xi\varphi_S+\overline\nabla^{(1)}S\approx
0$, $P_{S_A}(s-1,s-2;i_{[2]}\Xi):=\nabla_\Xi
S_A+\nabla_{(2)}S\approx0$,
$P_A(s-2,s-2;i_{[1]}\Xi):=\nabla_{\Xi}A+\nabla_{(1)}S_A\approx0$
and $P_S(s-1,s-1;i_{[3]}\Xi):=\nabla_{\Xi}S\approx 0$; and
Labastida-like field equations $F_\varphi(s,s;\Xi)\approx0$,
$F_{\varphi_A}(s,s-2;\Xi)\approx0$ and
$F_A(s-2,s-2;\Xi)\approx0\,$. The parameters $\e$, $\e_S$ and
$\e_A$, respectively, can be used to gauge away
$\nabla_{(2)}\varphi$, $\nabla_{(1)}\varphi_S$ and
$\nabla_{(2)}\varphi_A$, whereafter all fields are on-shell. The
on-shell gauge parameters can then be used to gauge away all
fields except $\varphi\,$.

\begin{figure}
\begin{center}
\unitlength=.6mm
\begin{picture}(150,150)(30,-40)
\put(-20,100){\line(1,0){220}}
\multiput(40,120)(40,0){4}{\line(0,-1){170}}
\put(20,110){$\mR_{-1}$} \put(60,110){$\mR_{_0}$}
\put(100,110){$\mR_{_1}$} \put(140,110){$\mR_{_2}$}
\put(180,110){$\mR_{_3}$} \put(-18,85){$R_{\a(0)}$}
\put(-18,65){$R_{\a(1)}$} \put(-18,45){$R_{\a(2)}$}
\put(-18,25){$R_{\a(3)}$} \put(-18,5){$R_{\a(4)}$}
\put(-18,-25){$R_{\a(s_{1,2})}$}
\multiput(0,0)(0,20){5}{\multiput(10,0)(40,0){5}{\line(1,0){20}}
\multiput(10,0)(40,0){5}{\line(0,1){15}}
\multiput(30,15)(40,0){5}{\line(-1,0){20}}
\multiput(30,15)(40,0){5}{\line(0,-1){15}}}
\put(12,89){$\bigstar$} \put(52,89){$\blacklozenge$
$\blacklozenge$}\put(52,82){$\blacklozenge$} \put(92,89){$\bullet$
$\bullet$}\put(92,81){$\bullet$}
\put(132,89){$\blacktriangle$}\put(172,89){$\blacktriangledown$}
\put(92,69){$\blacklozenge$}\put(92,62){$\blacklozenge$}
\put(132,69){$\bullet$ $\bullet$}\put(132,61){$\bullet$}
\put(92,-21){$\square$}
\multiput(20,-10)(40,0){5}{$\vdots$}\multiput(20,-40)(40,0){5}{$\vdots$}
\multiput(10,-30)(40,0){5}{\line(1,0){20}}
\multiput(10,-30)(40,0){5}{\line(0,1){15}}
\multiput(30,-15)(40,0){5}{\line(-1,0){20}}
\multiput(30,-15)(40,0){5}{\line(0,-1){15}}
\end{picture}
\end{center}
\caption{{\small{The $\s^-$ cohomology in the case of $h_1=1$: i)
The $\bigstar$ is the differential gauge parameter; ii) the
$\blacklozenge$ at $q=0$ are the dynamical fields; iii) the
$\bullet$ at $q=1$ are the Proca-like first-order field equations;
iv) the $\blacklozenge$ at $q=1$ are the
Einstein-Fronsdal-Labastida-like second-order field equations; v)
the $\bullet$ at $q=2$ are Noether/Bianchi identities; vi) the
$\blacktriangle$ and $\blacktriangledown$ are higher Bianchi
identities. The $\square$ is the primary Weyl tensor which
``glues'' the potential module to the Weyl zero$\,$-form module.
While it is not part of the total $\s-$ cohomology, it is part of
the $\s^-$-$\,$cohomology restricted to the potential module.} }}
\label{fig2}
\end{figure}
%
\subsection{\sc St\"uckelberg fields and flat limit}
\label{Sec:flatlimit}
%
In this Section we first look at some examples of the unfolded
module $\mR_{f^-_I}$ defined in \eq{mRfi} which we, based on the
analysis performed so far, claim consists of the ASV module plus
the unfolded St\"uckelberg fields minus the Weyl zero$\,$-form of
the St\"uckelberg field $\chi_{_I}$ associated with the $I$th
block --- see the discussion in Section \ref{Sec:Prop} --- that is
projected away by the subsidiary constraint \eq{construu}. We then
argue that $\mR_{f^-_I}$ has a smooth flat limit in the sense of
the BMV conjecture albeit with additional topological p-forms in
flat space coming from the St\"uckelberg sector in AdS.
%
\subsubsection{\sc The example of $\Th=(2,1)$}
%
The $\sigma^-$-cohomology of the triangular module associated with
$\mR_{f^-_{_I}}$ with $I=1$ is depicted in Fig.
\ref{Table:21case}, where we have assigned a new grading $g'$ (see
caption) to all the radially reduced unfolded variables, including
those associated with the various St\"uckelberg fields. All these
fields are thus various components of the
$\miso(2,D-1)$-irreducible Skvortsov module associated with
$\widehat{\varphi}(\widehat{\Th})\,$.

We note that the cohomology contains two antisymmetric rank-2
objects that could form a trivial pair, namely the cohomologically
nontrivial gauge parameter and the zero-form $Y^{[2]}_{\bf 0}$ in
$\mR^{\bf 0}$. These quantities do not form a trivial pair because
the field equation for $Y^{[2]}_{\bf 0}$ loses its source
precisely in the unitary critically massless limit. We interpret
the field $Y^{[2]}_{\bf 0}$ as a zero-mode for $\chi_{_1}[2]$ that
remains upon imposing the subsidiary condition on the primary Weyl
tensor $C_{\chi_{_1}}(2,1):=\overline
\nabla^{(1)}\chi_{_1}[2]-{\rm traces}\approx 0$.

\begin{figure}\begin{center}
\begin{tabular}{l||l|l|l|l|}
   $g'=g+2$&  $g'=0$ &  $g'=1$ &  $g'=2$ &  $g'=3$   \\
   \hline\hline
   $\mR_{_{-2}}$ & --- & $\eta_{\mathbf{0}}^{\bullet}$ &
     $\eta_{\mathbf{0}}^{[1]}$ &     \\ \hline
   $\mR_{_{-1}}$ & $\epsilon_{\mathbf{0}}^{\bullet}$ &
   $\epsilon_{\mathbf{0}}^{[1]}\,$, $\epsilon_{\mathbf{1}}^{\bullet}$ &
   $\epsilon_{\mathbf{1}}^{[1]}\,$,
   $\kappa_{\mathbf{0}}^{[2]}\,$ & $\kappa_{\mathbf{0}}^{[3]}\,$ \\ \hline
   $\mR_{0}$ & $Y_{\mathbf{1}}^{\bullet}$ &
   $Y_{\mathbf{1}}^{[1]}\,$, $X_{\mathbf{2}}^{^\bullet}\,$,
   $Y_{\mathbf{0}}^{[2]}\,$
   &
   $Y_{\mathbf{0}}^{[3]}\,$, $X_{\mathbf{2}}^{[1]}\,$,
   $X_{\mathbf{1}}^{[2]}\,$
   & $X_{\mathbf{1}}^{[3]}\,$ \\ \hline
\end{tabular}\end{center}
\caption{{\footnotesize The set of $p\,$-form fields obtained upon radial
reduction of the $p$-forms ($p>0$) associated with the Skvortsov
module starting from $\widehat\varphi(\L\!\!=\!0;\widehat \Th)$
with $\widehat\Th=([2;1],[1;1])\,$. All the fields take value in
Lorentz-irreducible shapes. The relation between the two different
gradings used in Section \ref{Sec:UnitarASV} and in Figure is
$g=g'-2\,$. The grading $g$ is associated with the $ASV$ potential
whereas the $g'$ grading is associated with all the fields
obtained upon radial reduction from $D+1$ to $D\,$.}
}\label{Table:21spectrum}
\end{figure}

\begin{figure}\begin{center}
\begin{tabular}{|l|l|l|l|l|l|}
  \hline
   {\rm grade}& $\mR_{-2}$&$\mR_{_{-1}}$ & $\mR_{_0}$ & $\mR_{_1}$ & $\mR_{_2}$
   \\\hline
  $g'=0$ & --- & --- & --- & --- & --- \\\hline
  $g'=1$ & --- & --- & $\YoungDdash11$ & $\YoungDdash21~\YoungUdash1$ &
  $\YoungDdash22$ $\YoungUdash2$ $\circ$ \\[10pt]\hline
  $g'=2$ & --- & $\YoungD11$ & $\YoungD21~\YoungU1$ &
  $\YoungU2~\bullet$ & --- \\[10pt]\hline
  $g'=3$ & --- & --- & --- & $\YoungD21$ & $\YoungD11~\YoungU2~\bullet$
  \\[10pt]\hline
\end{tabular}\end{center}
\caption{The $\s^-$-cohomology of the unitary $(2,1)$ gauge field
in $AdS_D\,$. The solid shapes represent the cohomology for the
dynamical field $\varphi(2,1)\,$. The dashed shapes represent the
cohomology for the closed Weyl zero$\,$-form $Y^{\bf
0}\widehat{[3]}\,$. For the definition of the grading $g'$, see
caption of Fig. \ref{Table:21spectrum}.} \label{Table:21case}
\end{figure}
\subsubsection{\sc The example of $\Th=(3,1)$}

Let us consider the subsector $\mR_f(\L;\Th)\subset \mR_f$
obtained by constrained radial reduction of the gauge field
$\widehat\varphi(\L\!\!=\!0;\widehat \Th)$ with
$\widehat\Th=([3;1],[1;1])\,$.

The $p-$form sector thus obtained consists of the fields listed in
Fig. \ref{Table:31spectrum}.

\begin{figure}\begin{center}
\begin{tabular}{l||l|l|l|l|l|l|}
   &  $g=0$ &  $g=1$ &  $g=2$ &  $g=3$ &  $g=4$ &  $g=5$    \\
   \hline\hline
   $\mR_{_{-2}}$ & --- & $\zeta_{\mathbf{0}}^{\bullet}$ &
     $\zeta_{\mathbf{0}}^{[1]}$ &  $\zeta_{\mathbf{0}}^{[1,1]}$ &
     --- &  ---    \\ \hline
   $\mR_{_{-1}}$ & $\epsilon_{\mathbf{0}}^{\bullet}$ &
   $\epsilon_{\mathbf{0}}^{[1]}\,$, $\epsilon_{\mathbf{1}}^{\bullet}$ &
   $\epsilon_{\mathbf{0}}^{[1,1]}\,$, $\epsilon_{\mathbf{1}}^{[1]}\,$,
   $\eta_{\mathbf{0}}^{[2]}\,$ & $\epsilon_{\mathbf{1}}^{[1,1]}\,$,
   $\eta_{\mathbf{0}}^{[2,1]}\,$, $\kappa_{\mathbf{0}}^{[2,1]}\,$,
   $\eta_{\mathbf{0}}^{[3]}\,$  &
   $\kappa_{\mathbf{0}}^{[2,2]}\,$,
   $\eta_{\mathbf{0}}^{[3,1]}\,$, $\kappa_{\mathbf{0}}^{[3,1]}\,$  &
   $\kappa_{\mathbf{0}}^{[3,2]}$ \\ \hline
   $\mR_{0}$ & $Y_{\mathbf{1}}^{\bullet}$ &
   $Y_{\mathbf{1}}^{[1]}\,$, $X_{\mathbf{2}}^{^\bullet}\,$
   &
   $Y_{\mathbf{1}}^{[1,1]}\,$, $X_{\mathbf{2}}^{[1]}\,$,
   $X_{\mathbf{1}}^{[2]}\,$,
   & $X_{\mathbf{1}}^{[2,1]}\,$,  $Y_{\mathbf{0}}^{[3,1]}\,$,
   $X_{\mathbf{2}}^{[1,1]}\,$, $X_{\mathbf{1}}^{[3]}\,$, &
   $\widetilde{X}_{\mathbf{1}}^{[2,2]}\,$,
   $\widetilde{X}_{\mathbf{1}}^{[3,1]}\,$,
   $\widetilde{Y}_{\mathbf{0}}^{[3,2]}\,$ & $X_{\mathbf{1}}^{[3,2]}$ \\
   &  & $Y_{\mathbf{0}}^{[2]}\,$ &  $Y_{\mathbf{0}}^{[3]}\,$,
   $Y_{\mathbf{0}}^{[2,1]}\,$, $\widetilde{Y}_{\mathbf{0}}^{[2,1]}\,$ &
   $\widetilde{X}_{\mathbf{1}}^{[2,1]}\,$, $\widetilde{Y}_{\mathbf{0}}^{[3,1]}\,$,
   $\widetilde{Y}_{\mathbf{0}}^{[2,2]}\,$
   & $X_{\mathbf{1}}^{[3,1]}\,$ &  \\ \hline
\end{tabular}\end{center}
\caption{{\small{The set of $p\,$-form fields obtained upon dimensional
reduction of $\widehat\varphi(\L\!\!=\!0;\widehat \Th)$ with
$\widehat\Th=([3;1],[1;1])\,$. All the fields take value in
Lorentz-irreducible shapes.}}}\label{Table:31spectrum}
\end{figure}

The irreducible module carrying the unitary representation
$\mD(D;(3,1))$ is given by
\bea \mR'&=&\left\{\underbrace{U^{\bf
1}\widehat{[3,2]}}_{g=0,1,2};\underbrace{C^{\bf 0}[2,2]}_{g=2}
;\underbrace{X^{\bf 0}[3,2],X^{\bf 0}(3,2)}_{g=3};\cdots\right\}\
.\eea

\begin{figure}
\begin{center}
\begin{tabular}{|l||l|l|l|l|l|}
  \hline
    & $\mR_{_{-2}}$ & $\mR_{_{-1}}$ & $\mR_{_0}$ &$\mR_{_1}$ &
  $\mR_{_2}$  \\ \hline\hline
  $g=0$ & --- & --- & --- & --- & ---   \\ \hline
  $g=1$ & --- & --- & --- & --- & ---  \\ \hline
  $g=2$ & --- & --- & --- & --- & ---  \\ \hline
  $g=3$ & --- & $\YoungD{2}{1}$ &
  $\YoungD{3}{1}\oplus\YoungU{2}\oplus\YoungD{1}{1}$
  & $\YoungU{3}\oplus\YoungU{1}$ & ---
  \\  &  &  &  &  &    \\ \hline
  $g=4$ & --- & --- & --- & $\YoungD{3}{1}\oplus\YoungD{1}{1}$ &
  $\YoungU{3}\oplus\YoungD{2}{1}\oplus\YoungU{1}$
  \\  &  &  &  &  &    \\ \hline
\end{tabular}
\end{center}
\caption{$\s^-$-cohomologies for the unitary, massless,
spin-$(3,1)$ field in $AdS_D$ spacetime.}\label{Table:Coho31}
\end{figure}
%
\subsubsection{\sc{Smooth flat limit}}
%
To repeat, the analysis so far shows that in the unitary case the
radial reduction \eq{constr1} followed subsidiary constraint
\eq{construu} lead to the following reducible $\mso(2,D-1)$-module:
\bea \mR_{f^-_1}(\L;\Th)&=&\mS_{f^-_1}(\L;\Th) \oplus \mR'_{f^-_1}(\L;\Th)\ ,\\[5pt] \mR'_{f^-_1}(\L;\Th)&=&\widetilde\mR^{\bf h_1}_{\rm ASV} \supsetplus \mC^{\bf 0}(\L;\overline M^2_1;\overline\Th_1)\ ,\eea
where $\mS(\L;\Th)$ is a massively contractible cycle (see Section
I.4.4.4) for $\L\neq 0$ containing the BMV St\"uckelberg fields as
well as the frozen (see Section I.5.2) St\"uckeberg fields
associated with $I$th block (see Section \ref{Sec:Prop}).

The above reducible module has the \emph{smooth limit}
\bea \mR_{f^-_1}(\L;\Th)&\stackrel{\l\rightarrow 0}{\longrightarrow}& \mR_{\rm extra}(\L\!\!=\!0;\Th)\ \cup\ \mR_{\rm BMV}(\L\!\!=\!0;\Th)\ ,\\[5pt]
\mR_{\rm BMV}(\L\!\!=\!0;\Th)&:=&
\bigoplus_{\Th^{\prime}\in\S^1_{\rm BMV}(\Th)} \mR_{\rm
Skv}(\L\!\!=\!0;\Th^{\prime})\ ,\label{flaatlimit}\eea
where $\mR_{\rm Skv}(\L\!\!=\!0;\Th^{\prime})$ are the Skvortsov
modules predicted by the BMV conjecture and the complement
$\mR_{\rm extra}$ contains a finite set of topological fields. For
a fixed $\Th$ and $\L=0$, one can show that $Y^{\bf h_1-1}$ and
$\widehat\x_{(h_1+1)} X^{\bf h_1}$ still form a massively
contractible cycle, so that the flat-space potential modules
$\widetilde\mR(\L=0;\Th')$ do not contain any elements of form
degree less than $h_1$, nor of form degree $h_1$ with first block
smaller than that of $\Th\,$.

The Weyl zero$\,$-form module $\mC^{\bf 0}(\L;\overline
M{}^2_1;\overline\Th_1)=\O^{\bf 0}(U)\otimes {\cal
T}(\L;M^2_1;\overline\Th_1)$ has the limit
\bea {\cal T}(\L;\overline
M{}^2_1;\overline\Th_1)&\stackrel{\l\rightarrow
0}{\longrightarrow}&\bigoplus_{\Th^{\prime}\in\S^1_{\rm BMV}(\Th)}
{\cal T}(\L\!\!=\!0;M^2\!\!=\!0;\overline\Th^{\prime})\
,\label{BMVcalT}\eea
which together with harmonic expansion shows that the unitary
massless lowest-weight space representation of $\mso(2,D-1)$
contracts to the direct sum of massless irreps of $\miso(1,D-1)$
in accordance with the BMV conjecture.

Finally, we note that it should be possible to project away the
aforementioned frozen field content for $\L\neq 0$ without
affecting the smoothness of the flat limit, which we leave for
future studies.

\section{\sc\large{Conclusion}}
\label{Sec:Conclusions}

In the present paper we studied the BMV conjecture
\cite{Brink:2000ag} at the level of the field equations by
extending the unfolding analysis carried out by Skvortsov in
\cite{Skvortsov:2008vs} to the $AdS_D$ background. To this end, we
reformulated the equations of \cite{Skvortsov:2008vs} by using an
oscillator formalism. Certain operators were constructed, the
so-called cell operators, which were found to be very useful for
an alternative proof of the consistency of Skvortsov's equations.

We then proceeded with the following steps: We started from the
reformulation of Skvortsov's unfolded equations for a
mixed-symmetry gauge field $\widehat{\varphi}(\widehat \Th)$ in
$(D+1)$-dimensional flat space with signature $(2,D-1)$ and
radially reduced the $(D+1)$-dimensional unfolded fields to
$AdS_D\,$. Then we constrained the Lie derivatives of the fields
along a radial vector field [see Eqs. \eq{constr1}, \eq{Deltap}
and \eq{Upsilonp}]. Next, we constrained the $(p_{_I}+1)$th row of the
internal indices carried by the zero$\,$-forms [see Eq.
\eq{construu}] and verified that the generalized Weyl tensor of
$\varphi(\Th)$ carries Metsaev's unitary representation
$\mD(e_{_0};\Th)\,$.

In particular, we were able to prove the BMV conjecture in the case
of mixed-symmetry gauge fields whose corresponding Young diagrams
possess at most four rows. The nontrivial consistency of the
constraints imposed on the generalized Weyl tensors is a good sign
that these constraints are correct for arbitrary mixed-symmetry
gauge fields. In a future work we would like to further study the
consistency of our constraints in the general case. For this, it is
crucial to have a better understanding of the cell operators and
their commutation relations. Also of interest is to study further
the $p\,$-form sector ($p>0$) of the unfolded system in $AdS_D$ that
we displayed in the present work, in particular the precise
expression of the constraints that would enable one to project out
the frozen St\"uckelberg fields.

In relation with the previous issue, it would be very interesting
\cite{WIP} to make a precise link between these $p\,$-forms and the
gauge fields needed for a first-order action formulation of
arbitrary mixed-symmetry fields in $AdS_D\,$ along the lines
proposed by Zinoviev, see \cite{Zinoviev:2008ve} for the cases where
the shape associated with the field is a long hook with one cell in
the second row.

As shown in Appendix \ref{App:Singleton}, \emph{generic} mixed-symmetry
fields cannot be seen as singleton composites, though certain
long-hook fields arise in tensor products of two spin-$1/2$
fermionic singletons \cite{Vasiliev:2004cm}. The oscillator
realization of the constraints in our radial-reduction
construction does not appear to allow for a strict factorization
in terms of subsets of unconstrained oscillators.

This non-factorization property maybe is an artefact of our
construction and it would be very important, we believe \cite{WIP},
to investigate about an abstract enveloping-algebra approach to the
lowest-weight modules corresponding to generic mixed-symmetry
fields.


\section*{Acknowledgments}
We are thankful to K.~Alkalaev and E.~Skvortsov for enlightening
comments and E.~Skvortsov for several useful remarks on an early
version of this paper. We also wish to thank G.~Barnich,
F.~Bastianelli, X.~Bekaert, A.~Campoleoni, P.P.~Cook, J.~Demeyer,
F.A.~Dolan, J.~Engquist, D.~Francia, M.~G\"unaydin, S.~Leclercq,
T.~Poznansky, A.~Sagnotti, E.~Sezgin, D.~Sorokin, Ph.~Spindel,
F.~Strocchi, E.~Tonni and M.A.~Vasiliev for discussions. A.~Sagnotti is also thanked for his encouragements and support. This work
is supported in part by the EU contracts MRTN-CT-2004-503369 and
MRTN-CT-2004-512194 and by the NATO grant PST.CLG.978785. The
research of P.S. was supported in part by Scuola Normale Superiore
and by the MIUR-PRIN contract 2007-5ATT78

\begin{appendix}

\section{\sc\large{Notation and Conventions}}\label{App:0}

The direct sum of two vector spaces is written as $\mathfrak
A\uplus \mathfrak B\,$. If $\ml$ is a Lie algebra (or more
generally an associative algebra) then the decomposition of an
$\ml$-module $\mR$ under a subalgebra $\mk\subseteq \ml$ is
denoted by $\mR|_{\mk}$. A module $\mathfrak R$ containing an
invariant subspace $\mathfrak I$, an ideal, is said to be either
(i) indecomposable if the complement of $\mI$ is not invariant in
which case one writes $\mathfrak R|_\ml=\mathfrak I\subsetplus
(\mathfrak R/\mathfrak I)\,$; or (ii) decomposable if both
$\mathfrak I$ and $\mathfrak R/\mathfrak I$ are invariant in which
case one writes $\mathfrak R|_\ml=\mathfrak I\oplus (\mathfrak
R/\mathfrak I)\,$.

Infinite-dimensional modules can be presented in many ways depending on how they
are sliced under various subalgebras.
If $\mk\subset \ml$ one refers to finite-dimensional $\mk$-irreps with non-degenerate bilinear forms as $\mk$-types, which we denote by
$\Th_\a$, $\Th_{\a_i}$ \emph{etc.} labeled by indices $\a$, $\a_i$ \emph{etc.}.
Correspondingly, if there exists a slicing $\mR|_\mk$ consisting of $\mk$-types
then we refer to such expansions as an $\mk$-typesetting of $\mR$. In particular,
we refer to finite-dimensional Lorentz-irreps as Lorentz types (that will be
tensorial in this paper). In unfolded dynamics one may view typesetting as local
coordinatizations of infinite-dimensional target spaces for unfolded sigma models.
We set aside issues of topology.

Young diagrams, or row/column-ordered shapes, with $m_i$ cells in the $i$th
row/column, $i=1,\dots,n$ are labeled by $(m_{_0},\dots,m_{n+1})$ and
$[m_{_0},\dots,m_{n+1}]$ where $m_i\geqslant m_{i+1}$ and $m_{_0}:=\infty$ and
$m_{n+1}:=0\,$. We let
$\mathbb P_{\Th}$ denote Young projections on shape $\Th\,$.
We also use the block-notation
\bea \left([s_{_1};h_{_1}],[s_{_2};h_{_2}],...,[s_{_B};h_{_B}]\right)\ :=
(\underbrace{m_1,\cdots,m_{h_1}}_{=s_{_1}},\underbrace{m_{h_1+1},
\dots,m_{h_1+h_2}}_{=s_{_2}}
\dots)\ ,\eea
for a shape with $B$ rectangular blocks of lengths $s_{_I}>s_{_{I+1}}$ and heights
$h_{_I}\geqslant 1$, $I=1,2,...,B\,$.
The space of shapes ${\cal S}$ forms a module,
the Schur module, for the universal Howe-dual algebra $\msl(\infty)\,$,
obtained as a formal limit of $\msl(\nu_\pm)$ acting in the spaces
${\cal S}^\pm_{\nu_\pm}$ of shapes with total height
$p_{_B}:=\sum_{I=1}^B h_{_I}\leqslant\nu_+$
($\msl(D)$-types in symmetric bases)
or widths $s_{_1}\leqslant \nu_-$ (($\msl(D)$-types in anti-symmetric bases).
Extension to traceless Lorentz tensors leads to Howe-dual algebras $\msp(2\nu_+)$ and
$\mso(\nu_-)\,$, with formal limits
$\msp(2\infty)$ and $\mso(2\infty)$, respectively.

The Schur module ${\cal S}$ can be treated \emph{explicitly} by using
``cell operators'' $\beta_{a,(i)}$ and $\bar\beta^{a,(i)}$ defined
(see Paper II) to act faithfully in ${\cal S}$ by removing or adding,
respectively, a cell containing the $\msl(D)$-index $a$ in the $i$th row.
Schematically,
\bea \bar\beta^{a,(i)}(m_1,\dots,m_i,\dots,m_n) & = &
(m_1,\dots,m_i+1,\dots,m_n) \ ,\nn\\[5pt]
\beta_{a,(i)}(m_1,\dots,m_i,\dots,m_n) & = &
(m_1,\dots,m_i-1,\dots,m_n) \ .\nn\eea
Similarly, $\beta_{a,[i]}$ and $\bar\beta^{a,[i]}\,$, respectively, remove and
add an $a$-labeled box in the $i$th column.

We let $\widehat\mg$ denote the real form of $\mso(D+1)$ with metric
$\eta_{AB}={\rm diag}(\s,\eta_{ab})$ where $\s=\pm 1$ and
$\eta_{ab}=(-1,\d_{rs})\,$, and with generators $\widehat M_{AB}$ obeying
the commutation rules
\bea [\widehat M_{AB},\widehat M_{CD}]&=&2i\,\eta_{C[B}\widehat
M_{A]D}-2i\,\eta_{D[B}\widehat M_{A]C}\ .\eea
We let $\mm:=\mso(1,D-1)$ and $\ms:=\mso(D-1)$ denote the ``canonical''
Lorentz and spin subalgebras, respectively, with generators $M_{ab}$ and
$M_{rs}\,$. We let $\mg_\l:=\mm\subsetplus \mpe$ where $\mpe$ is spanned by the
transvections\footnote{We are here abusing a standard terminology
used in the context of symplectic algebras, the only point being to make clear
the distinction between the cases where the generators
$\{P_a\}$ are commuting or not.} obeying
\bea [P_a,P_b]\ =\
i\l^2M_{ab}\ ,\quad [M_{ab},P_c]\ =\ 2i\eta_{c[b}P_{a]}\ .\eea
If $\l^2=0$ then $\mg_\l\cong \miso(1,D-1)$ and if $\l^2\neq 0$ then $\mg_\l\cong \widehat\mg\,$ with $\sigma=-\lambda^2/|\lambda^2|\,$, the isometry algebras of $AdS_D$ ($\s=-1$) and $dS_D$ ($\s=1$) with radius $L_{\rm AdS}:=L$ and $L_{\rm dS}:=-i\,L\,$, respectively, where $L:=\lambda^{-1}$ is assumed to be real for $AdS_D$ and purely imaginary for $dS_D\,$. The $\mg_\l$-valued connection $\Omega$ and curvature $\cal R\,$ are defined as follows
\bea \O& :=& e+\o  :=  -i(e^a\,P_a+\ft12\, \o^{ab}\,M_{ab})\
,\label{canonicalconnection}\\
{\cal R} & := & d\O+\O^2 \ = \ -i\left[T^a P_a +\ft12
(R^{ab}+\l^2e^a e^b)\,M_{ab}\right]\ ,\\[5pt]
T^a & := & de^a+ \o^a{}_b~e^b \ , \quad R^{ab} \ := \
d\o^{ab}+\o^a{}_c~\o^c{}_b\ ,\eea
and are associated with a cosmological constant $\L=-\frac{(D-1)(D-2)}{2}\,\l^2\,$.
The Lie derivative along a vector field $\xi\,$ is
${\cal L}_{\xi} := d~i_\xi + i_\xi~d$ and we use conventions where the exterior
total derivative $d$ and the inner derivative $i_\xi$ act from the left.
If the frame field $e^a$ is invertible we define the inverse frame
field $\th^a$ by $i_{\th^a} e^b=\eta^{ab}\,$.

We use weak equalities $\approx$ to denote equations that hold on the constraints
surface.
In the maximally symmetric backgrounds ${\cal R}\approx 0\,$
the connection $\Omega$ can be frozen to a fixed background value,
breaking the diffeomorphisms down to isometries
$\d_{\e(\x)}$ with Killing parameters $\e(\x)=i_\xi(e+\o)$ obeying
$\d_{\e(\x)}(e+\o)\approx {\cal L}_\x(e+\o)=0$
(one has ${\cal L}_\x e^a=\d_{\e(\x)}e^a+i_\x T^a$ where
$\d_{\e(\x)}e^a=\nabla \e^a-\e^{ab}e_b$ with $\e^a=i_\x e^a\,$,
$\e^{ab}=i_\x \o^{ab}\,$ and $\nabla:=d- \frac{i}{2}\,\omega^{ab}M_{ab}\,$).

We use
$\mD^{\pm}(\pm e_{_0};\Th_{_0})$ to denote lowest-weight ($+$) and
highest-weight ($-$) modules of $\mg_\l$ that are sliced under its maximal compact
subalgebra $\mh\cong\mso(2)\oplus\mso(D-1)$ into $\mh$-types
$\ket{e;\th}^{\pm}\,$. In compact basis, the $\mso(2,D-1)$ algebra reads
\bea M_{0r}&=& \ft12(L^+_r+L^-_r)\ ,\quad P_r\ =\ \ft{i\l}2(L^+_r-L^-_r)\ ,
\quad E\ =\ \l^{-1} P_0\ ,\label{mp}\\[5pt]
[L^-_r,L^+_s]&=& 2iM_{rs} +2\d_{rs} E\ ,\quad [E,L^\pm_r]\ =\ \pm
L^\pm_r\ , \quad [M_{rs},L^\pm_t]\ =\ 2i\delta_{t[s}L^\pm_{r]}\
.\label{algd-1}\eea
By their definition, the modules $\mD^{\pm}(\pm e_{_0};\Th_{_0})$ are the irreps
obtained by factoring out all proper ideals in the generalized Verma module
generated from a unique lowest-energy ($+)$ or highest-energy ($-$) state
$\ket{\pm e_{_0};\Th_{_0}}^{\pm}$ with $E$-eigenvalue $\pm e_{_0}\,$.
We let $\mD( e_{_0};\Th_{_0}):= \mD^+( e_{_0};\Th_{_0})$ and
$\ket{e;\th}:=\ket{e;\th}^+$. The generalized Verma module is irreducible for
generic values of $e_{_0}$, \emph{i.e.} singular vectors arise only for certain
critical values related to $\Th_{_0}$.

In unfolded field theory the mass-square $M^2$ of an unfolded Lorentz tensor field
$\phi(\Th)$ (dynamical field, Weyl tensor, ...) carrying a $\mg_\l$-irrep
($\L\neq 0$) with representation $\rho$, is the eigenvalue of
\bea -\rho(P^a P_a)\ \equiv \ \l^2
\rho(\frac12M_{AB}M^{AB}-\frac12M_{ab}M^{ab})\ .\eea
In the case of $\L<0$ one sometimes deals with harmonic expansions involving
 lowest-weight spaces where
\bea C_{_2}[\mg_\l|\mD(e_{_0};\Th_{_0})]&=&e_{_0}[e_{_0}-2(\e_{_0}+1)]+
C_{_2}[\ms|\Th_{_0}]\
,\quad \ms\ :=\ \mso(D-1)\ ,\quad \e_{_0}~:=~\frac12 (D-3)\ \quad\eea
leading to the mass formula
\bea L^2 M{}^2&=& e_{_0}[e_{_0}-2(1+\e_{_0})]+C_{_2}[\ms|\Th_{_0}]-C_{_2}[\mm|\Th]\ .
\label{masses}\eea

We let ${\cal T}^\pm_{(i)}(\Th^\pm)$ denote $\miso(1,D-1)$-irreps with (a) largest
and smallest $\mm$-types $\Th^+$ and $\Th^-$, respectively; and (b) translations
represented by $\rho^+_{(i)}(P_a)=\beta_{a,(i)}$ and
$\rho^-_{(i)}(P_a)=\bar\c^{a,(i)}\,$ (the trace-corrected cell creation operator) for
fixed $i\geqslant 1\,$.
As a special case 
${\cal T}^-_{(1)}(\Th^-)\cong {\cal T}^\ast(\L\!\!=\!0;\overline M{}2\!\!=\!0;\Th^-)\,$,
the dual of the twisted-adjoint representation containing a strictly massless
primary Weyl tensor\footnote{In a similar context, see also the recent 
work~\cite{Alkalaev:2008gi} where the unfolding of mixed-symmetry fields
in flat space was reformulated using BRST-cohomological methods.}.
We also let ${\cal T}^\pm_{(0)}(\Th):=\Th$, the irrep consisting of a single
$\mm$-type $\Th$ annihilated by $P_a\,$.

The translations are nilpotent in ${\cal T}^\pm_{(i)}(\Th^\pm)$ for $i\geqslant 2$
and in ${\cal T}^+_{(1)}(\Th^+)$. Factoring out ideals yields ``cut''
finite-dimensional modules ${\cal T}^\pm_{(i),N}(\Th^\pm)$ of ``depth''
$N\geqslant 0$ such that $\left(\rho^{\pm}_{(i),N}(P_a)\right)^{n}\equiv\!\!\!\!\!/\ 0$
iff $n\leqslant N\,$. For $i\geqslant 2$ the duals $\left({\cal T}^\pm_{(i)}(\Th^\pm)\right)^\ast\cong{\cal T}^\mp_{(i),N}(\Th^{\prime \mp})$
for some $N$ and $\Th^{\prime \mp}$ determined from the shape of $\Th^\pm\,$.
In particular,
$({\cal T}^\pm_{(i)}(\Th^\pm))^\ast\cong {\cal T}^\mp_{(i)}(\Th^\mp)$
iff the $i$th row does not form a block of its own in $\Th^+$ nor $\Th^-\,$.

The $\miso(1,D-1)$-irreps ${\cal T}^\pm_{(i)}(\Th^\pm)$ with $i\geqslant 2$
and ${\cal T}^+_{(1)}(\Th^+)$ are contractions of
$\mso(2,D-1)$-types as follows: the $\mso(2,D-1)$-type
$\widehat\Th$ with its canonical representation $\widehat M_{AB}$ is
isomorphic to twisted representations $\widehat\Th^\pm_{(i),\k;\l}$
with canonical $\rho^\pm_{(i),\k;\l}(M_{ab}):=\widehat M_{ab}$ and non-canonical
$\rho^+_{(i),\k;\l}(P_a):=\l\,\widehat\x^B \widehat M_{Ba}+\k\,\beta_{a,(i)}$ and
$\rho^-_{(i),\k;\l}(P_a):=\l\,\widehat\x^B \widehat M_{Ba}+\k\,\bar\gamma_{a,(i)}$
where $\widehat\x^2=-1$ (these are representations for $[P_a,P_b]=i
\l2 M_{ab}$ for all values of $\k$, $\l$ and $i$). The limit
$\l\rightarrow 0$ at fixed $\k$ yields a reducible $\miso(1,D-1)$
representation that decomposes into
${\cal T}^{\pm}_{(i)}$-plets if $\k\neq 0$ and ${\cal
T}^{+}_{(0)}$-plets if $\k=0\,$.

\section{\sc \large{Oscillator Realizations of Classical Lie Algebras}}
\label{App:A}

\subsection{\sc{Howe-dual Lie algebras}}

We denote the classical algebras by
\bea \ml&:=&(\mgl(D;\Comp),\mso(D;\Comp),\msp(D;\Comp))\ , \qquad
\e(\ml)\ =\ (0,+1,-1)\ ,\label{classalg}\eea
where $D$ is assumed to be even for $\e=-1\,$. Their
finite-dimensional representations can be realized using bosonic
$(+)$ and fermionic $(-)$ oscillators, corresponding to tensors in
manifestly symmetric or anti-symmetric bases for the Young
projector, respectively. Omitting the tensor-spinorial
representations of $\mso(D;\Comp)$, the oscillators obey
\bea [\a_{i,a},\bar
\a^{j,b}]&:=&\a_{i,a}\bar\a^{j,b}+(-1)^{\ft12(1\pm1)}
\bar\a^{j,b}\a_{i,a}\ =\ \delta^j_i\delta^b_a\ ,\eea
where $a,b=1,\dots,D$ transform in the fundamental representation
of $\ml$, and $i=1,2,\dots,\nu_\pm$ are auxiliary flavor indices.
The oscillator algebras are invariant under the canonical
transformations generated by arbitrary Grassmann even polynomials
$\varepsilon(\a,\bar\a)$, \emph{viz.}
\bea \delta_\varepsilon
\a_{i,a}&=&[\varepsilon(\a,\bar\a),\a_{i,a}]\ ,\qquad
\delta_\varepsilon \bar\a^{i,a}\ =\
[\varepsilon(\a,\bar\a),\bar\a^{i,a}]\ ,\eea
forming an infinite-dimensional Lie algebra with commutator
$[\d_\varepsilon,\d_{\varepsilon'}]=\d_{[\varepsilon,\varepsilon']_\star}\,$.
The linear homogeneous canonical transformations form the
finite-dimensional subalgebras
\bea \ml^+&:=& \msp(2D\nu_+;\Comp)\ ,\qquad \ml^-\ :=\
\mso(2D\nu_-;\Comp)\ .\eea
These contain $\ml$ together with its Howe dual\footnote{A Howe
dual pair of Lie algebras is a pair of Lie subalgebras in a Lie
algebra which are their mutual centralizers.} $\widetilde \ml^\pm$
which is defined to be the maximal subalgebra of $\ml^\pm$ that
commutes with $\ml\,$. One has
\bea \ml=\mgl(D;\Comp)&:& \widetilde \ml^\pm\ =\ \mgl(\nu_\pm)\ ,
\\[5pt]
\ml=\mso(D;\Comp)&:& \widetilde \ml^+\ =\ \msp(2\nu_+;\Comp)\
,\qquad \widetilde
\ml^-\ =\ \mso(2\nu_-;\Comp)\ ,\\[5pt] \ml=\msp(D;\Comp)&:& \widetilde \ml^+\ =\
\mso(2\nu_+;\Comp)\ ,\qquad \widetilde \ml^-\ =\
\msp(2\nu_-;\Comp)\ .\eea
The oscillator realization of the generators of $\ml$ reads
\bea \mgl(D;\Comp)&:& M^a_b\ =\ \bar\a^{i,a} \a_{i,b}\ ,\\[5pt]
\mbox{$\mso(D;\Comp)$ and $\msp(D;\Comp)$}&:& M_{ab}\ =\ 2i
\bar\a^{i,c} J_{c\langle a|}\a_{i,|b\rangle}\ , \eea
with the commutation rules
\bea [M^a_b,M^c_d]\ =\ \delta^c_b M^a_d-\d^a_d M^c_b\ , \qquad
[M_{ab},M_{cd}]\ =\ 4iJ_{\langle c|\langle b}M_{a \rangle | d
\rangle}\ ,\eea
where
\bea M_{ab}\ =\ M_{\langle ab\rangle}\ :=\ -\e M_{ba}\ ,\qquad
J_{ab}&=& \e J_{ba}\ , \qquad J^{ab}J_{ac}\ =\ \d^b_c\ ,\eea
and indices are raised and lowered according to the convention
$X^a=J^{ab}X_b$ and $X_a=X^bJ_{ba}\,$. For definiteness, we take
$J_{ab}=\eta_{ab}$ of some signature $(p,q)$, $p+q=D\,$ in the
case of $\mso(D;\Comp)$ ($\e=+1$), and
$J_{ab}=\Omega_{ab}=\left[\ba{cc}0&\bf 1\\[-8pt]-\bf 1&0\ea\right]$ in the case of
$\msp(D;\Comp)\,$ ($\e=-1$). The oscillator realization of the generators of
$\widetilde \ml^\pm$ reads
\bea N^i_j&:=& \ft12\{\bar\a^{i,a},\a_{j,a}\}\ \equiv \
\ft12(\bar\a^{i,a}\a_{j,a}+\a_{j,a}\bar\a^{i,a})\ ,\quad\ T_{ij}\ :=\
\a_{i,a}\a_{j,b}J^{ab}\ ,\quad \overline T^{ij}\ :=\ \bar\a^{i,a}
\bar\a^{j,b} J_{ab}\ .\hspace{1cm} \eea
Their commutation rules take the form
\bea [T_{ij},\overline T^{kl}]&=& 4 \,N^{\langle
k}_{\langle i}\delta^{l\rangle}_{j\rangle}\ ,\qquad [N^i_j,
N^k_l]\ =\ \d_j^k N^i_l-\d_l^i N^k_j\ ,
\\[5pt]
[N^i_j,T_{kl}] &=& -2 \,T_{j\langle l}\d^i_{k\rangle}\ ,\qquad [
N^i_j, \overline T^{kl}]\ =\ 2\,\overline T^{i\langle
l}\d^{k\rangle}_j\ ,\eea
where
\bea T_{ij}&=& T_{\langle ij\rangle}\ := \ \pm\, \e \,T_{ji}\
.\eea
In the cases of $\e(\ml)=\pm1$, the above bases exhibit explicitly
the three-grading
\bea \widetilde \ml^\pm&=&T^{-1}\oplus N^{0}\oplus \overline
T^{+1}\ . \label{A1grading}\eea
%


\subsection{\sc Generalized Schur modules}


The oscillator algebra can be realized in various
oscillator-algebra modules ${\cal M}^\pm\,$. For given ${\cal
M}^\pm$, the corresponding \emph{generalized Schur module}
\bea {\cal S}^\pm&:= &\bigoplus_{\widetilde\l^\pm}
\Comp\otimes \ket{\widetilde\l^\pm}\ ,\eea
where $\ket{\widetilde\l^\pm}$, which we shall refer to as the
\emph{Schur states}, are the ground states of $\widetilde\ml^\pm$
in ${\cal M}^\pm$ with Howe-dual highest weights
$\widetilde\l^\pm=\{\widetilde\l^\pm_i\}_{i=1}^{\n_\pm}\,$. By
making a canonical choice of the Borel subalgebra for
$\widetilde\ml^\pm$, the Schur states can be chosen to obey
\bea\forall \e &:&(N^i_j-\d^i_j \widetilde
\l^\pm_i)\ket{\widetilde\l^\pm}\ =\ 0\qquad \mbox{for $i\leqslant
j$} \quad\mbox{(no sum on}\;i\mbox{)} \ ,\label{A1gs1}
\\[5pt]\e=\pm 1&:& T_{ij}\ket{\widetilde\l^\pm}
\ =\ 0\ .\label{A1gs2}\eea
We also define the shifted Howe-dual highest weights (see also \eq{A1li+} and \eq{A1li-} below)
\bea \widetilde w^\pm_i&:=& \widetilde\l^\pm_i\mp \ft{D}2\ .\eea
The Schur states $\ket{\widetilde\l^\pm}$ generate lowest-weight
spaces ${\cal D}^\pm(\widetilde\l^\pm)$ of $\widetilde \ml^\pm$,
and
\bea \left.{\cal
M}^\pm\right|_{\widetilde\ml^\pm}&=&\bigoplus_{\widetilde\l^\pm\in\widetilde\L^\pm}{\rm
mult}(\widetilde\l^\pm){\cal D}^\pm(\widetilde\l^\pm)\ ,\eea
where ${\rm mult}(\widetilde\l^\pm)\in\mathbb N$ are
multiplicities. For simplicity, we assume that ${\cal M}^\pm$ has
a non-degenerate inner product and that the $\widetilde \ml^\pm$
action on $\ket{\widetilde\l^\pm}$ does not yield any singular
vectors.

By construction an invariant polynomial $C[\ml]\in{\cal U}[\ml]$,
the enveloping algebra of $\ml$, can be rewritten as an invariant
polynomial $C[\widetilde \ml^\pm]\in{\cal U}[\widetilde \ml^\pm]$,
and hence assumes a fixed value, $C[\ml|\widetilde\l^\pm]$ say, in
${\cal D}^\pm(\widetilde\l^\pm)\,$. Hence, ${\cal
D}^\pm(\widetilde\l^\pm)$ decomposes under $\ml$ into
\bea \left.{\cal
D}^\pm(\widetilde\l^\pm)\right|_{\ml}&=&\bigoplus_{\l\in\L(\widetilde\l^\pm)}
{\rm mult}^\pm(\l|\widetilde\l^\pm) {\cal
D}^\pm(\l|\widetilde\l^\pm)\ , \eea
where $\L(\widetilde\l^\pm)$ contains the labels $\l$ of all
$\ml$-irreps ${\cal D}^\pm(\l|\widetilde\l^\pm)$ obeying
$C[\ml|\l]=C[\ml|\widetilde\l^\pm]$ for all invariants $C$, and
${\rm mult}^\pm(\l|\widetilde\l^\pm)\in\{0,1,\dots\}$ are
multiplicities. Consequently,
\bea \left.{\cal S}^\pm\right|_{\ml}&=&
\bigoplus_{\widetilde\l^\pm} \bigoplus_{\l\in\L(\widetilde\l^\pm)}
\bigoplus_{\mu=1}^{{\rm mult}^\pm(\l| \widetilde\l^\pm)}
\Comp\otimes \ket{\l|\widetilde\l^\pm;\mu}\ ,\eea
where $\ket{\l|\widetilde\l^\pm;\mu}\in{\cal
D}^\pm(\l|\widetilde\l^\pm)$ are the Schur states and the index
$\mu$ labels the degeneracy of the construction. If ${\cal
D}(\widetilde\l^\pm)$ decomposes into finite-dimensional irreps of
$\ml$, then the spectrum of invariants $\{C\}$ is sufficiently
large to fix $\l$ uniquely in terms of $\widetilde\l^\pm$, and
hence only the multiplicity remains a free parameter. In what
follows, one useful Howe-duality relation is that of the quadratic
Casimir operators
\bea \hspace{-0.5cm} C_{_2}[\mgl(D;\Comp)]&:=&M^a_b M^b_a\ ,\quad
C_{_2}[\mso(D;\Comp)]\ :=\ \ft12 M^{ab}M_{ab}\ ,\quad
C_{_2}[\msp(D;\Comp)]\ :=\ \ft12 M^{ab} M_{ab}\qquad\eea
that assume the values
\bea C_{_2}[\ml|\widetilde\l^\pm]&=&\sum_{i=1}^{\nu_\pm}\widetilde
w^\pm_i(D-\e\pm(\widetilde w^\pm_i+1-2i))\ .\label{A1cas}\eea
%


\subsection{\sc Fock-space realizations}\label{Sec:Fock}


Acting with the oscillators on a state $\ket{0}$ obeying
$\alpha_{i,a}\ket{0}=0$ yields the standard Fock space
\bea {\cal F}^\pm_{D;\nu_\pm}&=& \bigoplus_{R=0}^\infty {\cal
F}^\pm_{D;\nu_\pm;R}\ , \eea
where ${\cal F}^\pm_{D;\nu_\pm;R}=\left\{ \ket{X}\ :\quad
\left(\sum_i N^i_i-R\right)\ket{X}\ =\ 0\right\}$ are subspaces of
states of fixed rank $R\,$. These spaces have dimensions
\bea \dim{\cal F}^\pm_{D;\nu_\pm;R}&=&{1\over R!}
D^\pm(D^\pm\pm1)\cdots (D^\pm\pm (R-1))\ ,\qquad D^\pm\ :=\
D\nu_\pm\ .\label{A1dim}\eea
We note that for fermionic oscillators, $\dim{\cal
F}^-_{D;\nu_-;R}=0$ if $R>D^-$ and $\dim {\cal
F}^-_{D;\nu_-}=2^{D^-}\,$. The Fock space ${\cal
F}^\pm_{D;\nu_\pm}$ decomposes under $\ml\times \widetilde
\ml^\pm$ as follows:
\bea \left.{\cal F}^\pm_{D;\nu_\pm}\right|_{\ml\times \widetilde
\ml^\pm}&=&\bigoplus_{\D} {\rm mult}^\pm_\e(\D|D; \nu_\pm)~{\cal
D}^\pm(\l(\D)|\widetilde\l^\pm(\D))\ ,\label{A1dec}\eea
where the sum runs over all possible Young diagrams $\D$ (including the trivial diagram)
and\footnote{The standard Fock space can be equipped with the
positive definite inner product. From $||\widehat
N^i_j\ket{\D}||^2\geqslant 0$ it follows that
$\widetilde\l^\pm_i-\widetilde\l^\pm_j\geqslant 0$ if $i<j$ with
equality iff $N^i_j\ket{\D}= N^j_i\ket{\D}= 0\,$. One also notes
that
\bea (M^a_b)^\dagger&=& M^b_a\ ,\qquad (M_{ab})^\dagger \ =\
M^{ab}\ \equiv
J^{ac}J^{bd} M_{cd}\ ,\\[5pt]
( N^i_j)^\dagger&=& N^j_i\ ,\qquad (T_{ij})^\dagger\ =\ \pm
\overline T^{ij}\ ,\eea
using $(J_{ab})^\ast=J^{ab}\,$. The Fock space thus
decomposes into unitary finite-dimensional tensorial
representations of the compact real form of $\ml$, \emph{i.e.}
$\mathfrak{u}(D)$, $\mso(D)$ with $\eta_{ab}=\d_{ab}$, and
$\musp(D)=\msp(D; \Comp)\bigcap \mathfrak{u}(D)\,$. For the
Howe-dual algebra one finds unitary infinite-dimensional
representations of the maximally split non-compact real form of
$\widetilde \ml^+$ and unitary finite-dimensional representations
of the compact real form of $\widetilde \ml^-$, \emph{i.e.}
\bea \ba{lll}\ml&\widetilde \ml^+&\widetilde \ml^-\\[8pt] \mathfrak{u}(D)&
\mathfrak{u}(\nu_+)&\mathfrak{u}(\nu_-)\\[5pt]
\mso(D)&\msp(2\nu_+)&\mso(2\nu_-)\\[5pt] \musp(D)\qquad& \mso(\nu_+,\nu_+)\qquad&
\msp(2\nu_-)\ea\ .\eea
Generalized Fock spaces can be built on anti-vacua that are
annihilated by $\bar\a^{i,a}$ for some values of $a\,$. In the
case of bosonic oscillators, these modules have a non-degenerate
inner product matrix with alternating signature that yields
unitary representations of the non-compact real forms of $\ml\,$.
The various Fock-space realizations are subsumed into the Moyal
quantization of the oscillator algebra (see, for example,
\cite{Iazeolla:2008ix}).}
\bea \l_i(\D)&=& w_i\ ,\qquad i=1,\dots,D\ ,\label{A1li}\\[5pt]
\widetilde\l^+_i(\D)&=& \ft{D}2+w_i\ ,\quad  w_i\ =\ \widetilde w^+_i
\ ,
\qquad i=1,\dots,\nu_+\ ,\label{A1li+}\\[5pt] \widetilde\l^-_i(\D)&=&-\ft{D}2+h_i\
,\quad  h_i\ =\ \widetilde w^-_i\ ,\qquad i=1,\dots,\nu_-\
,\label{A1li-}\eea
with $w_i=w_i(\D)$ and $h_i=h_i(\D)$ being the number of cells in
the $i$th row and column of $\D$, respectively. We shall say that
$\D$ contains a block of height $h$ between the $i$th and
$(i+h-1)$th rows if $\widetilde w^\pm_i=\cdots=\widetilde
w^\pm_{i+h-1}$, and we define the transpose $\D^{\rm T}$ of $\D$
to be the Young diagram with $h_i(\D^{\rm T})=w_i(\D)$ (and hence
$w_i(\D^{\rm T})=h_i(\D)$).

As we shall demonstrate below, the multiplicities
\bea {\rm mult}^\pm_\e(\D|D;\nu_\pm)&\in& \{0,1\}\
.\label{A1mult}\eea
The Fock space realization is thus completely free of degeneracy
in the sense that the correspondence $\l\leftrightarrow\widetilde
\l^\pm$ is one-to-one and each dual pair $(\l|\widetilde\l^\pm)$
arises exactly once. Correspondingly, the decomposition of the
Schur module reads
\bea {\cal S}^\pm_{D;\nu_\pm}&=& \bigoplus_{\D} {\rm
mult}^\pm_\e(\D|D;\nu_\pm)~( \Comp\otimes \ket{\D})\ , \qquad
\ket{\D}\ =\ \ket{\l(\D)|\widetilde\l^\pm(\D)}\ .\eea

Let us examine more carefully the determination of \eq{A1mult}.

\subsection{\sc{Schur modules for $\mgl(D;\Comp)$}}
\label{subsec:thecaseof}

In the case of $\ml=\mgl(D;\Comp)$, both $\ml$ and $\widetilde
\ml^\pm$ leave ${\cal F}_{D;\nu_\pm;R}$ invariant, and
\bea {\cal F}_{D;\nu_\pm;R}&=&\bigoplus_{\D: ~{\rm rank}(\D)=R}
{\rm mult}^\pm_{_0}(\D|D;\nu_\pm) ~{\cal D}^\pm(\l(\D)|\widetilde
\l^\pm(\D))\ , \label{A1dec2}\eea
where the highest-weights are given by \eq{A1li}--\eq{A1li-} and
the multiplicities
\bea {\rm mult}^+_{_0}(\D|D;\nu_+)&=&\mx{\{}{ll}{0&\mbox{if
$h_1>\min(\nu_+,D)$}\ ,
\\[5pt] 1&\mbox{else}}{.}\label{A1mult+}\\[10pt]
{\rm mult}^-_{_0}(\D|D;\nu_-)&=&\mx{\{}{ll}{0&\mbox{if $h_1>D$ or
$w_1>\nu_-$}\ ,
\\[5pt]1&\mbox{else}}{.}\ .\label{A1mult-}\eea
The vanishing conditions follow immediately from the statistics of
the oscillators. To show that the non-vanishing multiplicities are
equal to $1$, one may use dimension formulae or directly decompose
${\cal F}^\pm_{D;\nu_\pm;R}$ under $\mgl(D;\Comp)\,$.

\begin{center}{\it Calculation of Multiplicities Using Dimension
Formulae}\end{center}

The total dimension of the right-hand side of \eq{A1dec2} is given
by
\bea d^\pm_R(D;\nu_\pm)&=&\sum_{\D} {\rm
mult}^\pm_{_0}(\D|D;\nu_\pm)
~d^\pm(\D|D;\nu_\pm)\ ,\\[5pt] d^\pm(\D|D;\nu_\pm)&=&\dim(\mgl(D)|\D)
~\dim(\mgl(\nu_\pm)|\widetilde \D^\pm)\ ,\label{A1sum}\eea
where the dual Young diagrams
\bea \widetilde \D^+&=&\D\ ,\qquad \widetilde \D^-\ =\ \D^{\rm T}\
,\eea
and
\bea \dim(\mgl(N)|\D)&=& {\prod_{(i,j)\in \D} (N+i-j)\over |\D|}\
,\qquad |\D| \ =\ \prod_{(i,j)\in \D} (w_i+h_j-i-j+1)\ ,\eea
which vanishes in case the height of $\D$ exceeds $N$ (and it is
invariant under insertions and removals of columns of height $N$
although this property is not needed here). Thus
$d^\pm(\D|D;\nu_\pm)$ vanishes iff ${\rm
mult}^\pm_{_0}(\D|D;\nu_\pm)$ vanishes. Moreover, the denominators
on the right-hand side of \eq{A1sum} are equal, and
\bea d^\pm(\D|D;\nu_\pm)&=& {\prod_{(i,j)\in \D}(D-i+j)
(\nu_\pm\pm i\mp j)\over |\D|^2}\ =\ \sum_{m,n=0}^R
d^\pm_{m,n}(\D) D^m (\nu_\pm)^n\ ,\label{A1dimD}\eea
with $d^\pm_{R,R}(\D)=1/|\D|^2\,$. Thus, from the sum rule
\bea \sum_{\D}{1\over |\D|^2}&=& {1\over R!}\
,\label{A1sumrule}\eea
which is a consequence of the formula giving the decomposition of
the regular representation of the symmetric group $S_R$ in irreps
and of the fact that the dimension of the irrep associated with
$\Delta$ is $R!/|\Delta|\,$, it follows that the total dimension
$d^\pm_R(D;\nu_\pm)$ is a polynomial in $D$ and $\nu_\pm$ with
leading behavior given by
\bea d^\pm_R(D;\nu_\pm)&=&
\ft1{R!}(D^\pm)^R(1+\alpha)+(\mbox{terms of lower order in $D$ and
$\nu_\pm$})\ ,\eea
for some non-negative integer $\alpha\,$. Then, it results that
\eq{A1mult+} and \eq{A1mult-} must hold in order to reproduce the
leading behavior of (\ref{A1dim}), \emph{i.e.} $\alpha=0\,$. We
note that the sub-leading coefficients contain generalizations of
the sum rule \eq{A1sumrule}.

\begin{center}{\it Direct Decomposition of
${\cal F}^\pm_{D;\nu_\pm;R}$}\end{center}

In the case of bosonic oscillators, the monomial
\bea \ket{(m_1)\otimes\cdots\otimes
(m_{\nu_+})}&=&\bar\a^{1,a_1(m_1)}\cdots
\bar\a^{\nu_+,a_{\nu_+}(m_{\nu_+})}\ket{0}\ ,\qquad
\sum_{i=1}^{\nu_+} m_i=R\ ,\eea
where $\bar\a^{i,a_i(m_i)}=\bar\a^{i,a_{i,1}}\cdots
\bar\a^{i,a_{i,m_i}}$, decomposes under $\mgl(D;\Comp)$ into
\bea \ket{(m_1)\otimes\cdots\otimes
(m_{\nu_+})}&=&\sum_{\{p_{ij}\}} \prod_{1\leqslant i<j\leqslant
\nu_+}~(N^j_i)^{p_{ij}}\ket{\D}\ , \label{A1dec2bis}\eea
where: \emph{i)} $\ket{\D}$ are carry $\mgl(D;\Comp)$-irreps
labelled by admissible Young diagrams $\D$; \emph{ii)} $\ket{\D}$
are Schur states obeying \eq{A1gs1} with $\widetilde\l^+_i$ given
by \eq{A1li+}; and \emph{iii)} $\{p_{ij}\}$ are sets of integers
$p_{ij}\in\{0,1,2,\dots\}$ that parameterize the numbers of cells
that are lifted from the $j$th row to the $i$th row in applying
the Littlewood-Richardson rule to $(m_1)\otimes\cdots\otimes
(m_{\nu_+})$. It follows that
\bea w_i&= & m_i+\sum_{i<j}p_{ij}-\sum_{j<i}p_{ji}\ ,\qquad
i=1,\dots,\nu_+\ ,\eea

which imply that $w_i$ obey the admissibility conditions
\bea w_i&\geqslant&w_{i+1}\ ,\qquad w_i\ =\ 0\quad\mbox{for
$i>\min(D,\nu_+)$}\ . \eea
The states $\prod_{1\leqslant i<j\leqslant \nu_+}~(N^j_i)^{p_{ij}}
\ket{\D}$ belong to the $\D$-plet of $\mgl(D;\Comp)$ for all
admissible $\{p_{ij}\}$, while they are Schur states iff
$p_{ij}=0\,$. Hence the decomposition \eq{A1dec2bis} contains a
Schur state iff $m_1\geqslant m_{_2}\geqslant
m_{\min(D,\nu_+)}\geqslant 0$ and $m_i=0$ for $i>\min(D,\nu_+)$,
in which case its multiplicity is given by $1$, which shows
\eq{A1mult+}.

Similarly, the case of fermionic oscillators, the monomial
\bea \ket{[m_1]\otimes \cdots\otimes [m_{\nu_-}]}&=&
\bar\a^{1,a_1[m_1]}\cdots
\bar\a^{\nu_-,a_{\nu_-}[m_{\nu_-}]}\ket{0}\ ,\qquad
\sum_{i=1}^{\nu_-} m_i=R\ ,\eea
decomposes under $\mgl(D;\Comp)$ into
\bea \ket{[m_1]\otimes\cdots\otimes
[m_{\nu_-}]}&=&\sum_{\{p_{ij}\}}\prod_{1\leqslant i<j\leqslant
\nu_-} (N^j_i)^{p_{ij}}\ket{\D}\ ,\label{A1dec3}\eea
where $\ket{\D}$ carry the $\D$-plet of $\mgl(D;\Comp)$ and obey
\eq{A1gs1}, and
\bea h_i&=& m_i+\sum_{i<j}p_{ij}-\sum_{j<i}p_{ji}\ ,\qquad
i=1,\dots,\nu_+\ ,\eea
subject to the admissibility conditions
\bea D\geqslant h_i&\geqslant&h_{i+1}\ ,\qquad h_i\ =\
0\quad\mbox{for $i>\nu_-$}\ .\eea
Hence Schur states arise in \eq{A1dec3} iff $p_{ij}=0$, in which
case their multiplicity is given by $1$, from which \eq{A1mult-}
follows.

\subsection{\sc{Schur modules for $\mso(D;\Comp)$ and $\msp(D;\Comp)$}}

The actions of $\mso(D;\Comp)$ ($\e=+1$) and $\msp(D;\Comp)$
($\e=-1$) leave ${\cal F}^\pm_{D;\nu_\pm;R}$ invariant, while
their Howe duals act in representations that in general range over
more than one value of $R\,$. Correspondingly, for fixed $R$ the
$\mgl(D)$-irreps in ${\cal F}^\pm_{D;\nu_\pm;R}$ decompose into
$J$-traceless states obeying
\bea T_{ij} \ket{\D}&=&0\ ,\eea
and $J$-traces, \emph{i.e.} states in the image of $\overline
T^{ij}\,$. Using the fermionic oscillators, \emph{i.e.} the
anti-symmetric basis of Young projectors, one can show that
\bea h_i+h_j+2\,t_{ij}&\leqslant &D\quad \mbox{for}\quad
\mx{\{}{ll}{i\neq j&\mbox{if
$\ml=\mso(D;\Comp)$}\ \\[5pt]\mbox{all $i,j$}&\mbox{if $\ml=\msp(D;\Comp)$}}{.}\ ,
\label{A1trcond}\eea
where $t_{ij}$ denote the total number of traces that have been
inserted into columns $i$ and $j\,$. The same conditions must hold
also in the case of bosonic oscillators. Thus,
\bea {\cal F}^\pm_{D;\nu_\pm}&=& \bigoplus_{\D}{\rm
mult}^\pm_\e(\D|D;\nu_\pm) ~{\cal
D}^\pm(\l(\D)|\widetilde\l^\pm(\D))\ ,\eea
where the highest weights of $\ml$ and $\widetilde \ml^\pm$ are
given by \eq{A1li}--\eq{A1li-} and
\bea {\rm mult}^\pm_{\e}(\D|D;\nu_-)&=&{\rm
mult}^\pm_{_0}(\D|D;\nu_-) ~\th_\e(\D)\ , \label{A1mult2}\eea
with $\th_\e(\D)$ accounting for the condition \eq{A1trcond} in
the case that $t_{ij}=0$, \emph{i.e.}
\bea \th_\e(\D)&=& \mx{\{}{ll}{1&\mbox{if \eq{A1trcond} holds for
$t_{ij}=0$}\\[5pt]0&\mbox{else}}{.} \ .\eea
We note that for $\mso(D;\Comp)$ the highest weight
$\widetilde\l^-_1$ of $\widetilde \ml^-=\mso(2\nu_-)$ may become
negative, in which case one may redefine the \eq{A1grading} by
normal-ordering the Howe-dual generators with respect to
$\prod_{a=1}^D\bar\a^{1,a}\ket{0}$ (instead of $\ket{0}$), which
leads to an exchange of $h_1$ by $D-h_1$ and hence $\widetilde
\l^-_1$ by $-\widetilde\l^-_1\,$. We also note that if $\nu_-=2$
then the Schur states of ${\cal F}^-_{D;2;D\nu_-/2}$ are
annihilated by both $T_{12}$ and $\overline T^{12}$ and hence obey
$h_1+h_{_2}=D$, although they form singlets of $\widetilde \ml^-$
only if $h_1=h_{_2}=D/2$ and $D$ is even.
%

\section{\sc \large Radial reduction of the background connection}\label{App:reduc}

We denote the $\miso(2,D-1)$-covariant derivative on
$\widehat{\cal M}_{D+1}$ by
\bea \widehat {\cal D}&:=& d-i(\widehat E^A\widehat \Pi_A+\frac12
\widehat\O^{AB}\widehat M_{AB})\ ,\eea
where $\widehat \Pi_A$ are the translation generators,
$\widehat{E}^A$ and $\widehat{\Omega}^{AB}$ are the vielbein and
$\mso(2,D-1)$-valued connection, respectively. The connection is
flat if\footnote{Although not used here, we note that the flat
vielbein can be expressed locally as $\widehat E^A=\widehat\nabla
\widehat V^A$. In foliations with maximally symmetric leaves and
constant $\widehat\x^A$, the gauge function can be chosen to be
$\widehat V^A=\l^{-1} \widehat\x^A$. } $\widehat
T^A:=\widehat{\nabla}\widehat{E}^A\ :=\ d\widehat E^A
+\widehat{\O}^{AB} \widehat{E}_B\approx0$ and $\widehat R^{AB}
:=d\widehat{\Omega}^{AB} +\widehat{\Omega}^{AC}
\widehat{\Omega}_C{}^B \approx0$. A local foliation of $\widehat{{\cal
M}}_{D+1}$, as defined in Section I.3.7, induces a splitting
\bea \widehat E^A&:=&\widehat e^A+N\widehat\x^A\ ,\qquad
\widehat\O^{AB}\ :=\ \widehat\o^{AB}+N\widehat\L^{AB}\ ,\eea
where $\widehat\x^A:=i_\x \widehat E^A$ and $\widehat\L^{AB}:=i_\x
\widehat\O^{AB}$, which implies $i_\x\widehat e^A=0$ and
$i_\x\widehat\o^{AB}=0\,$. Upon defining
\bea \widehat D&:=&d-\frac i2\widehat\o^{AB}\widehat M_{AB}\ ,\eea
the flatness conditions decompose into components that are
transverse and parallel to $i_N$ as follows
\bea &&(\widehat D-N{\cal L}_\x) \widehat e^A\ \approx\ 0\ ,\qquad (\widehat D-N{\cal L}_\x)\widehat\x^A-{\cal L}_\x\widehat e^A-\widehat\L^{AB} \widehat e_B\ \approx\ 0\ ,\\[5pt]
&&d\widehat\o^{AB}+\widehat\o^{AC}\widehat\o_C{}^B-N{\cal
L}_\x\widehat\o^{AB}\ \approx\ 0\ ,\qquad (\widehat D-N{\cal
L}_\xi)\widehat\L^{AB}-{\cal L}_\x\widehat\o^{AB}\ \approx\ 0\
.\eea
There remains a manifest covariance under $O(2,D-1)$ gauge
transformations with parameters annihilated by ${\cal L}_\x\,$. We
denote this gauge group by $O(2,D-1)_{\rm leaf}\,$. Maximally
symmetric leaves arise from foliations obeying
\bea \widehat\L^{AB}&\stackrel{!}{=}&0\ ,\quad {\cal L}_\x \widehat e^A\
\stackrel{!}{=}\ \l(L)\widehat e^A\ ,\quad {\cal L}_\xi \widehat\x^A\
\stackrel{!}{=}\ 0\ ,\label{maxsymmcond}\eea
which implies $\widehat D\equiv \widehat\nabla$ and the local
relations
\bea &&\widehat D \widehat e^A\ \approx \l N\widehat e^A\ ,\quad \widehat\x_A\widehat e^A\ \approx\ 0\ ,\quad\widehat D\widehat\x^A\ \approx\ \l\widehat e^A\ ,\quad d\l+\l^2 N\ \approx\ 0\ ,\\[5pt]&&d\widehat\o^{AB}+\widehat\o^{AC}\widehat\o_C{}^B\ \approx\ 0\ ,\quad {\cal L}_\x\widehat\o^{AB}\ \approx\ 0\ ,\quad ({\cal L}_\x)^2\widehat e^A\ \approx\ 0\ .\eea
One may choose
\bea \l&=& L^{-1}\ ,\eea
and use local $O(2,D-1)_{\rm leaf}$-symmetry to bring $\widehat\x^A$ to
a locally constant vector, \emph{i.e.}
\bea d\widehat\xi^A&\stackrel{!}{=}&
0\qquad \mbox{(gauge-fix $O(2,D-1)_{\rm leaf}$)} \eea
whose residual local symmetry group we denote by $G_{\rm
leaf}({\widehat\x}^2)$. The global decomposition is
\bea\widehat{\cal M}_{D+1}&=&\widehat{\cal
M}^{~(-1)}_{D+1}\cup\widehat{\cal M}^{~(0)}_{D}\cup\widehat{\cal
M}^{~(1)}_{D+1}\ ,\eea
where $\widehat{\cal M}^{~(k)}_{D(k)}$ are regions of dimension
$D(k)$ foliated with maximally symmetric leaves with
$\widehat\x^2=k$ and local $G_{\rm leaf}(k)$ symmetry. In
$\widehat{\cal M}^{~(-1)}_{D+1}$ the projector
$\widehat\x_A\mathbb P^A_B:=0$, $\mathbb P^A_B\widehat\x^B:=0$
obeys $\mathbb P^A_B:=(0,\mathbb P^a_B)$ where the index $a$
transforms as a vector under residual local $G_{\rm leaf}(-1)\cong
O(1,D-1)_{\rm leaf}$ transformations. Defining
\bea \o^{AB}& :=& \widehat\o^{AB}+\l(\widehat e^A\widehat\x^B
-\widehat\x^A \widehat e^B)\ ,\eea
then the local relations imply that ($k=\widehat \x^2=-1$)
\bea \widehat\x_A\o^{AB}\ \approx\ 0\ ,\quad
d\o^{AB}+\o^{AC}\o_C{}^B+\l^2 \widehat e^A \widehat e^B\ \approx\ 0\
,\eea
and one identifies the leaves as $AdS_D$ spacetimes of radius $L$
with canonical flat $\mso(2,D-1)$-valued connections
\bea e^a&:=&i_L^\ast~ \mathbb P^a_A ~\widehat e^A\ ,\quad \o^{ab}\
:=\ i^\ast_L  ~\mathbb P^{a}_A ~\mathbb P^{b}_B ~\o^{AB} ,\eea
as defined in \eq{canonicalconnection}. Skvortsov's master-field
equations contain the $\miso(2,D-1)$-covariant derivatives
($i=1,\dots,\nu$)
\bea \widehat {\cal D}_{(i)}&:=&\widehat \nabla-i\widehat E_{(i)}\ =\ d-\frac{i}2 \widehat \O^{AB}\widehat M_{AB}-i\widehat E^A\widehat{\beta}_{A,(i)}\\[5pt]&=&d-\frac{i}2~\left(\omega^{AB}+2\l\widehat\x^A \widehat e^B\right)\widehat M_{AB}-i\widehat e^A\widehat\beta_{A,(i)}-iN\widehat\x_{(i)}\ ,\eea
where $\widehat\x_{(i)}:=\widehat\x^A\widehat{\beta}_{A,(i)}\,$. Radial
reduction can be analyzed directly on $\widehat{\cal M}_{D+1}$
using
\bea \widehat {\cal D}_{(i)}&:=& \widehat\nabla-i\widehat e_{(i)}
-iN\widehat\x_{(i)}\ ,\qquad e_{(i)}\ :=\  \widehat
e^A\widehat\b_{A,(i)}\ ,\eea
%
%
whilst the harmonic expansion and flat limit can be analyzed on
$AdS_D(L)$ using
\bea {\cal D}_{(i)}&:=&i_L^\ast\widehat{\cal D}_{(i)}\ :=\ \nabla
-i e^a P_{a,(i)}\ ,\qquad P_{a,(i)}\ :=\ \l\widehat\x^B\widehat
M_{Ba}+\widehat\b_{a,(i)}\ ,\eea
where $[\widehat\b_{A,(i)}, \widehat\b_{B,(i)}]=0$ and
$[P_{a,(i)},P_{b,(i)}]=i\l^2\widehat M_{ab}$.

\section{\sc \large Tensorial content of the $\s^-$-chains with $h_1=1$ and $q+g=2,3$ }\label{App:q+g=3}
%
The $\mm$-content of the $\s^-$-chain in the case of $h_1=1$,
$s_1-s_{_2}\geqslant 4$ is given for $q+g=2$ by
\bea X^{\bf 1}(R_{]a(2)})&\in&
\mx{[}{c}{s_1\\\Xi~\widetilde\otimes(2)}{]}\oplus
\mx{[}{c}{s_1-1\\\Xi~\widetilde\otimes(3)}{]}\oplus
\mx{[}{c}{s_1-1\\\Xi~\widetilde\otimes(2,1)}{]}\oplus
\mx{[}{c}{s_1-2\\\Xi~\widetilde\otimes(2)}{]}\nn\\[5pt]&&\oplus
\mx{[}{c}{s_1-1\\i_{[1]}\Xi~\widetilde\otimes(2)}{]}\oplus
\mx{[}{c}{s_1-1\\\Xi~\widetilde\otimes(1)}{]}\ ,\\[5pt]
R^{[2]}(R_{\a(1)})&\in&
\mx{[}{c}{s_1\\\Xi~\widetilde\otimes(2)}{]}\oplus
\mx{[}{c}{s_1\\\Xi~\widetilde\otimes[2]}{]}\oplus
\mx{[}{c}{s_1-1\\\Xi~\widetilde\otimes(2,1)}{]}\oplus
\mx{[}{c}{s_1-1\\\Xi~\widetilde\otimes[3]}{]}\nn\\[5pt]&&\oplus
\mx{[}{c}{s_1-1\\\Xi~\widetilde\otimes(1)}{]}\oplus
\mx{[}{c}{s_1-2\\\Xi~\widetilde\otimes(2)}{]}\oplus
\mx{[}{c}{s_1-2\\\Xi~\widetilde\otimes[2]}{]}\oplus
\mx{[}{c}{s_1\\i_{[1]}\Xi~\widetilde\otimes(1)}{]}\nn\\[5pt]&&\oplus
\mx{[}{c}{s_1\\\Xi}{]}\oplus
\mx{[}{c}{s_1-1\\i_{[1]}\Xi~\widetilde\otimes(2)}{]}\oplus
\mx{[}{c}{s_1-1\\i_{[1]}\Xi~\widetilde\otimes[2]}{]}\oplus
\mx{[}{c}{s_1-1\\\Xi~\widetilde\otimes(1)}{]}\oplus
\mx{[}{c}{s_1-2\\i_{[1]}\Xi~\widetilde\otimes(1)}{]}\nn\\[5pt]&&\oplus
\mx{[}{c}{s_1-2\\i_{[1]}\Xi}{]}\oplus
\mx{[}{c}{s_1-1\\i_{[2]}\Xi~\widetilde\otimes(1)}{]}\oplus
\mx{[}{c}{s_1-1\\i_{[1]}\Xi}{]}\ ,\\[5pt]
Z^{[3]}(R_{\a(0)})&\in&
\mx{[}{c}{s_1\\\Xi~\widetilde\otimes[2]}{]}\oplus
\mx{[}{c}{s_1-1\\\Xi~\widetilde\otimes[3]}{]}\oplus
\mx{[}{c}{s_1-2\\\Xi~\widetilde\otimes[2]}{]}\oplus
\mx{[}{c}{s_1-1\\\Xi~\widetilde\otimes(1)}{]}\nn\\[5pt]&&\oplus
\mx{[}{c}{s_1-1\\i_{[1]}\Xi}{]}\oplus
\mx{[}{c}{s_1-2\\i_{[1]}\Xi~\widetilde\otimes(1)}{]}\oplus
\mx{[}{c}{s_1\\i_{[2]}\Xi}{]}\oplus
\mx{[}{c}{s_1-1\\i_{[2]}\Xi~\widetilde\otimes(1)}{]}\nn\\[5pt]&&\oplus
\mx{[}{c}{s_1-1\\i_{[1]}\Xi~\widetilde\otimes[2]}{]}\oplus
\mx{[}{c}{s_1\\i_{[1]}\Xi~\widetilde\otimes(1)}{]}\ ,\eea
and for $q+g=3$ by
\bea X^{\bf
1}(R_{\a(3)})&\in&\mx{[}{c}{s_1\\\Xi~\widetilde\otimes(3)}{]}\oplus
\mx{[}{c}{s_1-1\\\Xi~\widetilde\otimes(4)}{]}\oplus
\mx{[}{c}{s_1-1\\\Xi~\widetilde\otimes(3,1)}{]}\oplus
\mx{[}{c}{s_1-2\\\Xi~\widetilde\otimes(3)}{]}\nn\\[5pt]&&\oplus
\mx{[}{c}{s_1-1\\i_{[1]}\Xi~\widetilde\otimes(3)}{]}\oplus
\mx{[}{c}{s_1-1\\\Xi~\widetilde\otimes(2)}{]}\ ,\\[5pt]
R^{[2]}(R_{\a(2)})&\in&
\mx{[}{c}{s_1\\\Xi~\widetilde\otimes(3)}{]}\oplus
\mx{[}{c}{s_1\\\Xi~\widetilde\otimes(2,1}{]}\oplus
\mx{[}{c}{s_1-1\\\Xi~\widetilde\otimes(3,1)}{]}\oplus
\mx{[}{c}{s_1-1\\\Xi~\widetilde\otimes[3,1]}{]}\nn\\[5pt]&&\oplus
\mx{[}{c}{s_1-1\\\Xi~\widetilde\otimes(2)}{]}\oplus
\mx{[}{c}{s_1-2\\\Xi~\widetilde\otimes(3)}{]}\oplus
\mx{[}{c}{s_1-2\\\Xi~\widetilde\otimes[2,1]}{]}\oplus
\mx{[}{c}{s_1\\i_{[1]}\Xi~\widetilde\otimes(2)}{]}\nn\\[5pt]&&\oplus
\mx{[}{c}{s_1-1\\i_{[1]}\Xi~\widetilde\otimes(3)}{]}\oplus
\mx{[}{c}{s_1-1\\i_{[1]}\Xi~\widetilde\otimes[2,1]}{]}\oplus
\mx{[}{c}{s_1\\\Xi~\widetilde\otimes(1)}{]}\oplus
\mx{[}{c}{s_1-1\\\Xi~\widetilde\otimes(2)}{]}\nn\\[5pt]&&\oplus
\mx{[}{c}{s_1-1\\\Xi~\widetilde\otimes[2]}{]}\oplus
\mx{[}{c}{s_1-2\\i_{[1]}\Xi~\widetilde\otimes(2)}{]}\oplus
\mx{[}{c}{s_1-2\\\Xi~\widetilde\otimes(1)}{]}\oplus
\mx{[}{c}{s_1-1\\i_{[2]}\Xi~\widetilde\otimes(2)}{]}\nn\\[5pt]&&\oplus
\mx{[}{c}{s_1-1\\i_{[1]}\Xi~\widetilde\otimes(1)}{]}\ ,\\[5pt]
Z^{[3]}(R_{\a(1)})&\in&
\mx{[}{c}{s_1\\\Xi~\widetilde\otimes(2,1)}{]}\oplus
\mx{[}{c}{s_1\\\Xi~\widetilde\otimes[3]}{]}\oplus
\mx{[}{c}{s_1-1\\\Xi~\widetilde\otimes[3,1]}{]}\oplus
\mx{[}{c}{s_1-1\\\Xi~\widetilde\otimes[4]}{]}\nn\\[5pt]&&\oplus
\mx{[}{c}{s_1-1\\\Xi~\widetilde\otimes(2)}{]}\oplus
\mx{[}{c}{s_1-1\\\Xi~\widetilde\otimes[2]}{]}\oplus
\mx{[}{c}{s_1-2\\\Xi~\widetilde\otimes[2,1]}{]}\oplus
\mx{[}{c}{s_1-2\\\Xi~\widetilde\otimes[3]}{]}\nn\\[5pt]&&\oplus
\mx{[}{c}{s_1\\i_{[1]}\Xi~\widetilde\otimes(2)}{]}\oplus
\mx{[}{c}{s_1\\i_{[1]}\Xi~\widetilde\otimes[2]}{]}\oplus
\mx{[}{c}{s_1-1\\i_{[1]}\Xi~\widetilde\otimes[2,1]}{]}\oplus
\mx{[}{c}{s_1-1\\i_{[1]}\Xi~\widetilde\otimes[3]}{]}\nn\\[5pt]&&\oplus
\mx{[}{c}{s_1-1\\i_{[1]}\Xi~\widetilde\otimes(1)}{]}\oplus
\mx{[}{c}{s_1-2\\i_{[1]}\Xi~\widetilde\otimes(2)}{]}\oplus
\mx{[}{c}{s_1-2\\i_{[1]}\Xi~\widetilde\otimes[2]}{]}\oplus
\mx{[}{c}{s_1-1\\\Xi}{]}\nn\\[5pt]&&\oplus
\mx{[}{c}{s_1-2\\\Xi~\widetilde\otimes(1)}{]}\oplus
\mx{[}{c}{s_1\\\Xi~\widetilde\otimes(1)}{]}\oplus
\mx{[}{c}{s_1-1\\\Xi~\widetilde\otimes[2]}{]}\oplus
\mx{[}{c}{s_1\\i_{[1]}\Xi}{]}\nn\\[5pt]&&\oplus
\mx{[}{c}{s_1-1\\i_{[1]}\Xi~\widetilde\otimes(1)}{]}\oplus
\mx{[}{c}{s_1\\i_{[2]}\Xi~\widetilde\otimes(1)}{]}\oplus
\mx{[}{c}{s_1-1\\i_{[2]}\Xi~\widetilde\otimes(2)}{]}\oplus
\mx{[}{c}{s_1-1\\i_{[2]}\Xi~\widetilde\otimes[2]}{]}\nn\\[5pt]&&\oplus
\mx{[}{c}{s_1-2\\i_{[2]}\Xi~\widetilde\otimes(1)}{]}\oplus
\mx{[}{c}{s_1-2\\i_{[1]}\Xi}{]}\oplus
\mx{[}{c}{s_1-1\\i_{[3]}\Xi~\widetilde\otimes(1)}{]}\oplus
\mx{[}{c}{s_1-1\\i_{[2]}\Xi}{]}\ ,\\[5pt]
Z^{[4]}_3(R_{\a(0)})&\in&
\mx{[}{c}{s_1\\\Xi~\widetilde\otimes[3]}{]}\oplus
\mx{[}{c}{s_1-1\\\Xi~\widetilde\otimes[4]}{]}\oplus
\mx{[}{c}{s_1-1\\\Xi~\widetilde\otimes[2]}{]}\oplus
\mx{[}{c}{s_1-2\\\Xi~\widetilde\otimes[3]}{]}\nn\\[5pt]&&\oplus
\mx{[}{c}{s_1\\i_{[1]}\Xi~\widetilde\otimes[2]}{]}\oplus
\mx{[}{c}{s_1-1\\i_{[1]}\Xi~\widetilde\otimes[3]}{]}\oplus
\mx{[}{c}{s_1-1\\i_{[1]}\Xi~\widetilde\otimes(1)}{]}\oplus
\mx{[}{c}{s_1-2\\i_{[1]}\Xi~\widetilde\otimes[2]}{]}\nn\\[5pt]&&\oplus
\mx{[}{c}{s_1\\i_{[2]}\Xi~\widetilde\otimes(1)}{]}\oplus
\mx{[}{c}{s_1-1\\i_{[2]}\Xi~\widetilde\otimes[2]}{]}\oplus
\mx{[}{c}{s_1-1\\i_{[2]}\Xi}{]}\oplus
\mx{[}{c}{s_1-2\\i_{[2]}\Xi~\widetilde\otimes(1)}{]}\nn\\[5pt]&&\oplus
\mx{[}{c}{s_1\\i_{[3]}\Xi}{]}\oplus
\mx{[}{c}{s_1-1\\i_{[3]}\Xi~\widetilde\otimes(1)}{]}\ .\eea

\section{\sc \large Mixed-symmetry gauge fields and singleton
composites}\label{App:Singleton}
%
The bosonic singletons $\mD_s\equiv D(\e_{_0}+s;([s;h]))$ consist
of states $\ket{e_n;([s+n;1],[s;h-1])}$, $n=0,1,\dots$, of energy
$e_n = \e_{_0} + s + n$ and $\mso(D-1)$ spin $([s+n;1],[s;h-1])$
where $\e_{_0}=h-1=(D-3)/2$ and  $s>0$ requires $D$ to be odd. The
tensor product $\mD_{s_1}\otimes\cdots \otimes\mD_{s_P}$ consists
of states with energy
\bea e &=& \sum_{i=1}^P (\e_{_0}+s_i + n_i)\ ,\eea
and spin
\bea (s_1+n_1 ,s_1,\dots,s_1) \otimes \cdots \otimes
(s_P+n_P,s_P,\dots,s_P) &= &\bigoplus_{t_1,\dots,t_\nu}
(t_1,\dots,t_\nu)\ ,\eea
where $t_1 \leqslant \sum_{i=1}^P ( s_i + n_i)\,$. Thus
\bea e  &\geqslant & P \e_{_0} + t_1\ .\eea
The ground states of unitary\footnote{The tensor product of singletons is unitary and hence cannot contain the non-unitary massless representations with $e<t_1 + D - 2 - h_1$.} massless representations have
\bea e &=& t_1 + D - 2 - h_1\ ,\eea
where $h_1$ is the height of the first block of
$(t_1,\dots,t_\nu)$, \emph{i.e.} $t_1=\cdots=t_{h_1}>
t_{h_1+1}\,$. Such states fit inside $P$-fold product only if
$P\e_{_0}+t_1\leqslant t_1+ D - 2 - h_1$, that is $P \leqslant 2(
D - 2 - h_1)/(D-3)\,$. Since $h_1\geqslant 1$ and $P\geqslant 2$ it follows that
\bea h_1&=&1\ ,\qquad P \ =\  2\ ,\eea
that is, only unitary mixed-symmetry massless fields with $h_1=1$
and with at most $6$ blocks can be singleton composites. Since
$h_1=1$ the corresponding ASV potentials are $1$-forms, for which
there could be a standard non-abelian closure of the gauge algebra.

\end{appendix}

\providecommand{\href}[2]{#2}\begingroup\raggedright\endgroup


\end{document}